\documentclass[12pt]{article}
\usepackage{amsmath,amsfonts,amssymb,psfrag}
\usepackage{mathrsfs, dsfont}
\usepackage{multirow,array,multirow,stmaryrd}
\usepackage{comment}
\usepackage{sectsty}
\usepackage[letterpaper,top=1in,bottom=1.15in,left=1in,right=1in]{geometry}
\usepackage[format=hang,margin=1.5cm,font=small,labelfont=bf]{caption}
\usepackage{subfigure}
\usepackage{color}

\usepackage{ifpdf}
\ifpdf
  \usepackage[pdftex]{graphicx}
  \usepackage{epstopdf}
\else
  \usepackage[dvips]{graphicx}
\fi
\usepackage[bookmarks,breaklinks]{hyperref}		
\usepackage[all,cmtip]{xy}

\numberwithin{equation}{section}
\numberwithin{table}{section}

\sectionfont{\large\bfseries}
\subsectionfont{\normalsize\bfseries}
\subsubsectionfont{\normalsize\it}

\newcommand{\beq}{\begin{equation}}
\newcommand{\eeq}{\end{equation}}
\newcommand{\be}{\begin{equation}} 
\newcommand{\ee}{\end{equation}}
\newcommand{\bea}{\begin{eqnarray}}
\newcommand{\eea}{\end{eqnarray}}   
\newcommand{\ben}{\begin{eqnarray*}}
\newcommand{\een}{\end{eqnarray*}}                  
\newcommand{\ba}{\begin{aligned}}
\newcommand{\ea}{\end{aligned}}
\newcommand{\bt}{\begin{tabular}}
\newcommand{\et}{\end{tabular}}
\newcommand{\bc}{\begin{center}}
\newcommand{\ec}{\end{center}}

\newcommand{\cO}{\mathcal{O}}
\newcommand{\cT}{\mathcal{T}}

\newcommand{\cC}{\mathcal{C}}

\newcommand{\cL}{\mathcal{L}}

\newcommand{\cN}{\mathcal{N}}
\newcommand{\cW}{\mathcal{W}}

\newcommand{\cR}{\mathcal{R}}

\newcommand{\I}{\text{Im}}

\DeclareMathOperator{\vol}{vol}

\newcommand{\dd}{d}

\newcommand{\bbZ}{\mathbb{Z}}
\newcommand{\bbR}{\mathbb{R}}
\newcommand{\bbC}{\mathbb{C}}
\newcommand{\bbP}{\mathbb{P}}

\newcommand{\nn}{\nonumber}

\newcommand{\cref}{{\bf [check ref]}}

\newcommand{\id}{\mathbf{1}}

\newcommand{\WV}{\mathcal{W}}      

\newcommand{\norm}[1]{\lVert #1\rVert}

\newcommand{\simga}{\sigma}

\psfrag{n1}{$\nu_1$}
\psfrag{n2}{$\nu_2$}
\psfrag{n1'}{$\nu_1'$}
\psfrag{n2'}{$\nu_2'$}
\psfrag{n9}{$\nu_9$}
\psfrag{n10'}{$\nu_{10}'$}
\psfrag{t1}{$\tilde{\nu}_1$}
\psfrag{t2}{$\tilde{\nu}_2$}
\psfrag{t9}{$\tilde{\nu}_9$}
\psfrag{t1'}{$\tilde{\nu}_1'$}
\psfrag{t2'}{$\tilde{\nu}_2'$}
\psfrag{t10'}{$\tilde{\nu}_{10}'$}

\entrymodifiers={+!!<0pt,\fontdimen22\textfont2>}		

\def\C{\mathds{C}}

\def\P{\mathds{P}}

\numberwithin{equation}{section}			

\makeatletter
\renewcommand*\env@matrix[1][*\c@MaxMatrixCols c]{%
  \hskip -\arraycolsep
  \let\@ifnextchar\new@ifnextchar
  \array{#1}}
\makeatother

\begin{document}

\baselineskip=17pt

\begin{titlepage}
\begin{flushright}
\parbox[t]{1.8in}{
BONN-TH-2010-10\\
MPP-2010-158}
\end{flushright}

\begin{center}

\vspace*{ 1.2cm}

{\large \bf Five-Brane Superpotentials, Blow-Up Geometries and\\[.1cm] 
            SU(3) Structure Manifolds
}

\vskip 1.2cm

\begin{center}
 \bf{Thomas W.~Grimm$^{a,b}$, Albrecht Klemm$^{a}$ and Denis Klevers$^{a}$} \footnote{\texttt{grimm@mppmu.mpg.de\\\hspace*{.4cm} aklemm, klevers@th.physik.uni-bonn.de}}
\end{center}
\vskip .2cm

\emph{$^a$ Bethe Center for Theoretical Physics, Universit\"at Bonn, \\[.1cm]
Nussallee 12, 53115 Bonn, Germany}
\\[0.25cm]
 \emph{$^{b}$ Max Planck Institute for Physics, \\ 
                       F\"ohringer Ring 6, 80805 Munich, Germany} 
\\[0.15cm]
\end{center}

\vskip 0.2cm

\begin{center} {\bf ABSTRACT } \end{center}

We investigate the dynamics of space-time filling five-branes wrapped on curves in 
heterotic and orientifold Calabi-Yau compactifications. We first study
the leading $\cN=1$ scalar potential on the infinite deformation space of the brane-curve
around a supersymmetric configuration. The higher order potential 
is also determined by a brane superpotential which we compute for a subset of light deformations. We argue  
that these deformations  map to new complex structure deformations
of a non-Calabi-Yau manifold which is obtained by blowing up the brane-curve into a four-cycle and by replacing the 
brane by background fluxes. This translates the original brane-bulk system into a unifying 
geometrical formulation. Using this blow-up 
geometry we compute the 
complete set of open-closed Picard-Fuchs differential equations and identify the brane superpotential 
at special points in the field space for five-branes 
in toric Calabi-Yau hypersurfaces. This has an 
interpretation in open mirror symmetry and enables us to list compact disk instanton 
invariants. As a first step towards promoting the 
blow-up geometry to a supersymmetric heterotic background we propose 
a non-K\"ahler $SU(3)$ structure and an identification of the three-form flux.

\hfill December, 2010
\end{titlepage}

\tableofcontents

\section{Introduction}

In recent years much effort has focused on the study of $\cN=1$ effective theories 
arising in string compactifications with space-time filling D- or 
NS5-branes~\cite{Blumenhagen:2005mu,Blumenhagen:2006ci}.
In particular, the presence of D-branes will introduce open string fields which are localized in the internal 
compact dimensions. For a stack of branes these fields are crucial to provide the 
matter and gauge fields in the four-dimensional effective theory. However, there 
is generically also a universal set of open string fields which describes the 
dynamics of the D-branes even in the absence of a non-trivial gauge theory. The study 
of such open string fields is crucial for both phenomenological as well as 
conceptional reasons. Firstly, the effective four-dimensional 
scalar potential for these fields will determine possible Type II or heterotic 
string vacua. Secondly, there are various dualities where the open string deformations 
are crucial to provide a complete picture.

Flux compactifications provide a mechanism to give a mass to the 
deformations of the compactification manifold~\cite{Douglas:2006es,Blumenhagen:2006ci,Denef:2008wq}.
It was shown that in compact models the fluxes cannot be chosen arbitrarily, but rather 
have to satisfy global consistency conditions. For example, tadpole cancellation conditions 
in $\cN=1$ orientifold compactifications with space-time filling D-branes often 
enforce the presence of background flux. Hence, the inclusion of branes and fluxes 
in string compactifications are closely linked by global consistency conditions. The potential 
induced by the flux quanta encodes obstructions to the deformations of the internal manifold, 
while the topological data specifying brane wrappings encode obstructions to the 
brane deformations. In $\cN=1$ compactifications both obstructions can be encoded 
by a superpotential or D-terms. Determining the $\cN=1$ characteristic data 
thus allows to identify flat directions and the shape of the potential determining 
the physics of the four-dimensional effective theory. 

In this work we study $\cN=1$ compactifications of Type IIB string theory 
with D5-branes and O5-planes, and compactifications of the 
heterotic string with NS5-branes and vector bundles. In these backgrounds
space-time filling five-branes wrap curves $\Sigma$ in the internal Calabi-Yau geometry $Z_3$. 
The fields included in the four-dimensional $\cN=1$ effective theory 
are the deformations around a supersymmetric vacuum configuration. 
In particular, we will focus on the closed string degrees of freedom corresponding 
to complex structure deformations of $Z_3$
and the open string degrees of freedom corresponding to brane deformations of the curve $\Sigma$. 
In the absence of fluxes and branes the complex structure deformations 
of the Calabi-Yau threefold $Z_3$ are unobstructed and correspond to 
flat directions of the classical potential. The corresponding infinitessimal 
massless deformations are given by elements of $H^{1}(TZ_3)$. Including 
the five-brane in a fixed background $Z_3$ one can consider 
first an infinite set of infinitesimal deformations of the embedding curve around 
a supersymmetric configuration. Most of these deformations are massive with a scalar potential determined 
in ref.~\cite{Simons,McLean}, as we will recall below. We show that this leading 
potential can be encoded by a superpotential when using the metric on the infinite 
dimensional field space. The deformations which remain unobstructed at leading order reside in 
$H^0(N\Sigma)$, where $N\Sigma$ is the normal bundle to $\Sigma$ in $Z_3$.
The higher order obstructions can be studied by computing the 
brane superpotential as a function of these deformations and the complex 
structure deformations of $Z_3$. It was shown in ref.~\cite{Witten:1997ep} that this  
superpotential is given by a chain integral of the holomorphic 
three-form over a three-chain ending on $\Sigma$. For local models this 
integral can be computed directly as shown in refs.~\cite{Aganagic:2000gs,Aganagic:2001nx}, or by 
open-closed Picard-Fuchs systems \cite{Mayr:2001xk,Lerche:2001cw}. Superpotentials for branes in compact 
Calabi-Yau manifolds were investigated in \cite{Walcher,KSch,Baumgartl:2007an,Jockers,Grimm:2008dq,
Alim:2009rf,Grimm:2009ef,Alim:2009bx,Aganagic:2009jq,Li:2009dz,Grimm:2009sy,Shimizu:2010us}.

In order to study the brane deformations and complex structure deformations
it is natural to look for a formalism where these deformations are treated on an equal footing. 
A very canonical procedure was proposed in ref.~\cite{Grimm:2008dq}, which suggests
to replace the pair $(\Sigma,Z_3)$ of the brane-curve and the Calabi-Yau threefold by 
an associated non-Calabi-Yau manifold $\hat Z_3$, that is obtained by blowing 
up along $\Sigma$ in $Z_3$. The purpose of the present paper is to make this proposal concrete. 
We show how to use $\hat{Z}_3$ to compute a complete open-closed Picard-Fuchs 
system for $(\Sigma,Z_3)$. In particular we determine the general structure 
of the Picard Fuchs differential system for Calabi-Yau hypersurfaces and brane-curves $\Sigma$, 
using complex structure deformations on $\hat{Z}_3$.
We demonstrate that its solutions are identified with the periods of the Calabi-Yau 
threefold $Z_3$ as well as with the five-brane superpotentials. 
The completeness of the system allows us to investigate the global structure of the open-closed
deformation space and to study the solutions at special loci in the open-closed deformation space, where we 
recover brane superpotentials for different brane geometries $\Sigma$. For example we obtain superpotentials for two-parameter 
deformations of rational curves or the curves mirror dual to involution branes. 
In particular we will apply open mirror symmetry at large volume to match the local disk instanton invariants 
of \cite{Aganagic:2001nx} and to obtain new predictions for integer disk invariants 
on compact Calabi-Yau manifolds. We note that the geometry of $\hat{Z}_3$ is compatible with the geometry of a compact 
Calabi-Yau fourfold for setups $(\Sigma,Z_3)$ where heterotic/F-theory duality applies 
as investigated in detail in \cite{Grimm:2009sy}. However the blow-up procedure 
applies in general. For an alternative treatment of the pair $(\Sigma,Z_3)$, in 
which the curve deformations of $\Sigma$ are studied by considering deformations of 
an auxiliary four-cycle moving with $\Sigma$ in $Z_3$, we refer the reader to 
refs.~\cite{Lerche:2001cw,Jockers,Alim:2009rf,Alim:2009bx}. 

As we will make more precise in the main text, the blow-up of 
each five-brane curve $\Sigma$ in $Z_3$ implies 
that complex structure deformations of $Z_3$ and brane deformations of $\Sigma$
are mapped to complex structure deformations of $\hat Z_3$. Furthermore, the 
obstructions are also matched in this procedure, as can be physically 
understood by the identification of the superpotentials. 
The new non-Calabi-Yau geometry $\hat Z_3$ is constructed by canonically 
attaching a two-sphere $\bbP^1$ at each point of $\Sigma$, which turns $\Sigma$ into an associated divisor $E$. 
After blow-up the class of the five-brane curve is represented by a two-form $F_2$
on this blow-up divisor $E$. This implies that the five-brane wave-function, which was sharply localized on $\Sigma$ in $Z_3$, has 
been smeared over the blow-up direction in $\hat{Z}_3$. However, the brane-wave function after the blow-up is still sharply localized on the divisor $E$.
More mathematically, one finds that the five-brane was described by a four-form delta-current 
localized on $\Sigma$ in $Z_3$, while on $\hat Z_3$ one is working with a two-form current localized 
on $E$ and a smooth flux $F_2$ on $E$.  While the exceptional divisor $E$ has
no deformations it will have a non-trivial complex geometry which alters when 
changing the complex structure of $\hat Z_3$. Hence, complex structure deformations 
change the shape of the blown-up wave-function, but do not move its position inside $\hat{Z}_3$.

The blow-up procedure applied to five-brane configurations not only 
turns out to be a powerful computational tool, we also aim to give a more physical interpretation 
of this procedure. Note that even on the blow-up space $\hat Z_3$ the 
fluxes $F_2$ on $E$ appear as the 
delta-source of three-form fluxes in both the heterotic and Type II 
theories. It is natural to look for a further delocalization. 
More precisely, instead of picking a wave-function sharply peaked on $E$ one chooses a smooth 
two-form matching its cohomology class. We propose that 
this process of further delocalization can be used to define an $SU(3)$ structure on $\hat Z_3$ 
 which identifies $\hat Z_3$ as a complex but non-K\"ahler manifold. 
The existence of such a structure is required when demanding the 
four-dimensional effective theory to have $\cN=1$ supersymmetry \cite{Strominger:1986uh}.
Hence, despite the existence of a K\"ahler structure on $\hat Z_3$, we 
argue for the usage of specific non-K\"ahler structure as required by supersymmetry.
This is similar to the logic used in the recent constructions of non-K\"ahler backgrounds 
presented in \cite{Larfors:2010wb,Chen:2010bn}.\footnote{Other explicit constructions of 
heterotic non-K\"ahler vacua have
appeared, for example, in \cite{Becker:2004qh,Becker:2006et}.}
To support this identification we then argue that the supersymmetry 
conditions on the brane imposed by the superpotential 
before the blow-up are naturally translated 
into the supersymmetry conditions on a compactification geometry 
$\hat Z_3$ with a non-trivial flux background as dictated by the Type II or heterotic 
string theory. This provides first evidence that $\hat Z_3$ with an 
appropriate $SU(3)$ structure comprises a dual description of the brane 
setup.\\[.2cm]
The paper is organized as follows:

In section \ref{N=1gensection} we discuss various 
aspects of five-branes in Calabi-Yau threefolds $Z_3$. We briefly introduce 
the notion of currents to discuss global consistency conditions in the $\cN=1$ heterotic 
and Type II settings under consideration. A variation of the brane volume functional 
leads to the derivation of the leading scalar potential.
We show that this scalar potential can be expressed through a superpotential
which can be extended to a chain integral. To prepare for setting up the 
blow-up proposal we also 
recall some crucial notions on the deformation and obstruction 
theory of curves in complex threefolds. 

In section \ref{5braneblowupsanddefs} the blow-up proposal is introduced. 
We explain in detail the geometric construction of $\hat Z_3$ in general 
and in toric constructions. It is subsequently argued that complex 
structure deformations of $\hat Z_3$ unify the brane deformations and complex 
structure deformations of $Z_3$. We argue how also obstructed brane deformations 
are mapped to $\hat Z_3$ and can be constraint by a flux superpotential. 
As a key object to study the open-closed system on $\hat Z_3$ we introduce 
the residue integral for the pull-back of the holomorphic $(3,0)$-form $\Omega$ from 
$Z_3$ to $\hat Z_3$. 

In section \ref{IIBBlowUp} we use the blow-up space $\hat Z_3$ to derive the 
open-closed Picard-Fuchs equations and determine their solutions for 
branes in torically realized Calabi-Yau hypersurfaces. In detail  
we analyze branes in the mirror quintic, and the mirror of the 
degree-18 hypersurface in weighted projective space $\bbP^{4}(1,1,1,6,9)$.
We cover examples which have several brane deformations as well as 
several complex structure deformations. Having a complete set of solutions 
we comment on their superpotential interpretation at special points in the 
field space. Open mirror symmetry allows us to list compact disk 
instanton invariants. 
 
In section \ref{su3structur} we propose to dissolve the 
brane further on $\hat Z_3$ by introducing an $SU(3)$
structure which delocalizes the brane sources. We recall 
some basic geometric constructions for $\hat Z_3$ endowing 
it with a K\"ahler structure. A non-K\"ahler structure 
is required for $\hat Z_3$ to be a heterotic vacuum with  
background fluxes. We thus introduce a non-K\"ahler twist 
in a neighborhood of the blow-up divisor. In addition 
we remove the zeros of the pull-back $\hat \Omega$ of the 
holomorphic $(3,0)$-form from $Z_3$ to $\hat Z_3$ by 
defining a new non-holomorphic three-form. The 
construction is performed such that the zeros of 
the original holomorphic three-form $\hat \Omega$ cancel
against the poles of $d \hat J$ arising due to the 
five-brane sources. This allows a match of the 
brane superpotential with a flux/non-Calabi-Yau superpotential
of the form 
determined in \cite{Behrndt:2000zh,LopesCardoso:2003af,Benmachiche:2008ma}.

\section{Five-Brane $\mathcal{N}=1$ Effective Dynamics} \label{N=1gensection}

In this section we discuss various aspects of five-brane dynamics. Our 
point of view will be geometrical and appropriate to formulate the blow-up 
proposal in section~\ref{5braneblowupsanddefs}. In subsection~\ref{N=1branes} 
we summarize heterotic and Type IIB string compactifications with five-branes 
focusing on global consistency conditions and the use of currents to 
describe localized sources.
The deformations of both the bulk complex structure and the brane
positions around a supersymmetric vacuum are discussed in subsection \ref{fivebranes}. 
In subsection \ref{fivebranesuperpotential} we summarize the superpotentials encoding the scalar
potentials for the closed and open deformations. Finally, in subsection \ref{N=1onalldef}, 
we show explicitly that the leading order scalar potential for the 
infinite set of normal deformations of a five-brane can be obtained from an 
$\cN=1$ superpotential. 

\subsection{Five-Branes in $\mathcal{N}=1$ Compactifications} \label{N=1branes}

In the following we consider four-dimensional $\mathcal{N}=1$ compactifications 
of Type IIB string theory and the heterotic string. The Type IIB setups will 
be orientifold compactifications with D5-branes and O5-planes on 
a  three-dimensional Calabi-Yau manifold $Z_3$ modded out by the orientifold involution.
In the heterotic string we consider NS5-branes and vector bundles on 
a smooth Calabi-Yau threefold $Z_3$. 
In both compactifications global consistency conditions restrict the choice of 
valid configurations and link the discrete data counting 
branes and fluxes via tadpole cancellation conditions.

Let us focus on Type IIB string theory first. We allow for the inclusion of
space-time filling D5-branes wrapping curves $\Sigma,\Sigma_i$, and 
O5-planes wrapping curves $\tilde \Sigma_\alpha$ in the Calabi-Yau $Z_3$. 
The Bianchi identity for the R-R field strength $F_3$ signals the presence 
of localized D5-brane and O5-plane sources by singular 
delta-forms\footnote{The analogue of delta-function as a $p$-form is properly called a delta-current, as introduced below.} $\delta_{\Sigma}, \delta_{\Sigma_i}, \delta_{\tilde \Sigma_\alpha}$,
\beq \label{dF_3}
    d F_3 = \delta_{\Sigma} + \sum_i \delta_{\Sigma_i}-2\sum_\alpha\delta_{\tilde \Sigma_\alpha}\,.
\eeq
Note that the O5-planes carry $-2$ times the charge of a D5-brane. 
Read as an equation in cohomology, i.e.~integrating \eqref{dF_3} over a basis 
of closed four-cycles, it requires that the cohomology class on the right hand side 
is trivial. This yields the global tadpole cancellation condition.
Thus it is necessary to include O5-planes 
on curves $\tilde{\Sigma}_\alpha$ in the same homology class as the brane 
curves in order to guarantee tadpole cancellation 
without breaking supersymmetry.

In the second setup we consider a heterotic compactification with a NS-five-brane. 
This five-brane is a source for the three-form field strength $H_3$ of 
the NS-NS B-field for which the Bianchi identity takes the form
\beq \label{dH_3het}
   d H_3 = \delta_{\Sigma} + \text{tr} \cR \wedge \cR - \tfrac{1}{30} \text{Tr} F\wedge F \ .
\eeq 
In addition to the delta-form signaling the presence of the five-brane 
there is a smooth contribution from the heterotic vector bundle.
The symbols ``tr'' and ``Tr'' denote the traces in the vector representation of $\text{O}(1,9)$
and the adjoint of $\text{SO}(32)$ or $E_8$ for the gauge fields $F$ in the heterotic theory, respectively. 
Note that global tadpole cancellation implies that the right-hand side of 
\eqref{dH_3het} has to vanish in cohomology,
\beq \label{tadpoleHet}
  [\delta_{\Sigma}] +  \tfrac{1}{60}c_2(E)  = c_2(Z_3)\ ,  
\eeq
where we expressed the Chern characters as $ch_2(Z_3)=-c_2(Z_3)$, $ch_2(E)=-c_2(E)$ 
for a compactification on a Calabi-Yau threefold $Z_3$. This tadpole in particular requires the inclusion of gauge background bundles $E$
over $Z_3$ with structure group contained in the ten-dimensional heterotic gauge group \cite{Candelas:1985en}. 
Note that in many heterotic compactifications the inclusion of five-branes is not a choice, 
but rather required by tadpole cancellation as demonstrated explicitly in the case
of elliptic threefolds $Z_3$ in \cite{Friedman:1997yq}.

In both theories the wavefunction of the five-brane is sharply peaked\footnote{The delta-form $\delta_\Sigma$ is the wavefunction of a brane in eigenstate of the position space operator.} at the curve $\Sigma$ which is reflected by the delta-function $\delta_\Sigma$ in the Bianchi identity. In contrast to the global 
tadpole condition \eqref{tadpoleHet} that fixes only the cohomology classes, the Bianchi identity fixes, up to gauge transformations, the actual forms $F_3$ and $H_3$ pointwise and implies that globally defined forms $C_2$ and $B_2$ with $F_3=dC_2$ and $H_3=dB_2$ do not exist\footnote{For a discussion of the global structure of $C_2$, $B_2$ in terms of  $\check{\text{C}}$ech de Rham complexes we refer to \cite{Freed:1998tg}.}. However, in a local patch we can solve the Bianchi identity for the field strength and the potential explicitly as we present in the following.
Let us note here that one crucial point that leads to the blow-up proposal below is the appropriate 
mathematical treatment of the actual forms $H_3$ or $F_3$ that become singular near the brane.

In the vicinity of a single brane source the Bianchi identity reads
\begin{equation}
	d \sigma_3=\delta_{\Sigma}\,,
\label{eq:general}
\end{equation}
where $\sigma_3$ is identified with the singular part in the field strength $H_3$, $F_3$ in both theories. 
For the setups we will consider the other 
localized sources do not interfere with the following local analysis and 
can be treated similarly. Furthermore, we will ignore the smooth part 
of the heterotic bundle $c_2(E)$. 
The equation \eqref{eq:general} is best treated in the theory of currents, see e.g.~\cite{Griffiths}. 
In this context $\sigma_3$ can be understood as the Poincar\'e dual of a chain $\Gamma$ in the following way. 
First we associate a functional $T_\Gamma$ to every three-chain $\Gamma$ with boundary $\partial \Gamma=\Sigma$ by two defining properties. For any smooth three- and two-form $\eta_3$, $\varphi_2$ we have 
\beq \label{eq:currentChain}
	T_{\Gamma}(\eta_3)=\int_{\Gamma}\eta_3\,,\qquad \quad dT_{\Gamma}(\varphi_2)=\int_{\Gamma}d\varphi_2=\int_\Sigma \varphi_2=T_{\Sigma}(\varphi_2)\,.
\eeq
Such a map from smooth forms to complex numbers is usually denoted as a current and 
is a generalization of distributions to forms. In this language \eqref{eq:currentChain} 
is usually written as $dT_\Gamma=T_{\Sigma}$. This is precisely the dual of the expression 
\eqref{eq:general} on the level of currents. Indeed we can use $\sigma_3$ to define a current $T_{\sigma_3}$ that also enjoys $dT_{\sigma_3}=T_{\Sigma}$ as follows
\bea
	T_{\sigma_3}(\eta_3) &=&\int_{Z_3}\sigma_3\wedge \eta_3\,,\\
	dT_{\sigma_3}(\varphi_2) &=& \int_{Z_3}\sigma_3\wedge d\varphi_2=\int_{Z_3}\delta_{\Sigma}\wedge\varphi_2=\int_{\Sigma}\varphi_2=T_{\Sigma}(\varphi_2)\,.\nn 
\eea
Thus we identify $\sigma_3$ and $\delta_\Sigma$ as the Poincare dual of $\Gamma$ and $\Sigma$ respectively.
However, both $\sigma_3$ and $\delta_{\Sigma}$ are not forms in the usual sense. $\delta_{\Sigma}$ fails to be a form 
similar to the fact that the delta-function fails to be a function. 
$\sigma_3$ is not a form on $Z_3$. However, it is a smooth form on 
the open space $Z_3-\Sigma$. This can be seen directly in a 
local analysis in the fiber of the normal bundle $N_{Z_3}\Sigma$, that is isomorphic to $\mathds{C}^2$. Let us summarize 
the essential results.

On $N_{Z_3}\Sigma\vert_p\cong \mathds{R}^4$ the form $\sigma_3$ is the unique rotationally invariant form on $\mathds{R}^4-\{0\}$ that is orthogonal to $dr$ and integrates to $1$ over a three-sphere $S^3_r$ of any radius $r$. In hyperspherical coordinates we obtain
\beq
\sigma_3=\frac{1}{2\pi^2}\vol_{S^3}\,,\quad \int_{S^3_r} \vol_{S^3}=1\,. 
\eeq
Thus, $\sigma_3$ is ill-defined at $r=0$ where the three-sphere $S^3_r$ degenerates. Consequently, we can deal with $\sigma_3$ rigorously by working on the open manifold $\mathds{R}^4-\{0\}$ where $\sigma_3$ is a smooth form and by taking boundary contributions into account in the following way. 

Whenever we have a bulk integral over $Z_3$ we replace it by an integral over the open manifold $Z_3-\Sigma$ as \cite{Freed:1998tg}
\begin{equation}
	\int_{Z_3}\mathcal{L}:=\lim_{\epsilon\rightarrow 0}\int_{Z_3-\mathcal{U}_\epsilon^{(4)}(\Sigma)}\mathcal{L}\,.
\label{eq:Z3-C}
\end{equation}
where we substract a tubular neighborhood $\mathcal{U}_{\epsilon}^{(4)}(\Sigma)$ of radius $\epsilon$ over $\Sigma$.
All integrands are regular when evaluated on this open manifold, even those including the 
singular form $\sigma_3$ in $H_3$, $F_3$. One has to consider two cases, either $\mathcal{L}$ is 
well-behaved in $\Sigma$ and thus the limit $\epsilon\rightarrow 0$ in \eqref{eq:Z3-C} 
just gives back the integral over $Z_3$. In the other case $\mathcal{L}$ contains the 
form $\sigma_3$ and a boundary term is produced by partial integration as follows
\begin{equation}
	\lim_{\epsilon\rightarrow 0}\int_{Z_3-\mathcal{U}_\epsilon^{(4)}(\Sigma)}\sigma_3\wedge d\varphi_2=\lim_{\epsilon\rightarrow 0}\int_{S^3_\epsilon(\Sigma)}\sigma_3\wedge \varphi_2=\int_{\Sigma}\varphi_2\,,
\label{eq:boundaryTerms}
\end{equation} 
where we used in the second equality that $\sigma_3$ is locally exact, $\sigma_3=\frac{1}{2\pi^2}\vol_{S^3}$, and  integrates $\sigma_3$ to $1$ in each $S^3_\epsilon$-fiber of the sphere bundle $S^3_\epsilon(\Sigma)=\partial \mathcal{U}_\epsilon^{(4)}(\Sigma) $ over $\Sigma$.

We conclude by discussing the global structure of this construction. Since the normal bundle $N_{Z_3}\Sigma$ is in general non-trivial, we have to take into account the effects of a non-trivial connection. As worked out in \cite{Freed:1998tg} the adequate globalization of $\sigma_3$ is related to the Thom-class $e_3/2$ of the normal bundle, see e.g.~\cite{BottTu} for a reference. The Thom class is the unique closed form $de_3=0$, that is gauge invariant under the SO$(4)$ structure on $N_{Z_3}\Sigma$ and that integrates to $1$ over any fiber $S^3_r$.  The basic idea is to smooth out the localized source of the five-brane \eqref{eq:general} using a smooth bump form $d\rho$ normalized to integral $1$ with $\rho(r)=-1$ around $r\sim 0$ and $\rho(r)=0$ for $r> 2\epsilon$ such that the support $supp(d\rho)\subset ]\epsilon,2\epsilon[$. Then
\beq
d\sigma_3=d\rho\wedge e_3/2\,
\eeq
approaches $\delta_{\Sigma}$ when taking the limit $\epsilon\rightarrow 0$. Thus, we identify the contribution of the five-brane as $\sigma_3=-d\rho\wedge e_2^{(0)}/2$ with $e_3=de_2^{(0)}$ locally, where a possible term $\rho\, e_3$ has been discarded since $e_3$ is not well-defined at the position $r=0$ of the brane. We note further that $e_2^{(0)}$ is not a global form since it is not gauge invariant under the SO$(4)$ action on the normal bundle. Then, we obtain the global expressions for the field strength $F_3$ and $H_3$ as
\beq \label{eq:globalForms}
	F_3=\left<F_3\right>+dC_2-d\rho\wedge e_2^{(0)}/2\,,\quad H_3=\left<H_3\right>+dB_2-d\rho\wedge e_2^{(0)}/2+\omega_3\,,
\eeq
where $\omega_3=\omega_3^{\rm{L}}-\omega_3^{\rm{G}}$ denotes the Chern-Simons form for $\text{tr} R^2-\frac{1}{30}\text{Tr} F^2$ and  $\left<F_3\right>$, $\left<H_3\right>$ are background fluxes in $H^3(Z_3,\mathds{Z})$. With these formulas at hand we immediately check that the reasoning of \eqref{eq:Z3-C} and the localization \eqref{eq:boundaryTerms} to the boundary of the open manifold $Z_3-\Sigma$ applies globally. Furthermore, the expansion \eqref{eq:globalForms} formally unifies the superpotentials as we will discuss in detail below in section \ref{fivebranesuperpotential}.

We conclude by noting that \eqref{eq:globalForms} implies that $C_2$ respectively $B_2$ have an anomalous transformation under the SO$(4)$ gauge transformations of $N_{Z_3}\Sigma$. This is necessary to compensate the anomalous transformation $\delta e_2^{(0)}$ so that $F_3$ respectively $H_3$ are gauge invariant. This anomalous transformation plays a crucial role for anomaly cancellation in the presence of five-branes \cite{Freed:1998tg}.

\subsection{Deformations and Supersymmetry Conditions}
\label{fivebranes}

In this section we discuss the light fields associated to geometric deformations of a five-brane. In many situations there is a superpotential for these fields that obstructs deforming the brane at higher order. As a preparation to understand obstructed deformations we briefly review the unobstructed case of the familiar example of complex structure deformations of a Calabi-Yau manifold as found in standard textbooks like \cite{Huybrechts} or the original work of \cite{Kodaira:cs}. This introduces the necessary concepts to understand the more complicated case of brane deformations and superpotentials.

\subsubsection{Bulk Deformations: Deformations of Complex Structures}
\label{bulkdeformations}

We consider a complex manifold $Z_3$ as a real manifold equipped with a background complex structure $I:TZ_3\rightarrow TZ_3$ with $I^2=-\id$ for which the Nijenhuis-tensor $N$ vanishes ensuring integrability\footnote{In the following we denote the (anti-)holomorphic tangent bundle just by ($\overline{TZ_3}$) $TZ_3$.} $[T^{(0,1)}Z_3,T^{(0,1)}Z_3]\subset T^{(0,1)}Z_3$. The complex structure determines the Dolbeault operator $\bar{\partial}$ and vice versa. Finite deformations of $I$ are described by elements $A(\underline{t})$ in $\Omega^{(0,1)}(Z_3,TZ_3)$ that are $(0,1)$-forms taking values in $TZ_3$ and depend on parameters $\underline{t}$ denoting coordinates on a parameter manifold\footnote{We choose $M$ so that the Kodaira-Spencer map $T_0M\rightarrow H^1(Z_3,TZ_3)$ is bijective at a point $0\in M$.} $M$. In fact, if we perturb $\bar{\partial}$ by $A$ and demand $(\bar{\partial}+A)^2=0$, the deformation $A$ has to obey the Maurer-Cartan equation
\begin{equation}\label{eq:MaurerCeq}
 	\bar\partial A+\frac12[A,A]=0\,.
\end{equation}
If we write $A$ as a formal power series in $t$, $A=A_1(\underline{t})+A_2(\underline{t})+\ldots$ we obtain 
\beq \label{eq:MaurerCorder}
	\bar{\partial} A_1=0\,,\qquad \bar{\partial} A_n+\frac12\sum_{i=1}^{n-1}[A_i,A_{n-i}]=0\,,\, n>1\,,
\eeq
where $A_n(\underline{t})$ denotes a homogeneous polynomial in $t$ of degree $n$.
Taking coordinate transformations into account that trivially change the complex structure $I$, we learn that first order or infinitesimal deformations of $I$ for which $t\sim 0$ are in one-to-one correspondence with classes $[v]=[A_1]$ in the cohomology group $H^1(Z_3,TZ_3)$, called the Kodaira-Spencer class of $A_1$. However, deformation classes $A_1$ lift to finite deformations only if we can recursively solve \eqref{eq:MaurerCorder} for the $A_n$ at every finite order in $n$. We immediately identify the necessary condition for $[A_1]=[v]$ being integrable with the integrability condition
\begin{equation}\label{eq:obstructionclass}
\bar \partial A_2=-\frac12[A_1,A_1],
\end{equation}
 which means that $[A_1,A_1]$ has to be $\bar\partial$-exact in order to find a solution. Thus, one associates to every class $v$ in $H^1(Z_3,TZ_3)$ the class $[v,v]$ in $H^2(Z_3,TZ_3)$ on the right hand side of \eqref{eq:obstructionclass} called the obstruction class. It necessarily has to vanish in order to define a finite $A(\underline{t})$ obeying \eqref{eq:MaurerCorder} with $[A_1]=[v]$. This is in particular the case if $H^2(Z_3,TZ_3)=0$. However, this is not a necessary condition for the existence of $A(\underline{t})$ since the obstruction classes for integrating an infinitesimal $A_1$ can be zero even for $H^2(Z_3,TZ_3)\neq 0$. Indeed, this is the generic case for Calabi-Yau manifolds. It is the content of the classical theorem by Tian and Todorov that for every Calabi-Yau manifold all commutators $\sum_i[A_i,A_{n-i}]$ in the recursive equations \eqref{eq:MaurerCorder} are exact so that a finite $A(\underline{t})$ exists for every infinitesimal deformation $[A_1]=[v]$. Thus, complex structure deformations of a Calabi-Yau manifold are generically unobstructed such that there is a global moduli space of complex structures of complex dimension\footnote{Here we use the isomorphism $H^1(Z_3,TZ_3)= H^{(2,1)}(Z_3)$ by contraction with the $(3,0)$-form $\Omega$.} $h^{(2,1)}$. In physics, this is reflected by the fact that, as long as background 
fluxes are absent, there is no scalar potential in the effective theory 
of a Calabi-Yau compactification for the fields $t(x)$ associated to 
complex structure deformations.

\subsubsection{Brane Deformations I: Infinitesimal Deformations of Holomorphic Curves}
\label{branedeformationsI}
Let us now present an analogous discussion for deformations of branes. We consider a five-brane 
wrapped on a curve $\Sigma$ in a given Calabi-Yau background $Z_3$. 

The five-brane will preserve $\mathcal{N}=1$ supersymmetry 
if $\Sigma$ is a holomorphic curve. A holomorphic curve can be specified 
as curve of minimal volume in its homology class. In the language of 
calibrations this condition reads
\beq \label{vol=J}
   \vol_{\Sigma }= J|_{\Sigma}\ 
\eeq 
using the calibration by the K\"ahler form $J$ on $Z_3$. 
In the effective four-dimensional theory the volume 
of the wrapped curve contributes terms to the scalar potential. 
However, the leading term for holomorphic curves is canceled by contributions 
from the supersymmetric O5-planes in Type IIB or bundle and curvature contributions 
in the heterotic string using \eqref{dF_3}, \eqref{dH_3het}. Thus, this part of the vacuum energy 
cancels which is a necessary condition for supersymmetry. This is easily 
seen, for example, in orientifold setups.
The orientifold compactification
preserve $\mathcal{N}=1$ supersymmetry in the effective 
theory if the geometric part of the 
orientifold projection is a holomorphic and isometric 
involution $\sigma$ acting on $Z_3$ as discussed in \cite{Blumenhagen:2005mu,Blumenhagen:2006ci}. 
Hence, the O5-planes, being the fix-point set of $\sigma$, 
wrap holomorphic curves inside $Z_3$ and are also calibrated 
with respect to $J$. Thus they 
contribute the same potential in the vacuum with opposite sign (see, e.g.~\cite{Grimm:2008dq}
for a more detailed discussion).

\begin{figure}[htb]
\begin{center}
\includegraphics[width=.5\textwidth]{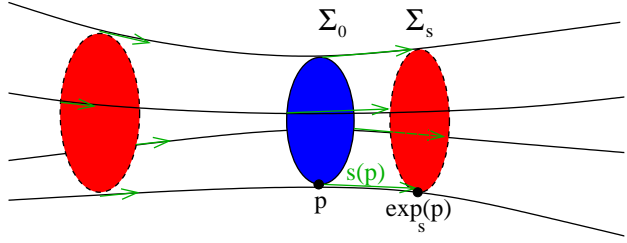}
\begin{quote}
\caption{Deformations of $\Sigma_0$ along geodesics in $Z_3$. The tangent vector $s(p)$ along the geodesic passing $p$ is a normal vector of $\Sigma_0$ at $p$.\vspace{-1.2cm}} \label{geodesicdef}
\end{quote}
\end{center}
\end{figure}

Let us now consider a general fluctuation 
of the supersymmetric $\Sigma \equiv \Sigma_0$ to a nearby curve $\Sigma_{s}$. From the above one 
expects the generation of a positive potential when deforming $\Sigma$ non-holomorphically. 
A deformation is described by a complex section $s$ of the normal bundle
$N_{Z_3} \Sigma \equiv N_{Z_3}^{1,0} \Sigma$.
The split of the complexified normal bundle has been performed in a background complex structure of $Z_3$.
Clearly, the space of such sections is infinite dimensional as is the 
space of all $\Sigma_{s}$. To make the identification between $\Sigma_{s}$
and $s$ more explicit, one recalls that in a sufficiently small neighborhood of 
$\Sigma_0$ the exponential map $exp_s$ is a diffeomorphism of $\Sigma_0$ onto $\Sigma_s$. 
Roughly speaking, one has to consider geodesics through each point 
$p$ on $\Sigma_0$ with tangent 
$s(p)$ and move this point along the geodesic 
for a distance of $||s||$ to obtain the nearby curve $\Sigma_s$ as depicted in figure \ref{geodesicdef}.
That the holomorphic curve $\Sigma_0$ is of minimal volume can now be seen 
infinitesimally. For any normal deformation $\Sigma_{\epsilon s}$ to $\Sigma_0$, 
i.e.~a deformation with infinitesimal displacement $\epsilon s$,
the volume increases quadratically \cite{Simons,McLean} as
\beq \label{eq:volvar}
 	\left.\frac{d^2}{d \epsilon^2}\text{Vol}(\Sigma_{\epsilon s})\right\vert_{\epsilon=0}=\frac12\int_{\Sigma}\norm{\bar{\partial} s}^2\, \vol_{\Sigma}\,,
\eeq 
where $\norm{\bar{\partial} s}^2$ denotes the contraction of indices, both on the curve as well as on its normal bundle $N_{Z_3}\Sigma$ via the metric. In the effective action, \eqref{eq:volvar} is the leading F-term potential when deforming the D5-brane curve non-holomorphically as we discuss in more detail in subsection \ref{N=1onalldef}. 
A quadratic term of the form \eqref{eq:volvar} in the scalar potential implies 
that the four-dimensional fields corresponding to these non-holomorphic deformations $s$ 
acquire masses given by the value of the integral \eqref{eq:volvar}\footnote{The first variation of $\vol(\Sigma_u)$ vanishes by the First Cousin Principle \cite{McLean}.}. 
When integrating out the massive deformations with $\bar \partial s \neq 0$
the remaining sections are elements of 
\beq
   H^{0}(\Sigma,N_{Z_3} \Sigma) \equiv H^{0}_{\bar \partial}(\Sigma,N^{1,0}_{Z_3} \Sigma) \ \subset \  C^\infty(N_{Z_3}\Sigma) \ .
\eeq
Reversely only holomorphic sections $s \in H^{0}(\Sigma,N_{Z_3} \Sigma)$ deforming $\Sigma$ 
into a nearby curve $\Sigma_{s}$ can lead to massless or light fields 
in the effective theory. It is crucial to note that even for an  $s \in H^{0}(\Sigma,N_{Z_3} \Sigma)$ 
the deformation might be obstructed at higher order and hence not 
yield a massless deformation. The higher order mass terms for these 
deformations can be studied by computing the superpotential as we 
will discuss throughout the next sections.

Before delving into the discussion of these holomorphic deformations, let us conclude with a discussion of the effect of complex structure deformations on the F-term potential \eqref{vol=J} and the hierarchy of masses of fields associated to brane deformations. 
Deformations generated by non-holomorphic vector fields $s$ do not obey the 
classical equations of motion. 
The main complication is  
that there are infinitely many such off-shell deformations and it would be very hard 
to compute their full scalar potential. 
In contrast to $\mathcal{C}^{\infty}(N_{Z_3}\Sigma)$, the space $H^0(\Sigma,N_{Z_3}\Sigma)$
is finite dimensional. However, there is a distinguished finite dimensional subset of $\mathcal{C}^{\infty}(N_{Z_3}\Sigma)$ that should not be integrated out in the effective action. This is related to the fact that the dimension of $H^0(\Sigma,N_{Z_3}\Sigma)$ is not 
a topological quantity and will generically jump when varying the complex structure of $Z_3$.
For example, this can lift some of the holomorphic deformations $s \in H^0(\Sigma,N_{Z_3}\Sigma)$ 
since the notion of a holomorphic section is changed. 
Indeed, by deforming $\bar{\partial}$ by $A$ in $H^1(Z_3,TZ_3)$ we obtain
\beq\label{eq:csinducedmass}
	\left.\frac{d^2}{d \epsilon^2} \text{Vol}(\Sigma_{\epsilon s})\right\vert_{\epsilon=0}
	=\frac12\int_{\Sigma}\norm{As}^2 \vol_{\Sigma}=
	\frac12 
	|t|^2\int_{\Sigma}\norm{A_1s}^2 \vol_{\Sigma}
	+\mathcal{O}(t^4)\, ,
\eeq
where $A_1$ is the first order complex structure deformation of $Z_3$ as introduced above.
Here we used that the complex structure on $\Sigma$ is induced from $Z_3$ and $s$ is 
in $H^0(N_{Z_3}\Sigma)$ in the unperturbed complex structure on $Z_3$, $\bar\partial s=0$. 
This result is clear from the point of view of the new complex structure $\bar{\partial}'=\bar{\partial}+A$, 
since $\bar\partial's=A s \neq 0$. Thus $s$ is a section in $\mathcal{C}^{\infty}(N_{Z_3}\Sigma)$ 
in the new complex structure unless $s$ is in the kernel 
of $A$. Similarly, the corresponding field acquires a mass given by 
the integral \eqref{eq:csinducedmass}. However, the main difference 
to a generic massive mode in $\mathcal{C}^{\infty}(N_{Z_3}\Sigma)$ 
with mass at the compactification scale, cf.~eq. \eqref{eq:Ftermpotfinal}, 
is the proportionality to the square of the VEV of $t$. Consequently 
the mass of this field can be made parametrically small tuning the value of $t$. 
Thus, we can summarize our approach to identify the light fields as follows:
(1) drop an infinite set of deformations $s$ which are massive via 
\eqref{eq:volvar} at each point in the complex structure moduli space, 
(2) include any brane deformation that has vanishing \eqref{eq:volvar} 
at some point in the closed string moduli space. 
These remaining deformations are 
not necessarily massless at higher orders in the 
complex structure deformations, or at higher $\epsilon$ order when expanding $\text{Vol}(\Sigma_{\epsilon s})$.
This induces a five-brane superpotential $W$ which can be computed using the blow-up proposal as we will 
show for a number of examples in section \ref{IIBBlowUp}.\footnote{The critical locus of $W$ will either set the VEV $t$ back to zero promoting $s$ to a unobstructed deformation or will leave a discrete set of holomorphic curves.}

\subsubsection{Brane Deformations II: Analytic Families of Holomorphic Curves}
\label{branedeformationsII}

Let us now present the standard account on deformations of holomorphic curves \cite{Kodaira:curves}. The basic question in this context is, as in the case of complex structure deformations, whether a given infinitesimal deformation can be integrated. Mathematically, finite deformations are described by the existence of an analytic family of compact submanifolds, in our context of curves. An analytic family of curves is a fiber bundle over a complex base or parameter manifold $M$ with fibers of holomorphic curves $\Sigma_u$ in $Z_3$ over each point $u\in M$.

Given a single curve $\Sigma$ in $Z_3$ one can ask the reverse question, namely under which conditions does an analytic family of curves exist? 
The answer to this question was formulated by Kodaira \cite{Kodaira:curves}. In general an analytic family of holomorphic curves\footnote{Kodaira considered the general case of a compact complex submanifold in an arbitrary complex manifold.} exists if the obstructions $\psi$, that are elements in $H^1(\Sigma,N_{Z_3}\Sigma)$, vanish at every order $m$, which is of course trivially the case if $H^1(\Sigma,N_{Z_3}\Sigma)=0$. Then $\underline{u}$ are coordinates of points $u$ in $M$ and a basis of holomorphic sections in $H^0(\Sigma_u,N_{Z_3}\Sigma_u)$ is given by the tangent space of $TM_u$ via the isomorphism\footnote{This map is called the infinitesimal displacement of $\Sigma_u$ along $\frac{\partial}{\partial u^a}$ \cite{Kodaira:curves}.}
\begin{equation} \label{eq:Kodairaopen}
  	\varphi^z_*:\, \frac{\partial}{\partial u^a}\ \longmapsto \ \frac{\partial \varphi^i(z^i;u)}{\partial u^a}
\end{equation}
at every point $u$ in $M$. Here, $\varphi^i$, $i=1,2$, are local normal coordinates to $\Sigma_u$, cf.~eqn. \eqref{eq:displacedcurve}. In other words, in this case every deformation $H^0(\Sigma,N_{Z_3}\Sigma)$ corresponds to a finite direction $u^a$ in the complex parameter manifold $M$ of the analytic family of curves. 

This theorem can be understood locally \cite{Kodaira:curves} but is somewhat technical. Starting with the single holomorphic curve $\Sigma$ we introduce patches $U_i$ on $Z_3$ covering $\Sigma$ with coordinates $y^{i}_1$, $y^{i}_2$, $z^{i}$. Then $\Sigma$ is described as $y^{i}_1=y^{i}_2=0$ and $z^{i}$ is tangential to $\Sigma$. A deformation $\Sigma_{\underline{u}}$ of $\Sigma=\Sigma_0$ is described by finding functions $\varphi_l^{i}(z^{i};\underline{u})$, $l=1,2$, with the boundary condition $\varphi_l^{i}(z^{i};0)=0$ such that $\Sigma_{\underline{u}}$ reads
\begin{equation} \label{eq:displacedcurve}
 	y^{i}_1=\varphi^{i}_1(z^{i};\underline{u})\,,\quad y^{i}_2=\varphi^{i}_2(z^{i};\underline{u})\,
\end{equation}
upon introducing parameters $\underline{u}$ for convenience chosen in polycylinders $||\underline{u}||<\epsilon$. Furthermore, the first derivatives $\frac{\partial}{\partial u_a}\varphi^{i}_k\vert_{\underline{u}=0}$ should form a basis $s^a$ of $H^0(\Sigma,N_{Z_3}\Sigma)$. In addition, these functions have to obey specific consistency conditions, that we now discuss. As in the complex structure case, these functions are explicitly constructed as a power series 
\begin{equation} \label{eq:powerseriesopen}
 \varphi^{i}(z^i;\underline{u})=\varphi^{i}(0)+\varphi^{i}_{1}(\underline{u})+\varphi^{i}_{ 2}(\underline{u})+\ldots,\quad \norm{\underline{u}}<\epsilon\,,
\end{equation}
where we suppress the dependence on $z^{i}$ and further denote a homogeneous polynomial in $u$ of degree $n$ by $\varphi^{i}_n(\underline{u})$. 
The first order deformation is defined as
\begin{equation}
 	\varphi^{i}_{1}(\underline{u})=\sum_a u^a s^{(i)}_a(z^{i})\,,
\end{equation}
where $a=1,\ldots, h^0(N\Sigma)$ in the basis $s_a$ of $H^0(\Sigma,N_{Z_3}\Sigma)$ so that \eqref{eq:Kodairaopen} is obviously an isomorphism. 

Then the $m^{\text{th}}$ obstructions $\psi^{ik}(z^{k};\underline{u})$ are homogeneous polynomials of order $m+1$ taking values in $\check{\text{C}}$ech 1-cocycles on the intersection $U_i\cap U_j\cap U_k$ of the open covering of $\Sigma$ with coefficients in $N_{Z_3}\Sigma$. This means that the collection of local section $\psi^{ik}(z^{k};\underline{u})$ defines an element in the $\check{\text{C}}$ech-cohomology $H^1(\Sigma,N_{Z_3}\Sigma)$. It expresses the possible mismatch in gluing together the $\varphi^{i}(z^{i};\underline{u})$ defined on open patches $U_i\cap \Sigma_{\underline{u}}$ consistently to a global section on $\Sigma_{\underline{u}}$ at order $m+1$ in $\underline{u}$. In other words if the obstruction at $m^{\text{th}}$ order is trivial and we consider \eqref{eq:displacedcurve} on $U_i$ and $U_k$,
\begin{equation}
 	U_i:\, y^{i}_l=\varphi^{i}_l(z^{i};\underline{u})\,, \qquad U_k:\,y^{k}_l=\varphi_l^{k}(z^k;\underline{u})\,,\qquad (l=1,2)\,,
\end{equation}
then there exist functions $f^{ik}$ and $g^{ik}$ with $y^{i}_l=f_l^{ik}(\underline{y}^{k},z^{k})$, $z^{i}=g^{ik}(\underline{y}^{k},z^{k})$ so that 
\begin{eqnarray}
       y^{i}&=&\varphi^{i}(z^{i};\underline{u})=\varphi^{i}(g^{ik}(\underline{y}^{k},z^{k});\underline{u})=\varphi^{i}(g^{ik}(\varphi^{k}(z^{k};\underline{u}),z^{k});\underline{u})\nonumber\\
       y^{i}&=&f^{ik}(\underline{y}^{k},z^{k})=f^{ik}(\varphi^{k}(z^{k};\underline{u}),z^{k})\nonumber\\
&\Rightarrow&\varphi^{i}(g^{ik}(\varphi^{k}(z^{k};\underline{u}),z^{k});\underline{u})=f^{ik}(\varphi^{k}(z^{k};\underline{u}),z^{k})
\end{eqnarray}
holds at order $m+1$ in $\underline{u}$. Here we suppressed the index $l$ labeling the coordinates $y^i_1$, $y^i_2$. Then $\psi^{ik}(z;\underline{u})$ is the homogeneous polynomial of degree $m+1$
\begin{equation}
 	\psi^{ik}(z^k;\underline{u}):=\big[\varphi^i(g^{ik}(\varphi^k(z^k;\underline{u}),z^k);\underline{u})-f^{ik}(\varphi^k(z^k;\underline{u}),z^k)\big]_{m+1}
\end{equation}
where we expand $\varphi^i$, $\varphi^k$ to order $m$ in $\underline{u}$. It can be shown to have the transformation 
\begin{equation}
 	\psi^{ik}(z^k;\underline{u})=\psi^{ij}(z^j;\underline{u})+F^{ij}(z^j)\cdot\psi^{jk}(z^k;\underline{u})\,,
\end{equation}
where $F^{ij}(z^j)$ is the complex $2\times 2$ transition matrices on $N_{Z_3}\Sigma$ at a point $z^j$ in $\Sigma_u$ that acts on the two-component vector $\psi^{jk}\equiv(\psi_1^{jk},\psi_2^{jk})$.
This equation identifies the $\psi^{ik}$ as elements in $H^1(N_{Z_3}\Sigma)$ which can be identified by the Dolbeault theorem with $\bar{\partial}$-closed $(0,1)$-forms taking values in $N_{Z_3}\Sigma$, $H^1(N_{Z_3}\Sigma)=H^{(0,1)}_{\bar\partial}(NZ_3)$. Assuming that all $\psi^{ik}$ are trivial in cohomology and further proving the convergence of the power series \eqref{eq:powerseriesopen}, the analytic family of holomorphic curves is constructed. 

In principle one can calculate the obstructions $\psi^{ik}$ according to this construction at any order $m$. However, the obstructions are precisely encoded in the superpotential of the effective theory of a five-brane on $\Sigma$. This superpotential is in general a complicated function of both the brane and bulk deformations. Thus, determining the superpotential is equivalent to solving the deformation theory of a pair given by the curve $\Sigma$ and the Calabi-Yau threefold $Z_3$ containing it. It is this physical ansatz that we will take in the following.

\subsection{The Five-Brane Superpotential}
\label{fivebranesuperpotential}

In the following we discuss the perturbative superpotentials both in the type IIB as well as in the heterotic theory. Using the expansion \eqref{eq:globalForms} the superpotential can conveniently be written as
\begin{equation} \label{eq:unifiedSuperpots}
	W^{\rm IIB}=\int_{Z_3}\Omega\wedge F_3\,\qquad W^{\rm Het}=\int_{Z_3}\Omega\wedge H_3\,.
\end{equation}
We put special emphasis on the part contributed by the brane $W_{\rm brane}$ and prove that the volume variation \eqref{eq:volvar} is the leading order F-term potential. 

The brane superpotential has the following properties. It depends holomorphically on the complex structure moduli of $Z_3$, as well as on the (obstructed) deformations corresponding to holomorphic sections of $N_{Z_3}\Sigma$. More precisely, we expect a superpotential $W_{\rm brane}=u_a^{n+1}$ if the deformation along the direction $s^a$ is obstructed at order $n$ \cite{Kachru:2000ih,Kachru:2000an}. Furthermore, the F-term supersymmetry conditions of $W_{\rm brane}$ correspond to holomorphic curves, in particular \eqref{eq:volvar} is reproduced at second order as we will see below. The appropriate functional with these properties was found in
\cite{Witten:1997ep} in the context of M-theory on a Calabi-Yau threefold $Z_3$ with a
spacetime-filling M5-brane supported on a curve $\Sigma$,
\begin{equation}
\label{eq:chain}
 	W_{\rm brane}=\int_{\Gamma(u)}\Omega(z)\,.
\end{equation}
Here $\Gamma(u)$ denotes a three-chain bounded by the deformed curve $\Sigma_u$ and the reference curve $\Sigma_{\rm rev}$
that is in the same homology class. It depends on both the
moduli $u$ of the five-brane on $\Sigma$ as well as the complex structure moduli $z$ of
$Z_3$ due to the holomorphic three-form $\Omega$. Alternatively, $W_{\rm
  brane}$ can be directly deduced by dimensional reduction of the D5-brane
action \cite{Grimm:2008dq}. The chain integral 
can be rewritten in the form \eqref{eq:unifiedSuperpots} using the language of currents and the expansion \eqref{eq:globalForms} as
\begin{equation}
	W_{\rm
          brane}=\int_{Z_3-\mathcal{U}_{\epsilon}^{(4)}(\Sigma)}\Omega\wedge
        \rho e_3.
\label{chainonopenset}  
\end{equation}
We note that this is gauge invariant under SO$(4)$ gauge transformations on $N_{Z_3}\Sigma$ and has only support on a small neighborhood of $\Sigma$ so that it localizes on $\Sigma$ as expected. 

The other terms in \eqref{eq:unifiedSuperpots} contribute the heterotic holomorphic Chern-Simons functional 
\begin{equation}
	W_{\rm CS}=\int_{Z_3}\Omega\wedge(A\wedge\bar\partial A+\frac{2}{3}A\wedge A\wedge A)
\end{equation}
by inserting $\omega_3^{\rm YM}$ and the
flux superpotential $W_{\rm flux}$ which is present for non-trivial background fluxes $G_3=\left<F_3\right>-\left<\tau H_3\right>$ in type IIB respectively $\left<H_3\right>$ in the heterotic string. It takes the form \cite{Gukov:1999ya}
\begin{equation}
\label{eq:fluxpot}
 	W_{\rm{flux}}=\int_{Z_3}\Omega(z) \wedge G_3  = M_iX^i(z)-N^iF_i(z)\,,
\end{equation}
where we expanded $G_3=N^i\alpha_i-M_i\beta^i$ and $\Omega=X^i\alpha_i-F_i\beta^i$ in the integral basis 
$\alpha_i$, $\beta^i$ of $H^3(Z_3,\mathds{Z})$ with integer flux numbers $(N^i,M_i)$ and periods $(X^i(z),F_i(z))$,
respectively. 
The complete complex structure dependence of $W_{\rm flux}$ is 
encoded in these periods. It is the great success of algebraic geometry
that $(X^i,F_i)$ can be calculated explicitly for a wide range of
examples, see~\cite{Polchinski:1995sm} and \cite{Douglas:2006es,Denef:2008wq} 
for reviews.  This is due to the fact that the periods obey differential equations, the so-called Picard-Fuchs
equations\footnote{See \cite{Hosono:1993qy} for a classic reference.}, that
can be solved explicitly and thus allow to determine the complete moduli
dependence of $W_{\rm{flux}}$ once the flux numbers are given. This explicit
analysis is based on the algebraic representation 
of the holomorphic three-form $\Omega$ and its periods $\Pi^k$ by the residue integral expressions 
\begin{equation} \label{eq:residueZ3}
 	\Omega(\underline{z})=\int_{S^1_P}\frac{\Delta_{\mathbb{P}_{\Delta}}}{P(\underline{x},\underline{z})}\,,\quad \Pi^k(\underline{z})=\int_{\Gamma^k\times {S^1_P}}\frac{\Delta_{\mathbb{P}_{\Delta}}}{P(\underline{x},\underline{z})}\,,
\end{equation}
where $Z_3$ is given as the zero locus of a polynomial constraint $P(\underline{x},\underline{z})$ in coordinates $\underline{x}$ 
in a toric ambient space $\mathbb{P}_{\Delta}$ with holomorphic top-form $\Delta_{\mathbb{P}_{\Delta}}$.
Here $\Gamma^k$, $k=1,\ldots, b_3$ with $b_3=2h^{2,1}+2$, denote a basis of  $H_3(Z_3,\mathds{Z})$ and $S^1_P$ denotes 
a small $S^1$ surrounding the zero locus of $P(\underline{x},\underline{z})$ in the normal 
direction. The $\underline{z}$ are the complex structure parameters of $\hat Z_3$, i.e. physically speaking the closed string moduli. 
Then one can perform the so-called Griffiths-Dwork
reduction method to obtain differential operators $\mathcal{D}_a$ with
\begin{equation}
\label{uptoexact}
 	\mathcal{D}_a(\underline{z})\Omega(\underline{z})=d\alpha_a
\end{equation}
for an two-form $\alpha_a$. Upon integration over $\Gamma^k$ this yields 
homogeneous linear differential equations for the periods $\Pi^k(\underline{z})$. 

In the subsequent sections we generalize this geometric description and technique to non-Calabi-Yau threefolds that we obtain by blowing up along the five-brane curve $\Sigma$. In particular, the generalized flux superpotential will contain both the closed superpotential $W_{\rm flux}$ as well as the brane superpotential $W_{\rm brane}$ and can be effectively calculated by solving Picard-Fuchs equations obtained from residues on SU$(3)$-structure manifolds.

\subsection{The $\cN=1$ Scalar Potential on the Full Deformation Space} \label{N=1onalldef}

In this concluding paragraph we prove the statement that the second variation of the volume \eqref{eq:volvar} is a part of the F-term potential of the D5-brane effective action. We first obtain this potential by dimensional reduction of the the DBI-action of the D5-brane. Then we use the D5-brane superpotential \eqref{eq:chain} and a generalization of the K\"ahler metric in \cite{Grimm:2008dq} to the infinite dimensional space $\mathcal{C}^\infty(\Sigma,N_{Z_3}\Sigma)$ to deduce the same potential as an F-term potential when gravity is decoupled.

We start from the ten-dimensional Dirac-Born-Infeld (DBI) action of a single D5-brane in the string frame given by
\beq \label{eq:DBI}
 S^{\text{SF}}_{\text{DBI}}=-\mu_5\int_{\mathcal{W}}d^6\xi
        e^{-\phi}\sqrt{-\text{det}\left(\iota^{\ast}\left(g_{10}+B_2\right)-\ell F\right)}\,,
\eeq
where $\mathcal{W}$ denotes the world-volume of the D5-brane that is embedded into $Z_3$ via $\iota$. Using this embedding the ten-dimensional metric $g_{10}$ and the NS-NS two-form $B_2$ are pulled back onto $\mathcal{W}$. $F$ denotes the D5-brane gauge field and $\ell=2\pi\alpha'$. We perform the dimensional reduction for the background of a D5-brane wrapping a holomorphic curve $\Sigma$. However, in contrast to the usual lore of dimensional reduction and following the logic of section \ref{branedeformationsI} we take into account general fluctuations of the D5-brane curve corresponding to sections $s$ in $\mathcal{C}^{\infty}(\Sigma,N_{Z_3}\Sigma)$. In a background with vanishing $B$-field and gauge flux $F$ the action \eqref{eq:DBI} is just the volume of the wrapped curve. Thus, the variation of $S_{\text DBI}^{\text SF}$ under a deformation along $s$ is captured, up to second order in the variation parameter $\epsilon$, by \eqref{eq:volvar} and reads  
\beq \label{eq:VDBI}
 	V_{\text{DBI}}\supset \frac{\mu_5e^{3\phi}}{\mathcal{V}^2}\left.\frac{d^2}{d \epsilon^2}\text{Vol}(\Sigma_{\epsilon s})\right\vert_{\epsilon=0}=\frac{\mu_5e^{3\phi}}{2\mathcal{V}^2}\int_{\Sigma}\norm{\bar{\partial} s}^2\, \vol_{\Sigma}\,.
\eeq 
Here we used the formula \eqref{eq:volvar} and further a Weyl-rescaling to the four-dimensional Einstein-frame to obtain the right factors of the dilaton $\phi$ and the compactification volume $\mathcal{V}$.

In the following we deduce this potential from the $\mathcal{N}=1$ formulation of the D5-brane effective action. Indeed, the term \eqref{eq:VDBI} is an F-term potential of the form
\begin{equation}
	V_{\text F}=e^K K^{a\bar b}\partial_{u_a}W_{\rm brane}\partial_{\bar{u}_{\bar b}}\bar{W}_{\rm brane}
\label{eq:Ftermpot}
\end{equation}
for the fields $u^a(x)$ associated to the expansion $s=u^as_a$ in a basis of $\mathcal{C}^{\infty}(\Sigma,N_{Z_3}\Sigma)$.
In order to evaluate $V_F$ we need the K\"ahler metric for the modes $u^a$ as well as a more tractable form of the brane superpotential $W_{\rm brane}$.  The K\"ahler metric for the $u^a$ as deformations in the infinite dimensional space $\mathcal{C}^{\infty}(\Sigma,N_{Z_3}\Sigma)$ is a straight forward generalization of the K\"ahler metric of \cite{Grimm:2008dq} originally considered for the modes counted by $H^0(\Sigma, N_{Z_3}\Sigma)$. It reads
\begin{equation}
	K_{a\bar b}=\frac{-i\mu_5e^\phi}{4\mathcal{V}}\int_{\Sigma}s_a\lrcorner \bar{s}_{\bar b}\lrcorner (J\wedge J)=\frac{i\mu_5e^\phi}{\int\Omega\wedge \bar\Omega}\int_{\Sigma}(\Omega_a)_{ij}(\bar{\Omega}_{\bar b})^{ij} \iota^*(J)\,,
\label{eq:Kahlermetric}
\end{equation}
where we introduced the abbreviation $\Omega_a=s_a\lrcorner \Omega$.
For details of this equality we refer to appendix \ref{app:potcalc}.
First we Taylor expand $W_{\rm brane}$ to second order in the brane deformations $u^a$ around the holomorphic curve $\Sigma=\Sigma_0$
\beq \label{eq:Wsecondorder}
	W_{\rm brane}=\int_{\Gamma_0}\Omega+\frac12u^au^b\int_{\Sigma}s_a\lrcorner ds_b\lrcorner \Omega+\mathcal{O}(u^3)
\eeq
where $s\lrcorner$ denotes the interior product with $s$ and $s_a$ denotes a section of $N_{Z_3}\Sigma$ that is not required to be holomorphic. $\Gamma_0$ is a chain ending on the holomorphic curve $\Sigma$, $\partial \Gamma_0=\Sigma-\Sigma_{\rm rev}$.
Second, introducing the abbreviation $\Omega_a=s_a\lrcorner \Omega$ the variation of \eqref{eq:Wsecondorder} with respect to $u_a$ reads
\beq
 	\partial_{u_a}W_{\rm brane}=-\mu_5\int_{\Sigma}\bar\partial s\lrcorner \Omega_a\,.
\eeq  
In addition we rescaled the superpotential $W_{\rm brane}\mapsto \mu_5 W_{\rm brane}$ to restore physical units as in \cite{Grimm:2008dq}.
In order to evaluate the contraction \eqref{eq:Ftermpot} we have to exploit that the $\Omega_a$ form a basis of sections of a specific bundle on $\Sigma$. Indeed, the isomorphism of $KZ_3\vert_{\Sigma}=T^* \Sigma\otimes N^*\Sigma$ which is a consequence of the normal bundle sequence of $\Sigma$ tells us that the $\Omega_a$ form a basis of sections of $\Omega^{(1,0)}(\Sigma,N^*\Sigma)$ with the property that $s_a\lrcorner \Omega_a=0$. We can use this basis to represent any other section. In particular, the contraction $	\bar\partial s\lrcorner J$ is a section of $\Omega^{(0,1)}(\Sigma,\overline{N^*\Sigma})$ that we can expand in the basis $\bar{\Omega}_{\bar a}$ as
\begin{equation} \label{eq:basisexpansion1}
	\bar\partial s\lrcorner J=\frac{-i\mu_5e^\phi}{\int\Omega\wedge \bar\Omega}\,\Omega_a\,K^{a\bar b}\int_{\Sigma}\bar{\partial} s\lrcorner \bar\Omega_{\bar b}\,.
\end{equation} 
Again we refer to appendix \ref{app:potcalc} for the details of this calculation.
Finally, we calculate the F-term potential \eqref{eq:Ftermpot} as
\begin{eqnarray} \label{eq:Ftermpotfinal}
	V_{\rm F}=e^K\mu_5^2\int_{\Sigma}\bar\partial s\lrcorner \big(\Omega_a K^{a\bar b}\int_{\Sigma}\partial \bar{s}\lrcorner \bar{\Omega}_b\big)
	= \frac{\mu_5e^{3\phi}}{2\mathcal{V}^2}\int_{\Sigma}||\bar\partial s||^2 \vol_{\Sigma}\,.
\end{eqnarray}
Here we used in the second equality the identity \eqref{eq:basisexpansion1} as well as $e^K=\frac{ie^{4\phi}}{2\mathcal{V}^2\int\Omega\wedge\bar\Omega}$, cf.~appendix \ref{app:potcalc}. The norm $||\cdot||^2$ denotes as before the contraction of all indices using the metric. 

This F-term potential is in perfect agreement with contribution \eqref{eq:VDBI} to the scalar potential $V_{\rm DBI}$ that we obtain from the reduction of the DBI-action \eqref{eq:DBI} using the variation \eqref{eq:volvar} of the calibrated volume.

\section{Five-Brane Blow-Ups and Unification of Open and Closed Deformations}
\label{5braneblowupsanddefs}

In section \ref{N=1branes} we started from the Bianchi 
identities for $F_3$ and $H_3$ and explained how five-brane 
sources on curves $\Sigma$ are properly described 
by delta-currents. Using the language of currents it was further 
explained, how to relate period and chain integrals on 
$Z_3$ to regularized integrals on the open manifold 
$Z_3-\Sigma$. 

The crucial point of the blow-up proposal \cite{Grimm:2008dq} presented in this section
is to replace the open manifold $Z_3-\Sigma$ by a physically equivalent 
geometry $\hat{Z}_3$ with a distinguished divisor $E$. In addition the blow-up
proposal naturally yields a flux $F_2=[\Sigma]$ on $E$ which can be understood as the 
partially dissolved five-brane charge. Furthermore, this implies an embedding of the open
and closed deformations of the geometry $Z_3$ and the brane on $\Sigma$ into pure 
complex structure deformations on $\hat{Z}_3$.

In the first part of this section, section \ref{geometricblowups}, 
we construct the manifold $\hat Z_3$ by blowing up a $\mathbb{P}^1$-bundle along $\Sigma$. 
This introduces the new divisor $E$ in $\hat{Z}_3$, the exceptional divisor. Outside of the 
exceptional divisor $E$, by construction a ruled surface over $\Sigma$, 
the open manifolds $Z_3-\Sigma$ and $\hat Z_3-E$ are biholomorphic. Using this fact and the formalism of  
section \ref{N=1branes} we can evaluate the open 
integrals on $\hat{Z}_3-E$. Furthermore it becomes possible 
to extend all open integrals, in particular $W_{\rm brane}$ in \eqref{chainonopenset}, the forms $H_3$ and $F_3$ as well as the closed 
periods $\Pi^k(\underline{z})$ from $\hat{Z}_3-E$ to $\hat Z_3$ by constructing local completions of these quantities in the vicinity of the divisor $E$. 
As explained in section \ref{unificationofdeformations} for 
the case in which $\Sigma$ is given as a complete intersection, 
our blow-up proposal unifies the description of the closed 
and open deformations on $Z_3$, which become now both complex 
structure deformations on $\hat Z_3$. Of particular importance 
for the superpotential \eqref{eq:unifiedSuperpots} and deformation theory on $\hat{Z}_3$ is the pullback  
of the holomorphic three-form $\Omega$ on $Z_3$ to $\hat Z_3$, 
which we construct in section \ref{hatomega}. 
In section \ref{potentialhatZ3minusD} we describe how 
the superpotentials are concretely mapped to the blow-up $\hat{Z}_3$. 

Ultimately, as explained in section \ref{su3structur}, on $\hat Z_3$ 
the flux and the brane superpotentials of section 
\ref{fivebranesuperpotential} are unified to a flux superpotential on $\hat Z_3$.  
The latter structure requires in addition to the extension 
of $\Omega$ also the extension of the K\"ahler form $J$ and the flux $H_3$ 
from $Z_3$ to $\hat Z_3$. Our formalism as presented in this section 
can be understood as the first step in the full geometrization 
of the five-brane and prepares the approach 
of section \ref{su3structur} to consider the flux-geometry  
$\hat Z_3$, $F_2$ as a string background with 
$SU(3)$ structure.

\subsection{Geometric Properties of the Blow-Up along $\Sigma$}
\label{geometricblowups} 

Given a $k$-dimensional complex submanifold $\Sigma_k$ 
in an $n$-dimensional complex manifold $Z_n$, it is a standard technology 
in algebraic geometry~\cite{Griffiths} to blow-up along $\Sigma_k$ to 
obtain a new $n$-dimensional complex manifold $\hat Z_n$. This directly applies 
to a supersymmetric five-brane on a holomorphic curve\footnote{For the use of the blow-up proposal to analyze non-holomorphic deformations of $\Sigma$ cf.~section \ref{unificationofdeformations}.} $\Sigma=\Sigma_1$ inside a Calabi-Yau threefold $Z_3$.  

\begin{figure}[htb]
\begin{center}
\includegraphics[width=.5\textwidth]{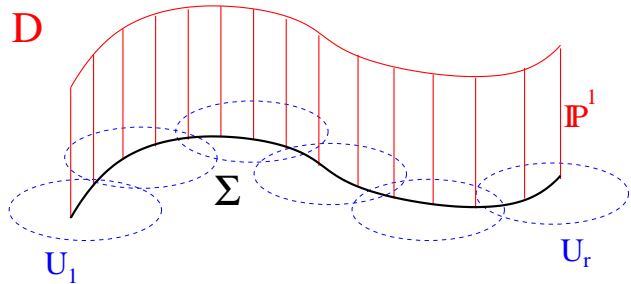}
\begin{quote}
\caption{Blow up of the curve $\Sigma$ to the ruled surface $E$} \label{blow-up}
\end{quote}
\end{center}
\end{figure}

Since the blow-up is a local operation, it can be described near the complex submanifold within neighborhoods 
$U_\alpha$ with the topology of a disk, which cover $\Sigma$. Let $y_{\alpha,\,i}$, $i=1,\ldots,3$ 
be local coordinates\footnote{Given the submanifold locally by $h_i(\underline{x})=0$, $i=1,2$ in generic $x_1,\ldots,x_3$ coordinates of $U_\alpha$ this choice is fixed by the inverse function theorem 
stating that for every point $x_0\in \C^3$ with 
$(\partial_k h_1\partial_l h_2-\partial_l h_1\partial_k h_2)\vert_{x_0}\neq0$ 
for $k,l\neq j$, there exists a local parameterization 
of $\cC$ near $x_0$ as a graph over $x_j$. 
In particular, the blow-up is independent of the coordinates 
used, cf.~p. 603 \cite{Griffiths}.} on $U_\alpha$ in which $\Sigma$ is 
specified by  the intersection of two divisors 
$D_{\alpha,\, 1}\cap D_{\alpha,\,2}$ i.e. by $y_{\alpha,\, i}=0, i=1,2$. 
  
The local blow-up can be described as a hypersurface constraint~\cite{Griffiths} 
\begin{equation}  \label{def-hatU}
\hat U_\alpha=\{(y_{\alpha,1},y_{\alpha,2}, y_{\alpha,3},
(l_{\alpha,1}:l_{\alpha,2}))\subset U_\alpha\times \mathbb{P}^1:
y_{\alpha,2} l_{\alpha,1}-y_{\alpha,1} l_{\alpha,2} =0\}\ .
\end{equation}  
Here $(l_{\alpha,1}:l_{\alpha,2})$ are projective coordinates of the $\mathbb{P}^1$ over the local patch $U_\alpha$. 
We define a projection map $\pi_\alpha:\hat{U}_\alpha\rightarrow U_\alpha$ by discarding the direction of the $\P^1$, 
\beq
	\pi_\alpha(y_{\alpha,1},y_{\alpha,2}, y_{\alpha,3},
(l_{\alpha,1}:l_{\alpha,2}))=(y_{\alpha,1},y_{\alpha,2}, y_{\alpha,3})\,.
\eeq
Obviously, $\hat U_\alpha-\pi^{-1}_\alpha(\Sigma)$ is biholomorphic to $U_\alpha-\Sigma$, 
as we can eliminate the $l_{\alpha,i}$ in the hypersurface \eqref{def-hatU} outside of the locus $y_{\alpha,1}=y_{\alpha,2}=0$ that defines $\Sigma$.  
Conversely, the set $E_\alpha:=\pi^{-1}_\alpha(0,0,y_{\alpha,3})=\pi^{-1}_{\alpha}(\Sigma)$ is described as follows. 
Over a point $(0,0,y_{\alpha,3})\in\Sigma$ in $U_\alpha$ the fibres of the projection $\pi_\alpha$ 
are canonically, i.e. independently of the coordinate system, identified with lines in 
the projectivized normal bundle $\mathbb{P}(N_{U_\alpha}\Sigma)=\mathbb{P}(\mathcal{O}(D_{\alpha,1})\oplus \mathcal{O}(D_{\alpha,2}))$, 
\begin{equation} 
(0,0,y_{\alpha,3}, (l_{\alpha,1}:l_{\alpha,2}))\,\mapsto\, l_{\alpha,1}
\frac{\partial}{\partial y_{\alpha,1}} + l_{\alpha,2} \frac{\partial}{\partial y_{\alpha,2}} \ . 
\end{equation}
This allows to glue the open sets $\hat{U}_\alpha$ to obtain $\hat{Z}_3$ and the $E_\alpha$ to obtain a divisor $E=\pi^{-1}(\Sigma)$ which is identified with
\beq \label{eq:Eglobal}
	E=\mathbb{P}(N_{Z_3}\Sigma)\,.
\eeq
Similarly we obtain a unique global projection map $\pi$ that is trivially extended to all open sets $\hat U_{\alpha}$ and thus to $\hat{Z}_3$ as the identity map on $\hat{Z}_3-E=Z_3-\Sigma$. 
The divisor $E$ is the exceptional divisor and it is a $\mathbb{P}^1$-ruled surface over $\Sigma$ by \eqref{eq:Eglobal}.

Essential facts that are intensively used in this paper are the biholomorphism 
\begin{equation}
\label{biholomorphism}
\pi:(\hat Z_3 - E)\rightarrow (Z_3- \Sigma)
\end{equation}
and the statement, that all relevant aspects of the blow-up can be 
analyzed locally in patches near $\Sigma$, except 
for the non-triviality of $N_{Z_3} \Sigma$, which 
is captured by the Thom class $\frac{e_3}{2}$. 

The relation between the local and global construction is 
particularly easy if $Z_3$ is given by a family hypersurface 
$P(\underline{x},\underline{z}) =0$ and $\Sigma$ is constructed  as a complete intersection of $P(\underline{x},\underline{z})=0$ and  
divisors $D_i$ given by $h_{i}(\underline{x},\underline{u})=0$, $i=1,2$ in $\mathbb{P}_\Delta$. Here the ambient space is in general 
a toric variety $\mathbb{P}_\Delta$ with homogeneous coordinates $\underline{x}$ and the variables $\underline{z}$, $\underline{u}$ parameterize the complex structure and brane moduli respectively. 
Then we can choose in any patch $U_\alpha$ coordinates so that $y_{\alpha,1}=h_1(\underline{x})|_{U_\alpha}$, 
$y_{\alpha,2}=h_2(\underline{x})_{U_\alpha}$ and $y_{\alpha,3}$ 
is a coordinate along $\Sigma$. Now $N_{U^{\alpha}}\Sigma$ is globally given as the sum of two line bundles, $N_{Z_3}\Sigma =\mathcal{O}(D_1)\oplus\mathcal{O}(D_2)$, and
$\hat Z_3$ is given globally as the complete intersection in the total space of the
projective bundle
\begin{equation} \label{eq:Wambient}
\mathcal{W}=\mathds{P}(\mathcal{O}(D_1)\oplus\mathcal{O}(D_2)).
\end{equation} 
Indeed, using the projective coordinates $(l_1,l_2)\sim
\lambda(l_1,l_2)$ on the $\mathds{P}^1$-fiber of the blow-up, $\hat{Z}_3$ can be
written as
\begin{equation}
        P(\underline{x},\underline{z})= 0\ ,\qquad 	Q\equiv l_1 h_2(\underline{x},\underline{u}) - l_2 h_1(\underline{x},\underline{u}) = 0 \, ,
\label{eq:blowup}
\end{equation}
in the projective bundle $\mathcal{W}$. 

We conclude by summarizing the basic geometrical properties of the blow-up $\hat{Z}_3$ of the Calabi-Yau threefold $Z_3$ along the curve $\Sigma$~\cite{Griffiths, Grimm:2008dq}.
We note that the intersection ring on the blow-up has the following 
relations on the level of intersection curves 
\beq
   E^2 = - \pi^* \Sigma  - \chi(\Sigma) F\ , \qquad E\cdot \pi^* D = (\Sigma \cdot D) F\ ,
\eeq
where $D$ is any divisor in $Z_3$, $F$ is the class of the $\P^1$-fiber of $E$, 
and $\chi(\Sigma)$ is the Euler number of the blow-up curve $\Sigma$.
The intersection numbers read
\beq
  E^3 = \chi(\Sigma)\ , \quad  F  \cdot E= -1\ , \quad F \cdot \pi^* D = 0 \ , \quad 
   E \cdot \pi^* \tilde \Sigma  = F \cdot \pi^* \tilde \Sigma = 0\ ,
\eeq
where $\tilde \Sigma$ is any curve in $Z_3$.
Since the $\mathbb{P}^1$-fibration of $E$ does not degenerate, 
the Hodge numbers of $E$ are equal to those of 
$\mathbb{P}^1\times \Sigma$ and hence  
\beq \label{eq:hodgenumbersblowup}
h^{(0,0)}=
h^{(2,2)}=1\,,\quad h^{(1,1)}=2\,,\quad h^{(1,0)}=h^{(1,2)}=g\,,\quad h^{(2,0)}=0\,, 
\eeq
where $g$ is the genus of $\Sigma$. In particular this introduces $g$ new classes in $H^{(2,1)}(\hat{Z}_3)$ that can not be obtained from $H^{(2,1)}(Z_3)$ via $\pi^*$. However, since $\hat{Z}_3$ does not meet the Calabi-Yau condition, this does not necessarily lead to $g$ new complex structures on $\hat{Z}_3$ as $H^{(2,1)}(\hat{Z}_3)$ is not necessarily isomorphic to $H^{1}(\hat{Z}_3,T\hat{Z}_3)$ which is the appropriate cohomology group counting complex structure deformations, as reviewed in section \ref{bulkdeformations}. In fact, the first Chern class of the blow-up is $c_1(\hat Z_3)=-K_{\hat Z_3}=\pi^*(c_1(Z_3))-[E]=-[E]\neq 0$ where we used that $Z_3$ is a 
Calabi-Yau manifold in the last equality. Hence, $\hat Z_3$ 
is not a Fano variety. Physically, it seems thus not possible 
to obtain a globally consistent string compactification by adding seven brane 
charges as in $F$-theory. For Calabi-Yau blow-ups  
the second Chern class is given by $c_2(\hat Z_3)=\pi^*(c_2(Z_3)+\eta_\Sigma)$, where 
$\eta_\Sigma\in H^4(Z_3)$ is the class dual to $\Sigma$.

It is a crucial feature of the blow-up procedure, that no \textit{new} degrees of freedom associated to deformations of $E$ are introduced.
Since by construction the normal bundle to $E$ in $\hat Z_3$ 
is the tautological bundle\footnote{The tautological bundle can be defined on any 
projectivization of a vector bundle, like $E=\mathbb{P}(N_{Z_3}\Sigma)$, by the defining property that $T$ restricted to each fiber agrees with the universal bundle $\mathcal{O}(1)$ on projective space \cite{Griffiths}.} $T$, which is a negative bundle on
$E$, $E$ has no 
deformations,
\beq \label{eq:Eisolated}
	H^0(E,N_{\hat Z_3}E)=\emptyset\,,
\eeq
i.e.~$E$ is isolated. Furthermore, it can be shown mathematically rigorously that all deformations of $Z_3$ and the curve $\Sigma$, that are deformations of complex structures of $Z_3$ and deformations of $\Sigma$ in $Z_3$, map to complex structure deformations of $\hat{Z}_3$.\footnote{We thank Daniel Huybrechts for a detailed explanation of the equivalence of the two deformation theories.} The K\"ahler sector of $Z_3$ maps to that of $\hat{Z}_3$ that contains one additional class of the exceptional divisor $E$.

\subsection{Unification of the Open and Closed Deformations Spaces}
\label{unificationofdeformations}  

In the first part of this section we present the key points of the blow-up proposal suggested in \cite{Grimm:2008dq} and its use to analyze the geometrical dynamics of five-branes. Then, in a second part we study the deformation space of branes wrapping rational curves via the complete intersection curves \eqref{eq:blowup} and the blow-up of the latter. 

\subsubsection{Matching Deformations and Obstructions: a Mathematical Proposal}
\label{unificationofdeformationsI}

We have described  in section \ref{branedeformationsII} that the 
infinitesimal elements $\varphi_*^z(\frac{\partial}{\partial u_a})=\partial_{u_a} \varphi(z;\underline{u})$ span 
the tangent space to the open deformations space  and 
live in $H^0(\Sigma,N_{Z_3} \Sigma)$, while it was reviewed 
in section \ref{bulkdeformations} that deformation of 
the closed complex structure deformations of a 
manifold $M$ live\footnote{Most statements about the complex structure
deformations apply to $Z_3$ and $\hat Z_3$. We denote both complex manifolds
by $M$ in the following.} in $H^1(M,T_{M})$. 
Up to global automorphisms of the toric ambient space $\mathbb{P}_\Delta$, 
which are compatible with the torus action, this cohomology  can be represented by 
the infinitesimal deformations $\delta_z$ of the parameters $\underline{z}$  
multiplying monomials in $P(\underline{x},\underline{z})= 0$ 
of the hypersurface. Likewise for the 
complete intersection (\ref{eq:blowup}) elements in $H^1(M,T_{M})$ 
can be represented by infinitesimal deformations $\delta_{(z,u)}
=:\delta_{\hat z}$ of the parameters in (\ref{eq:blowup}), 
modulo global automorphisms of $\cal W$.  
Using these facts, it is easy to check 
for the complete intersections  description 
(\ref{eq:blowup}) that  the moduli $\underline{u}$ of $\Sigma$ 
described by the coefficients of the monomials in 
$h_i(\underline{x},\underline{u})$, $i=1,2$ turn into 
complex structure moduli $\hat{\underline{z}}$ of $\hat{Z}_3$ since
$h_i(\underline{x},\underline{u})$, $i=1,2$  
enter the defining equations of $\hat{Z}_3$ 
via $Q$ in \eqref{eq:blowup}.

\begin{figure}[htb]
\begin{center}
\includegraphics[width=0.9\textwidth]{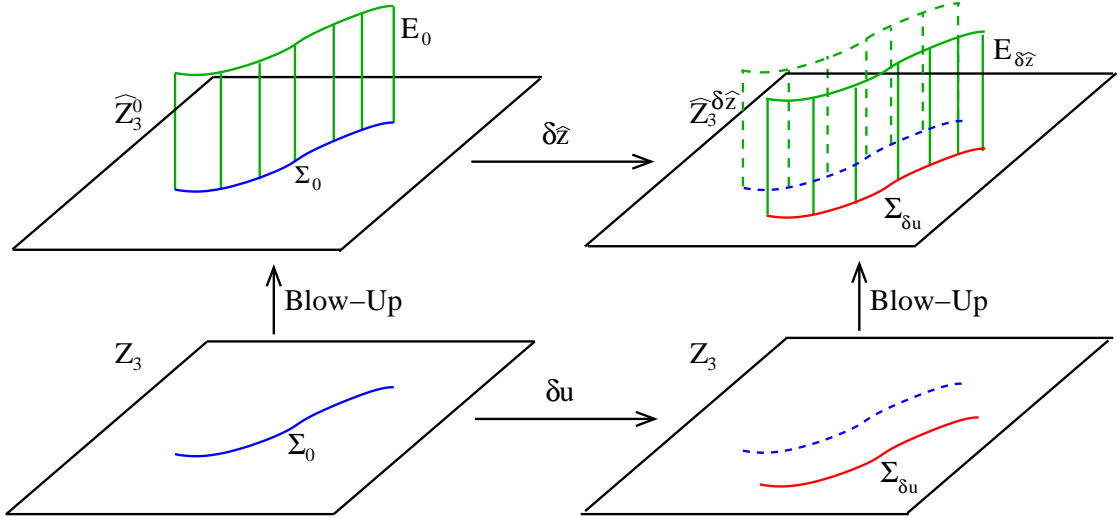}
\begin{quote}
\caption{Map of brane to complex structure deformations. A displacement $\Sigma_{\delta u}$ of a holomorphic curve $\Sigma_0$ yields a new blow-up $\hat{Z}^{\delta \hat{z}}_3$ with induced complex structure deformation $\delta \hat{z}$ and divisor $E_{\delta \hat{z}}$. The divisor $E_0$, which is holomorphic in $\hat{Z}_3^{0}$, is not holomorphic in $\hat{Z}_3^{\delta \hat z}$.} \label{fig:defsblow-up}
\end{quote}
\end{center}
\end{figure}

As noted below \eqref{eq:Eisolated}, the divisor $E$ is isolated in $\hat Z_3$, i.e. on $\hat Z_3$ there are no 
deformations associated to $E$. From this follows more illustratively, 
that blowing up along $\Sigma$ for different values of $\underline{u}$ yields 
diffeomorphic blow-ups $\hat{Z}_3$ which just differ by a choice of complex structure. The situation is visualized in figure \ref{fig:defsblow-up}. 
Mathematically, the equivalence even of the full deformation theory 
of $(Z_3,\Sigma)$ and $\hat Z_3$ are expected in general. This means that
not only the deformations of $(Z_3,\Sigma)$, counted by elements in $H^{1}(Z_3,TZ_3)$
and $H^{0}(\Sigma,N_{Z_3}\Sigma)$, agree with the complex structure deformations of $\hat{Z}_3$, that are in
$H^{1}(\hat{Z}_3,T\hat{Z}_3)$, but also the obstruction problems of both geometries. In particular, this implies an order-by-order match of the obstructions on both sides of the correspondence. 

As detailed in the section \ref{bulkdeformations} and
\ref{branedeformationsII} possible obstructions to the closed 
deformation space live in $H^2(M,T_M)$ while possible obstructions 
to the open deformations space live in $H^1(\Sigma,N_{Z_3} \Sigma)$. 
While one can conclude from the vanishing of these homology groups 
that the corresponding deformations are unobstructed, it is not 
necessarily true that the deformation problems are obstructed, if 
these homology groups do not vanish. In particular the complex 
structure deformations of Calabi-Yau spaces, such as $Z_3$, are unobstructed 
despite the fact that $H^2(Z_3,T_{Z_3})\neq 0$.
However the deformations of the curve $\Sigma$ can in general be obstructed by
elements $H^{1}(\Sigma,N_{Z_3}\Sigma)$ at some order. Given the equivalence of the
obstruction problems we expect that these are precisely matched by the obstructions to complex 
structure deformations on $\hat{Z}_3$ in $H^2(\hat{Z}_3,T\hat{Z}_3)$. 
In physical terms, the obstruction problem is in general expressed by a superpotential, which in the case
of the obstruction problem of $(Z_3,\Sigma)$ is given by the superpotential \eqref{eq:unifiedSuperpots}.
Our strategy to investigate the possible obstructions of $(Z_3,\Sigma)$ will be to match 
the calculation of the superpotential \eqref{eq:unifiedSuperpots} before and after the blow-up. 
Here it will be crucial to understand the lift of the brane superpotential $W_{\rm brane}$ of \eqref{eq:chain} under the blow-up, that
will be replaced by a specific flux superpotential on $\hat{Z}_3$. This flux superpotential induces obstructions to complex structure
deformations on $\hat{Z}_3$ that are equivalent to the original obstructions on moving the brane on $\Sigma$ expressed by $W_{\rm brane}$.
We will discuss the match of the superpotentials in a two step procedure in sections 
\ref{potentialhatZ3minusD} and \ref{su3structur}. In particular the blow-up $\hat{Z}_3$
will yield an easy calculational scheme of the superpotential as explained and applied to specific examples in section \ref{IIBBlowUp}.

\subsubsection{Matching Deformations and Obstructions: Concrete Examples}
\label{unificationofdeformationsII}

Before we proceed we have to explain how we use the blow-up $\hat{Z}_3$ constructed as the complete 
intersection (\ref{eq:blowup}) to calculate the superpotentials $W_{\rm brane}$ for five-branes on rational curves. This is crucial since the families 
of holomorphic curves themselves defined by the complete intersection of complex equations $h_1=h_2=0$ are  
unobstructed. Similarly on $\hat Z_3$ 
the corresponding complex structure deformations are unobstructed and the deformation problem and 
the corresponding superpotentials are trivial.

The general statement for the moduli 
space of holomorphic curves\footnote{In the following we will use the term of a 'moduli space' of holomorphic curves in $Z_3$ to denote an analytic family of holomorphic curves as introduced in section \ref{branedeformationsII}.} on Calabi-Yau threefolds is 
that its virtual deformation space is zero-dimensional~\cite{Kachru:2000ih, Kachru:2000an, katz, mirrorbook}.
Naively this could be interpreted as the statement, that generically holomorphic curves in a Calabi-Yau threefold never 
occur in families.
However, this conclusion is not true as one can learn already from the case of 
rational curves\footnote{A rational curve is birationally equivalent to a line i.e.~a curve of genus zero which is a $\P^1$.} in the quintic as explained in~\cite{katz}.
Rational curves in a generic Calabi-Yau manifold $Z_3$, like the quintic with a constraint $P=0$ including $101$ 
complex structure parameters $\underline{z}$ at generic values, are isolated and have a moduli space consisting 
of points, which we denote by ${\cal M}^{\underline z}(\mathbb{P}^1)=pts.$  However, at special loci  
$\underline{z}_0$ in the complex structure moduli space, which correspond to specially symmetric 
Calabi-Yau constraints $P=0$ like the Fermat point
\begin{equation} \label{eq:Fermatlocus}
	P=x_1^5+x_2^5+x_3^5+x_4^5+x_5^5\,
\end{equation}
of the quintic, a family of
curves parametrized by a finite dimensional moduli space 
${\cal M}^{{\underline  z}_0}(\mathbb{P}^1)$ can appear. 
Physically this means that the open superpotential 
becomes a constant of the brane moduli and  the scalar potential has a flat direction 
along ${\cal M}^{{\underline z}_0}(\mathbb{P}^1)$.
However it can be generally argued~\cite{Kachru:2000ih} 
that in the vicinity of the special loci in the closed string 
deformation space $\underline{z}_0$ a superpotential develops 
for the rational curves. In agreement with (\ref{eq:csinducedmass}) the superpotential starts linear in the closed 
string deformation $t\propto \delta_{z}$ away from $\underline{z}_0$ and is of arbitrary 
order in the open string moduli so that it has
$(-1)^{{\rm dim}({\cal M}^{{\underline  z}_0}(\mathbb{P}^1))} 
\chi({\cal M}^{\underline{z}_0}(\mathbb{P}^1))$ minima\footnote{This formula 
follows from complex deformation invariance of the BPS numbers associated 
to holomorphic curves~\cite{KatzKlemmVafa}.}. 
We note that this is precisely the most interesting physical situation, 
as $t$ can be made parametrically small against the compactification scale, as explained below equation (\ref{eq:csinducedmass}).

There is one important caveat in order when working with concrete algebraical curves. A given family of holomorphic curves in a specific algebraic representation $P=0$ of $Z_3$ can become obstructed due to the presence of non-algebraic complex structure deformations, 
i.e.~those $\underline{z}$ that are not contained in $P=0$. For example, this situation occurs in the 
Calabi-Yau hypersurfaces $Z_3$ of degree $2+2(n_1+n_2+n_3)$ in weighted projective spaces 
of the type $\mathbb{P}^4(1,1,2 n_1, 2 n_2, 2 n_3)$ 
with $n_i\in \mathbb{Z}$ as discussed in~\cite{Kachru:2000ih}.
This realization of the Calabi-Yau manifold $Z_3$ contains always 
a ruled surface, i.e. an $\mathbb{P}^1$ fibered over a (higher genus) 
Riemann surface\footnote{The same Riemann surface is identified with the moduli space ${\cal M}^{{\underline z}_0}(\mathbb{P}^1)$ in this case.}. The embedding of $Z_3$ 
in this particular ambient space is such that the generic 
obstructed situation, which corresponds to a non-vanishing 
superpotential, is not accessible using the algebraic 
deformations, i.e.~upon tuning the parameters in the Calabi-Yau constraint $P=0$.
The absence of these deformations as algebraic deformations in $P=0$ happens since 
the corresponding monomials are not compatible with the symmetries of the 
ambient space.

Let us next describe the obstructed deformation problem of rational curves and the relation to the complete intersection curves and the blow-up (\ref{eq:blowup}). 
The basic idea is to map the obstructed deformations of the rational curves to the 
algebraic moduli space of the complete intersection in the following way.
As mentioned in (\ref{eq:blowup}) the algebraic 
deformations parametrized by the closed moduli 
$\underline{z}$ and the open moduli ${\underline u}$ are unobstructed. 
Let us denote the corresponding open and closed  
moduli space of the complete intersection $\Sigma$, 
defined by $P(\underline{z})=0$ and  $h_{i}(\underline{x},\underline{u})=0$, $i=1,2$,
by ${\cal M}(\Sigma)$ and the open moduli space of 
$\Sigma$ for fixed closed moduli $\underline{z}$ by ${\cal M}^z(\Sigma)$. 
The generic dimension of this open moduli space 
$h^0(N \Sigma)$ is positive. 
The idea is to consider a representation of $Z_3$ which is 
compatible with a discrete symmetry group $G$. This symmetry group 
allows us to identify lower degree and genus curves with the complete intersection $\Sigma$ at a special 
sublocus of the moduli space ${\cal M}(\Sigma)$. In our main examples in section \ref{ToricBraneBlowup} 
and \ref{ToricBraneBlowupII} these are rational curves, i.e.~curves of degree one. Let us denote this sublocus by ${\cal M}_{\mathbb{P}^1}(\Sigma)$. This sublocus is determined by the requirement that the algebraic constraints $P$, $h_i$ degenerate so that they can be trivially factorized as powers of linear constraints,
\begin{equation} \label{eq:sublocus}
 	{\cal M}_{\mathbb{P}^1}(\Sigma):\quad P(\underline z)=h_i(\underline{u})=0\,\quad \Leftrightarrow\quad \prod_k\sum_l a^{(s)}_{lk} x_l=0\,,\,\, s=1,2,3\,,
\end{equation}
where the different linear factors $L^{(s)}_k=\sum_l a^{(s)}_{lk} x_l$ are identified by the discrete group $G$, $L^{(s)}_{k_1}\leftrightarrow L^{(s)}_{k_2}$. Then the right hand side of this identification describes rational curves
\begin{equation} \label{eq:rationalcurves}
  L^{(1)}_{k_1}=L^{(2)}_{k_1}=L^{(3)}_{k_1}=0
\end{equation}
modulo $G$ in the ambient space and in $Z_3$ since $P=0$ is trivially fulfilled. For a concrete situation we refer to section \ref{ToricBraneBlowup}.
In particular this identification embeds the moduli space 
${\cal M}^{{\underline z}_0}(\mathbb{P}^1)$ into ${\cal M}^{{\underline z}_0}(\Sigma)$ 
and more trivially the (discrete) ${\cal M}^{{\underline z}}(\mathbb{P}^1)$ 
into ${\cal M}^{{\underline z}}(\Sigma)$. 

\begin{figure}[htb]
\begin{center}
\includegraphics[width=.7\textwidth]{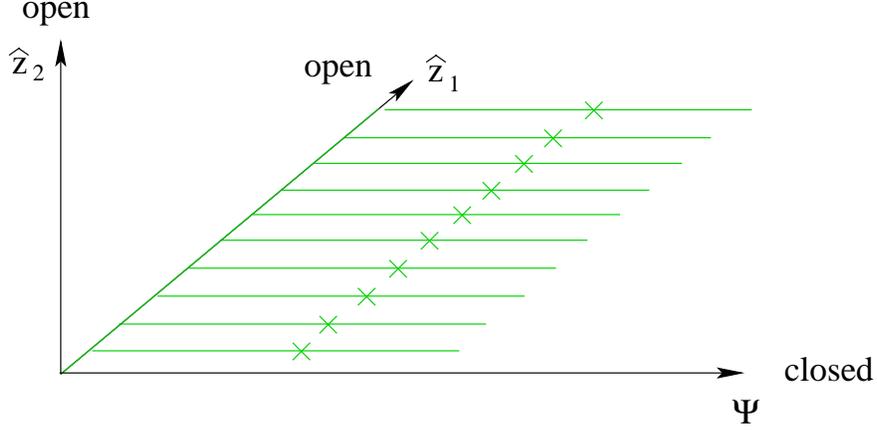}
\begin{quote}
\caption{Picture of the moduli space of ${\cal M}(\Sigma)$ given by the quintic 
modulo $\mathbb{Z}_5^3$ and the loci ${\cal M}_{\mathbb{P}^1}(\Sigma)$ where $\Sigma$ degenerates to (holomorphic) rational curves. All lines in the $\hat{z}_2=0$ plane correspond to the embedding of the moduli space of rational curves ${\cal M}_{\mathbb{P}^1}(\Sigma)\subset {\cal M}(\Sigma)$, cf.~(\ref{eq:RSquintic}).
At $\Psi=0$ the $\hat{z}_1$ direction opens up as a modulus of a family of $\mathbb{P}^1$'s over 
the genus $6$ curve ${\cal M}^{\Psi=0}(\mathbb{P}^1)$. For generic values of $\Psi$ 
only $(-1)^{{\rm dim}({\cal M}^{\Psi=0}(\mathbb{P}^1))} \chi({\cal M}^{\Psi=0}(\mathbb{P}^1))=10$ 
points belong to ${\cal M}^{\Psi}(\mathbb{P}^1)$. 
Away from ${\cal M}_{\mathbb{P}^1}(\Sigma)$ the holomorphic configuration in $Z_3$ is an irreducible higher genus curve, 
which corresponds to non-holomorphic $S^2$'s in $Z_3$.} \label{fig:moduli}
\end{quote}
\vspace{-1cm}
\end{center}
\end{figure}

More generally, i.e.~away from the sublocus ${\cal M}_{\mathbb{P}^1}(\Sigma)$ defined by \eqref{eq:sublocus}, this embedding implies that
the \textit{obstructed} deformation space of the 
rational curves \eqref{eq:rationalcurves} is identified with the \textit{unobstructed} moduli space ${\cal M}(\Sigma)$. This can be compared to the method presented in \cite{Jockers,Alim:2009rf,Alim:2009bx} where the obstructed deformations of a curve are identified with the unobstructed moduli of an appropriate divisor. For the curves we consider we depict the embedding of 
the deformation spaces of the rational curves into the moduli space ${\cal M}(\Sigma)$ of complete intersection $\Sigma$ in figure \ref{fig:moduli}, where we introduce new open moduli $\hat{z}^1$, $\hat{z}^2$ that are functions of the $u^i$.
The point is that away from ${\cal M}_{\mathbb{P}^1}(\Sigma)$ 
the identification \eqref{eq:sublocus} of the complete intersection curve $\Sigma$ with the holomorphic rational curves in 
the locus $P=0$ fails. We can understand this failure in two different ways emphasizing different aspects of the identification of the deformation space of rational curves with the true moduli space ${\cal{M}}(\Sigma)$. If we analyze the 
identification of $\Sigma$ with holomorphic rational 
curves infinitesimally close to ${\cal M}_{\mathbb{P}^1}(\Sigma)$  
one can either keep the linear constraints \eqref{eq:rationalcurves} and relax the condition that these rational 
curves lie identically on the $P=0$ locus 
or we linearize the equations $P=h_i=0$ such that the rational curves keep lying in the $P=0$ locus. 
However, since the trivial factorization \eqref{eq:sublocus} and the identification of the linear factors $L^{(s)}_k$ modulo $G$ fails, the latter possibility introduces non-holomorphic equations 
with nontrivial branching. This implies that the rational curve away from ${\cal M}_{\P^1}(\Sigma)$ inside ${\cal M}(\Sigma)$ is not holomorphic in $Z_3$. In particular, the process of turning on 
$t=\delta_{z}$ around $\underline{z}_0$ for values of the open moduli, 
that are only close to ${\cal M}^{ \underline{z}_0+\delta_{z} }(\mathbb{P}^1)$, can 
be understood as deforming the analytic family  ${\cal{M}}^{\underline{z}_0}(\P^1)$ of rational curves inside the $P=0$ locus to rational curves, which are non-holomorphic. For concreteness, for the later example (\ref{eq:RSquintic}) of the quintic, we identify $\delta z=\Psi$
with the complex structure of the mirror quintic deforming the Fermat locus \eqref{eq:Fermatlocus} as
\begin{equation}
 	P=x_1^5+x_2^5+x_3^5+x_4^5+x_5^5-5\Psi x_1x_2x_3x_4x_5\,.
\end{equation}
Then, $\delta z\neq 0$ deforms away from the one-dimensional moduli space of rational curves 
${\cal M}^{ \underline{z}_0}(\mathbb{P}^1)$ in $P=0$, that exists at the Fermat locus $\underline{z}_0=0$.

As noted before, a brane on a non-holomorphic curve is not supersymmetric and thus violates the F-term supersymmetry 
condition that is expressed by the 
superpotential $W_{\rm brane}$. In the following, in particular in the examples of sections \ref{ToricBraneBlowup} and \ref{ToricBraneBlowupII}, we will consider five-branes on rational curves in a given Calabi-Yau threefold $Z_3$ and their excitation about the supersymmetric minimum that correspond, in geometric terms, to possibly obstructed deformations about holomorphic curves. The brane excitations we consider correspond on the one hand to light fields that are generically obstructed and that become, as discussed in section \ref{branedeformationsI}, massless at some point in the complex structure moduli space. On the other hand we include fields that parameterize non-holomorphic deformations for all values of the closed moduli. In figure \ref{fig:moduli} the first type of fields corresponds to $\hat{z}_1$, which becomes massless at $\Psi=0$, and the second type of fields corresponds to $\hat{z}_2$. However, the crucial point for the consideration of this deformation space, as noted above, is the identification with the moduli space ${\cal M}(\Sigma)$ of complete intersection curves $\Sigma$. This identification and the unification of the open-closed moduli space of $(Z_3,\Sigma)$ in the blow-up \eqref{eq:blowup} of the complete intersection curve $\Sigma$ will enable us to calculate the superpotential $W_{\rm brane}$ for branes wrapping rational curves. We determine the periods on the complex structure moduli space of the blow-up $\hat{Z}_3$ that physically describe the closed and open superpotential $W_{\rm brane}$ upon turning on an appropriate flux on $\hat{Z}_3$. For this purpose we will describe $W_{\rm brane}$ explicitly by chain and flux integrals on the blow-up $\hat{Z}_3$ respectively in section \ref{potentialhatZ3minusD} and section \ref{su3structur}. Finally, we note that the periods on $\hat{Z}_3$ can equivalently be understood as a definition of the concept of periods on the \textit{brane moduli space} ${\cal M}(\Sigma)$, extending the familiar notion of periods on the complex structure moduli space of a Calabi-Yau manifold.

\subsection{Probing the Open-Closed Deformation Space: the Pullback of $\Omega$}
\label{hatomega}

The key in  describing  the deformations of complex structures on $\hat
Z_3$ are the construction and the properties of the pull-back $\hat \Omega=\pi^*(\Omega)$ of 
the  holomorphic three-form $\Omega$ from the Calabi-Yau threefold $Z_3$ to $\hat
Z_3$. Since the blow-up is a local procedure $\hat \Omega:= \pi^*(\Omega)$ will first be constructed in the local patches $\hat{U}_\alpha$ and then be globalized as a residue integral for the complete intersection \eqref{eq:blowup}. From this we obtain differential equations, the Picard-Fuchs equations, which determine the full complex structure dependence of $\hat{\Omega}$ and its periods.

Let us summarize the results of the actual calculation, which is done 
in appendix \ref{App:Local}. As in section \ref{geometricblowups} we assume that $\Sigma$ is represented as a complete intersection of divisors $D_i$, $i=1,2$, in $Z_3$ given by constraints $h_i(\underline{x},\underline{u})=0$ for coordinates $\underline{x}$. If we consider a patch $U_\alpha$ on $Z_3$ near $\Sigma$, then the holomorphic three-form $\Omega$ is locally given by 
\begin{equation} \label{eq:omegaLocal}
 \Omega=\dd x_1\wedge \dd x_2 \wedge \dd x_3 =\det J^{-1} \dd y_{\alpha,1} \wedge \dd y_{\alpha,2} \wedge \dd
y_{\alpha,3}\,,
\end{equation}
where $J$ is the Jacobian matrix for choosing coordinates $y_{\alpha,i}=h_i(\underline{x})$, $i=1,2$, and $y_{\alpha,3}=x_3$, 
starting with generic coordinates $\underline{x}$. This expression is pulled back via the projection map $\pi:\hat{Z}_3\rightarrow Z_3$ to the patch $\hat U_\alpha$ defined in  \eqref{def-hatU}. We introduce coordinates
\begin{equation}
l_1\neq 0\,:\quad z_{\alpha,1}^{(1)}=y_{\alpha,1}\,,\quad z_{\alpha,2}^{(1)}=\frac{l_2}{l_1}=\frac{y_{\alpha,2}}{y_{\alpha,1}}\,,\quad  
z_{\alpha,3}^{(1)}=y_{\alpha,3} 
\end{equation}
on $\hat{U}_\alpha$ for $l_1\neq 0$ to obtain  
\begin{equation} \label{eq:pullbacklocal}
\hat \Omega=\pi^*(\Omega)=z_{\alpha,1}^{(1)} \det
J^{-1} \dd z_{\alpha,1}^{(1)} \wedge 
\dd z_{\alpha,2}^{(1)} \wedge \dd z_{\alpha,3}^{(1)}\,.
\end{equation}
Here the subscript $*_\alpha$ and the superscript $*^{(1)}$ label the patches $U_\alpha$ on $Z_3$ as well as the patch $l_1\neq 0$ on the exceptional $\P^1$ with projective coordinates $(l_1:l_2)$\footnote{We drop the label $\alpha$ on the coordinates $l_i$ in order to shorten our formulas.}.
We obtain a similar expression on the second patch $l_2\neq 0$ of $\P^1$ using local coordinates $z_{\alpha,1}^{(2)}=\frac{l_1}{l_2}$, $z_{\alpha,2}^{(2)}=y_{\alpha,2}$ and  $z_{\alpha,3}^{(2)}=y_{\alpha,3}$. 
    
Now one can show that the pull-back map $\pi^*$ on $\Omega$ can be written as the residue
\begin{equation} \label{eq:HatOmegaLocal}
\hat{\Omega}=\int_{S^1_Q}\frac{h_i}{l_i}\frac{\Delta_{\P^1}}{Q}\wedge \Omega\,,\quad i=1,2\,.
\end{equation}
On can easily check that this is globally well-defined on both patches $l_i\neq 0$, $i=1,2$, covering the $\mathbb{P}^1$ using to the blow-up constraint $Q$ in \eqref{def-hatU} respectively \eqref{eq:blowup}. Here we insert the local expression \eqref{eq:omegaLocal} for $\Omega$ and 
\begin{equation} 
\Delta_{\mathbb{P}^1}=l_1 \dd l_2- l_2 \dd l_1\,,
\end{equation}
which is the invariant top-form on $\P^1$.  
In fact, the residuum \eqref{eq:HatOmegaLocal} specializes correctly to the local expressions \eqref{eq:pullbacklocal} of $\hat{\Omega}$ in every chart. This ensures that the residuum expression on $\hat{U}_\alpha$ can be globalized to $\hat{Z}_3$. We use the
standard residuum expression for the holomorphic three-form 
$\Omega$ given in (\ref{eq:residueZ3}) to replace the local expression \eqref{eq:omegaLocal} by
\beq
\label{eq:ResZhat}
\hat \Omega =
\int_{S^1_P}\int_{S^1_Q} \frac{h_1}{l_1}\frac{\Delta}{P Q}=\int_{S^1_P}\int_{S^1_Q} \frac{h_2}{l_2}\frac{\Delta}{P Q}\ ,
\eeq
where $P,Q$ are the two constraints of \eqref{eq:blowup}. The five-form $\Delta$ denotes an invariant holomorphic top-form on the five-dimensional ambient space
$\mathcal{W}$ defined in \eqref{eq:Wambient} and $S^1_P$, $S^1_Q$ are small loops around $\{P=0\}$, $\{Q=0\}$ encircling only the corresponding poles. 
The measure $\Delta$ is given explicitly in section \ref{ToricBranes}.
For the example of a trivial fibration it takes the schematic form 
\begin{equation}
 	\Delta=\Delta_{\mathbb{P}_{\Delta}}\wedge \Delta_{\mathbb{P}^1}\, .
\end{equation}
where $\Delta_{\mathbb{P}_{\Delta}}$ denotes the invariant top-form on the toric basis $\mathbb{P}_{\Delta}$. 

Let us now discuss the essential properties of $\hat{\Omega}$ and of the residue integral expression \eqref{eq:ResZhat}. By construction of $\hat{Z}_3$, in particular by the isomorphism $H^{(3,0)}(\hat{Z}_3)\cong H^{(3,0)}(Z_3)$, $\hat{\Omega}$ is the unique generator of $H^{(3,0)}(\hat{Z}_3)$ \cite{Grimm:2008dq}.
In general $\hat{\Omega}$ varies under a deformation of the complex structure on $\hat{Z}_3$. This is due to the fact that the notion of holomorphic and anti-holomorphic coordinates changes when changing the complex structure. More rigorously, this is described by the variation of Hodge structures, where the split 
\begin{equation}
  H^3(\hat{Z}_3)=\bigoplus_{i=0}^3 H^{(3-i,i)}(\hat{Z}_3)
\end{equation}
by the Hodge type $(p,q)$ is analyzed over the complex structure moduli space ${\cal M}(\hat{Z}_3)$ of $\hat{Z}_3$. Then $\hat{\Omega}$ is a holomorphic section of the locally constant vector bundle $H^3(\hat{Z}_3)$ over ${\cal M}(\hat{Z}_3)$ and is of type $(3,0)$ at a fixed point $\underline{\hat{z}}_0$ in ${\cal M}(\hat{Z}_3)$. When moving away from $\underline{\hat{z}}_0$ by an infinitesimal displacement $\delta_{z}$ the Hodge type of $\hat{\Omega}$ changes according to the diagram
\begin{equation} \label{eq:HodgeClosed}
         H^{(3,0)}(\hat{Z}_3)={\mathcal F}^3\, \overset{\delta_z}{\longrightarrow} \,
         \mathcal F^2 \, \overset{\delta_z}{\longrightarrow} \, 
        \mathcal F^1 \, \overset{\delta_z}{\longrightarrow} \,
                {\mathcal F}^0=H^3(\hat{Z}_3)
\end{equation}
where we define the holomorphic vector bundles $\mathcal{F}^p$ over $\underline{\hat{z}}$ in ${\cal M}(\hat{Z}_3)$ by
\begin{equation} \label{eq:Hodgefilt}
 	\mathcal{F}^p\vert_{\underline{\hat{z}}}=\bigoplus_{i\geq p}H^{(i,3-i)}(\hat{Z}_3)\vert_{\underline{\hat{z}}}\,.
\end{equation}
Most importantly, this implies the existence of differential equations, the Picard-Fuchs equations, on ${\cal M}(\hat{Z}_3)$ since the diagram \eqref{eq:HodgeClosed} terminates at fourth order in $\delta_z$. These can be explicitly obtained from the residue integral representation \eqref{eq:ResZhat} of $\hat{\Omega}$ by applying the Griffiths-Dwork reduction method. The Picard-Fuchs system in turn determines the full moduli dependence of $\hat{\Omega}$ and its periods.

Our general strategy will be to calculate all integrals 
relevant for the evaluation of the superpotential, discussed in section \ref{fivebranesuperpotential}, on $\hat Z_3$ using this Picard-Fuchs system. In fact, in the toric examples of sections \ref{ToricBraneBlowup}, \ref{ToricBraneBlowupII} we obtain a GKZ-system whose solutions are the periods of $\Omega$ and the brane superpotential $W_{\rm brane}$. This, as explained in \ref{unificationofdeformations}, unifies the closed and open deformations of $(Z_3,\Sigma)$, but, as we will see later in more detail,  
also the expression for individual pieces of the superpotential into a flux superpotential on $\hat{Z}_3$.  
Indeed, the unification of open-closed deformations, as mentioned before, as complex structure deformations on $\hat{Z}_3$ guarantees, that the study of variations of pure Hodge structures \eqref{eq:HodgeClosed} is sufficient \cite{Grimm:2008dq}. This is true despite the fact that $\hat{\Omega}$ vanishes\footnote{This can be directly seen, using the fact that $E$ is given in local coordinates by $z^{(1)}_{\alpha,1}=0$ respectively $z^{(2)}_{\alpha,2}=0$, from the local expression \eqref{eq:HatOmegaLocal} and its global counterpart (\ref{eq:ResZhat}).} as a section of 
$K\hat{Z}_3=E$ along the exceptional divisor $E$. Consequently $\hat{\Omega}$ naturally defines an element in open cohomology $H^3(\hat{Z}_3-E)\cong H^3(\hat{Z}_3,E)$, which in general carries a mixed Hodge structure \cite{Voisin}. However, the analysis of variations of this mixed Hodge structure reduces to the variation of the pure Hodge structure \eqref{eq:HodgeClosed} on the graded weight $Gr_3^WH^3(\hat{Z}_3-E)=H^3(\hat{Z}_3)$ since $E$ has no deformation\footnote{In the diagram of variations of mixed Hodge structures, the downward arrows corresponding to deforming $E$ do not exist, cf.~equation (4.40) of \cite{Grimm:2008dq}.}. To derive the superpotential as a solution of the Picard-Fuchs equations we also have to use an appropriate chain integral or current as explained below.

We conclude by discussing the expected structure of the Picard-Fuchs equations on ${\cal M}(\hat{Z}_3)$ from the residue \eqref{eq:ResZhat}. For a detailed discussion along the lines of concrete examples we refer to sections \ref{ToricBraneBlowup} and \ref{ToricBraneBlowupII}. In general all periods of $\hat{\Omega}$ over closed three-cycles in $\hat{Z}_3$ are solutions to this Picard-Fuchs system. 
First, we note that the Calabi-Yau constraint $P(\underline{x},\underline{a})$ appears both in the residues (\ref{eq:ResZhat}) on $\hat{Z}_3$ as in well as in (\ref{eq:residueZ3}). The parameters $\underline{a}$ multiplying monomials in $P$ are identified, modulo the symmetries of the toric ambient space $\mathbb{P}_{\Delta}$, with complex structure moduli $\underline{z}$. Since monomials in $Q(\underline{x},\underline{b})$ are multiplied by independent parameters $\underline{b}$, the Picard-Fuchs operators $\mathcal{L}_k(\underline{a})$ annihilating $\Omega$, expressed by derivatives w.r.t.~$\underline{a}$, annihilate $\hat{\Omega}$ as well. Second, since $\mathbb{P}_{\Delta}$ is the basis of the $\P^1$-fibration ${\cal W}$ the toric symmetries of $\mathbb{P}_{\Delta}$ are contained in those of ${\cal W}$. They act on $P(\underline{x},\underline{a})$ in the same way in ${\cal W}$ as in $\mathbb{P}_{\Delta}$, but also act on the parameters $\underline{b}$ in the constraint $Q$. Therefore, the operators $\mathcal{Z}_i(\underline{a})$, expressing the toric symmetries of $\mathbb{P}_{\Delta}$ on $Z_3$, are lifted to $\hat{Z}_3$ as
\begin{equation}	\hat{\mathcal{Z}}_i(\underline{a},\underline{b})=\mathcal{Z}_i(\underline{a})+\mathcal{Z}'_i(\underline{b})\,,
\label{eq:Zslift}
\end{equation}
where the first operator is as before on $Z_3$ and the second operator just contains differentials of the $\underline{b}$. Then, \eqref{eq:Zslift} is easily solved by choosing the coordinates $\underline{z}$ as on $Z_3$. The new torus symmetries of $\mathcal{W}$, that correspond to its $\P^1$-fiber, do not involve the $\underline{a}$, but only the $\underline{b}$. They merely fix the variables $\underline{u}$ in terms of the parameters $\underline{a}$ and $\underline{b}$. Consequently, dividing out all toric symmetries of $\mathcal{W}$, the form $\hat\Omega$ depends on $\hat{\underline{z}}\equiv(\underline{z},\underline{u})$ only,
\begin{equation}
	\hat{\Omega}(\underline{a},\underline{b})\equiv \hat\Omega(\underline{z},\underline{u})=\hat{\Omega}(\hat{\underline{z}})
\end{equation}
and the differential operators $\mathcal{L}_k(\underline{a})$ take, possibly after a factorization to operators $\mathcal{D}_k(\underline{\hat{z}})$ of lower degree, the schematic form
\begin{equation}
	\mathcal{D}_k({\underline{\hat{z}}})=\mathcal{D}_k^{Z_3}(\underline{z})+\mathcal{D}'_k(\underline{z},\underline{u})
	\label{eq:LInZHat}
\end{equation}
in these coordinates. Here $\mathcal{D}_k^{Z_3}(\underline{z})$ are the Picard-Fuchs operators of $Z_3$ in the coordinates $\underline{z}$ and $\mathcal{D}'_k(\underline{z},\underline{u})$ are at least linear in derivatives w.r.t.~$\underline{u}$.
It follows immediately that the Picard-Fuchs system for $\hat \Omega(\hat{\underline{z}})$ contains the Picard-Fuchs system for $\Omega(\underline{z})$, as determined by the $\mathcal{D}_k(\underline{\hat{z}})$, as a closed subsystem. 
Consequently the periods $\Pi^l(\underline{z})$ of $Z_3$ over 
closed three-cycles, that fulfill the differential equations $\mathcal{L}_k^{Z_3}(\underline{z})$, are also solutions to
(\ref{eq:LInZHat}). We note that there are new operators ${\mathcal{L}}_m(\hat{\underline{z}})$ due to the constraint $Q$ that do not have a counterpart on $Z_3$. They form, together with \eqref{eq:LInZHat}, a complete differential system on ${\cal M}(\hat{Z}_3)$. However, the ${\mathcal{L}}_m(\hat{\underline{z}})$ are again at least linear in differentials of $\underline{u}$, thus act trivially on functions independent of $\underline{u}$, such as $\Pi^l(\underline{z})$.

Geometrically the lift of the periods of $\Omega$ to $\hat{Z}_3$ is a consequence of the 
isomorphism of $\pi:Z_3- \Sigma\rightarrow \hat 
Z_3- E$ and the fact that $\hat \Omega$ 
vanishes on $E$.          
The analog lift of the 
more interesting open periods on $Z_3$, in particular $W_{\rm brane}$ is discussed in the next section, section  
\ref{potentialhatZ3minusD}, where we provide the corresponding 
expressions for the lifted superpotential on $\hat{Z}_3$. 

We conclude by mentioning that the structure \eqref{eq:LInZHat} allows to directly determine the inhomogeneous Picard-Fuchs equations obeyed by the domain wall tension $\mathcal{T}(\underline{z})$ between two five-branes,
\begin{equation}
 	\mathcal{D}_k^{Z_3}(\underline{z})\mathcal{T}(\underline{z})=f_k(\underline{z})\,.
\end{equation}
The tension $\mathcal{T}$ is obtained
as $\mathcal{T}(\underline{z})=W_{\rm brane}(\underline{z},\underline{u}^{rm c})-W_{\rm brane}(\underline{z},\underline{u}^{rm c}_0)$ and the inhomogeneity $f(\underline{z})$ upon evaluating $f_k(\underline{z})=\mathcal{D}'_k(\underline{z},\underline{u})W_{\rm brane}(\underline{z},\underline{u})\vert_{\underline{u}^{\rm c}}$, where $\underline{u}^{\rm c}$, $\underline{u}^{\rm c}_0$ are critical points of $W_{\rm brane}$. This inhomogeneous system was obtained from residues in \cite{Walcher}.

\subsection{Lift of the  Superpotentials} 
\label{potentialhatZ3minusD}
  
We have seen at the end of the previous section from the discussion of the Picard-Fuchs equations on $\hat{Z}_3$ that the 
periods $\Pi^l(\underline{z})$ over closed three-cycles lift from $Z_3$ 
to $\hat Z_3$. Illustratively this is clear because these integrals depend only on the geometry of $Z_3$ with the five-brane removed, cf.~section \ref{N=1branes}. Thus, in order to lift the flux 
superpotential (\ref{eq:fluxpot}), which is just a linear combination of the periods $\Pi^l(\underline{z})$, we just need to 
lift the flux data.  This is straightforward as the 
third cohomology of $\hat{Z}_3$ is given by \cite{Grimm:2008dq} 
\begin{equation}
	H^3(\hat{Z}_3)=\pi^* H^3(Z_3)\oplus H^3(E)
\label{eq:thirdcohom}
\end{equation}
so that any flux $G_3$ on $Z_3$ has a counterpart 
$\hat{G}_3=\pi^* G_3$ on $\hat{Z}_3$. Thus we readily 
obtain the lift of the flux superpotential to $\hat{Z}_3$ as
\begin{equation}
	W_{\rm flux}=\int_{\hat{Z}_3}\hat{\Omega}\wedge \hat{G}_3.
\label{eq:hatWflux}
\end{equation} 
Again the precise integral basis of cycles on $\hat{Z}_3$ for which this integral 
can be expanded in terms of periods of $\hat{\Omega}$ with integral coefficients has to be obtained by matching the classical terms at large radius and by assuring that the monodromy acts by integral transformations. Obviously, on 
$\hat Z_3$ flux configurations associated to the new three-cycles 
in $E$, which are not promoted from fluxes on $Z_3$ can be considered.

In order to lift the five-brane superpotential $W_{\rm brane}$ let us first 
make a local heuristic analysis, which casts already much of the 
general structure. Locally on a patch $U_\alpha$ we can write $\Omega=\dd \omega$
and evaluate the integral over the chain $\Gamma(u)$ leading to the five brane superpotential localized to the boundary $\partial \Gamma(u)=\Sigma-\Sigma_0$, 
\begin{equation}
	W_{\rm brane}=\int_{\Sigma}\omega\ .
\end{equation} 
Here we suppress the integral over the fixed reference curve $\Sigma_0$ in the same homology class as $\Sigma$, as it gives only rise to an irrelevant constant in $W_{\rm brane}$. 
Then we use the fact that the original curve $\Sigma$ is contained in the second cohomology $H^2(E)$ by its Poincare dual class 
\cite{Grimm:2008dq}. Thus we can write
\begin{equation}
	W_{\rm brane}=\int_E\pi^*(\omega)\wedge F_2
\label{intoverD}
\end{equation}
where $[F_2]$  is the class of $\Sigma$ in 
the exceptional divisor $E$, i.e.
\begin{equation} 
[F_2]=\Sigma \ \ \ \text{in} \ \  E\ 
\end{equation} 
We note that at the supersymmetric minimum $F_2$ is equal to the K\"ahler form of the Fubini study metric on the $\mathbb{P}^1$, 
i.e.~$F_2=\omega_{FS}$. 
Locally (\ref{intoverD}) can be written as an integral 
on $\hat Z_3$
\begin{equation}
	W_{\rm brane}=\int_{\Gamma_5}\hat{\Omega}\wedge F_2
	\label{eq:hatWbrane}
\end{equation}
over a five-chain $\Gamma_5$ with $\partial\Gamma_5=E-E_0$ on which we extend $F_2$. $E_0$ denotes a reference divisor in the same homology as $E$, e.g. the blow-up of $\Sigma_0$, to match the constant contributions.

To prove this more rigorously it is instructive to consider the lift of the Bianchi identity to $\hat Z_3$. The formalism is equal for the three-form 
R-R field strength $F_3$ in type IIB and the three-form NS-NS field
strength $H_3$ in the heterotic string. For the following 
analysis of the Bianchi identities, which is local along the 
curve $\Sigma$, one can focus on the source term $\delta_\Sigma$ of one five-brane neglecting the other terms in (\ref{dF_3}) respectively (\ref{dH_3het}).
The only aspect of the geometry which cannot be seen locally 
in a patch near a point in $\Sigma$ is the non triviality of 
the bundle $N_{\hat Z_3} E$, which is captured by its Thom 
class $\frac{e_1}{2}$. 
As can be calculated explicitly by evaluating the pull-back $\pi^*$ to the blow-up $\hat{Z}_3$, the form $H_3$ in \eqref{dH_3het} is replaced by\footnote{Note that mathematically 
$\delta_E \wedge F_2 \equiv \cT(F_2)$ is given by the Thom isomorphism $\cT: H^\bullet(E) \rightarrow H^{\bullet +2}(Y)_{cpt}$ 
of the normal bundle $N_{\hat Z_3}E$ in $\hat Z_3$. $\cT$ maps cohomology classes on $E$ to compactly supported classes in $\hat Z_3$.} 
\beq \label{dhatH_3}
  d\hat H_3 = \delta_E \wedge F_2=
\dd \rho \wedge \frac{e_1}{2} \wedge F_2 \ ,
\eeq
where the limit $\epsilon\rightarrow 0$ is implicit and 
we used 
\beq
\label{regdeltaD}
\lim_{\epsilon\rightarrow 0}\dd \rho \wedge \frac{e_1}{2}=\delta_E\ .
\eeq
 Formally (\ref{dhatH_3}) can be integrated in the language of currents to 
\beq
\label{hatH3}
\hat H_3=
\dd \rho \wedge \frac{e^{(0)}_{0}}{2} F_2+\dd B_2\ ,
\eeq 
where $e_1=\dd e^{(0)}_0$ and a possible term $\rho \frac{e_1}{2}\wedge F_2$ is neglected by the requirement of regularity of $\hat{H}_3$. Thus, by pulling back both $H_3$ and $\Omega$ to $\hat{Z}_3$ we lift the superpotential \eqref{eq:unifiedSuperpots} as
\begin{equation} 
\label{liftedbranesuperpotetial}
W_{\rm brane}=\lim_{\epsilon\rightarrow 0}\int_{\hat Z_3} \hat \Omega \wedge  \hat H_3=\lim_{\epsilon\rightarrow 0}\int_{\hat{Z}_3}\hat{\Omega}\wedge F_2\wedge \rho e_1 
\end{equation}
where we restrict to the singular part \eqref{hatH3} of $\hat{H}_3$ only. By (\ref{regdeltaD}) and the identity \eqref{hatH3} for $\hat{H}_3$ we see 
that this is equivalent to (\ref{intoverD}) and hence to 
(\ref{eq:hatWbrane}). We note that we can easily switch between the open manifold $\hat{Z}_3-E$ and $\hat{Z}_3$ in \eqref{liftedbranesuperpotetial} since $\hat{\Omega}\vert_E=0$.
Mathematically, this match of (\ref{liftedbranesuperpotetial}) and the original superpotential $W_{\rm brane}$ in (\ref{chainonopenset}) follows more 
geometrically by the canonical identification $\hat \Omega|_{\hat Z_3- E}=\Omega|_{Z_3- \Sigma}$ under the biholomorphism $Z_3-\Sigma\cong \hat{Z}_3-E$, by $\hat{\Omega}\vert_E=0$ and the lift of the Thom classes, $\rho e_3\cong \rho e_1\wedge F_2$. 

Finally, we conclude by arguing that $W_{\rm brane}$ is a solution to the Picard-Fuchs system on $\hat{Z}_3$. In fact, this is confirmed for the examples of sections \ref{ToricBraneBlowup}, \ref{ToricBraneBlowupII} using the corresponding open-closed GKZ-system on $\hat{Z}_3$. 
All we have to ensure is that integration of $\hat{\Omega}$ over $\hat{H}_3$ given in \eqref{hatH3}, which is not a closed form, commutes with the application of the Picard-Fuchs operators, that annihilate $\hat{\Omega}$. In addition, the whole integral has to be annihilated as well. Since $\hat{H}_3$ does not depend on the complex structure on $\hat{Z}_3$, all differential operators indeed commute with integration. Furthermore, for the GKZ-system of the form discussed below \eqref{eq:Zslift}, \eqref{eq:LInZHat}, the operators $\mathcal{L}_k(\underline{\hat{z}})$, $\hat{\mathcal{L}}_m(\underline{\hat{z}})$ annihilate $\hat{\Omega}$ identically and consequently also $W_{\rm brane}$ in \eqref{liftedbranesuperpotetial}. However, the operators $\hat{\mathcal{Z}}_i(\underline{a},\underline{b})$ expressing the toric symmetries of $\mathcal{W}$ obey in general \cite{Batyrev,CoxKatz}
\begin{equation}
 	\hat{\mathcal{Z}}_i(\underline{a},\underline{b})\hat{\Omega}(\underline{a},\underline{b})=d\hat{\alpha}\,
\end{equation}
for a two-form $\hat{\alpha}$. This can potentially lead to a non-zero result in $W_{\rm brane}$ by partial integration since $d\hat{H}_3\neq 0$. In fact we exploit \eqref{dhatH_3} to rewrite this as in integral over $E$,
\begin{equation}
 	\hat{\mathcal{Z}}_i(\underline{a},\underline{b})W_{\rm brane}=-\int_E \iota^*(\hat{\alpha})\wedge F_2\,,
\end{equation}
where $\iota:E\hookrightarrow \hat{Z}_3$ denotes the embedding of $E$. Fortunately, the pull-back $\iota^*(\hat{\alpha})$ vanishes on $E$ by the fact that we are dealing with variations of pure Hodge structures \eqref{eq:HodgeClosed} and thus can not reach the cohomology of $E$\ \footnote{As noted below \eqref{eq:Hodgefilt}, the cohomology of $E$ is included in the variations of mixed Hodge structures on $H^3(\hat{Z}_3-E)$ and can only be reached by deformations of $E$. These do not exist in our case since $E$ is rigid.}. Thus we have argued that $W_{\rm brane}$ is a solution of the GKZ-system on $\hat{Z}_3$. This will be further confirmed for the examples in section \ref{IIBBlowUp}. 

\section{Blowing-Up Five-Branes in Compact Calabi-Yau Threefolds}
\label{IIBBlowUp}

In this section we apply the blow-up proposal to a selection of examples of five-branes in compact Calabi-Yau threefolds. We consider two different Calabi-Yau threefolds, the one parameter example of the quintic and the two parameter Calabi-Yau in $\P^4(1,1,1,6,9)$, where the latter was the main example in \cite{Grimm:2009ef}. The type of branes we consider have the geometric interpretation of five-branes on rational curves, i.e.~holomorphic curves $\Sigma=\P^1$, at special loci in the open-closed deformation space. The number of open deformations is two. It is the common feature of both geometries that there is for generic values of the closed string moduli only a discrete number of such lines \cite{katz}, however, for special values of the moduli, at the Fermat point, an one parameter family of holomorphic curves. 

We begin our discussion in section \ref{ToricBranes} by reviewing aspects of toric mirror symmetry in the closed and open string case and the construction of toric GKZ-systems.
Then we start our main discussion in section \ref{toricbranesP11111} for the case of the quintic Calabi-Yau threefold. We show that five-branes on rational curves in the quintic can be described as toric branes $\Sigma$ at a special sublocus ${\cal M}(\P^1)$ of their moduli space $\mathcal{M}(\Sigma)$.
This establishes the mapping of the moduli space $\mathcal{M}(\Sigma)$ with the obstructed deformations space $\tilde{\P}^1$ of rational curves as discussed in section \ref{unificationofdeformationsII}. This is crucial since it enables us to work with a well-defined complex moduli space of $(Z_3,\Sigma)$ to describe the space of \textit{anholomorphic}, obstructed deformations of $\tilde{\P}^1$. 

From this description of the rational curve as a toric brane $\Sigma$ we readily construct the blow-up. We study the complex structure moduli space of $\hat{Z}_3$ in section \ref{toricbranesP11111-GKZ} by exploiting that $\hat{Z}_3$ is also governed by new toric data that is canonically related to the toric data of $(Z_3,\Sigma)$. First we construct the pull-back form $\hat{\Omega}$ on $\hat{Z}_3$ along the lines of section \ref{hatomega}. Then we read of a toric GKZ-system that is associated to the underlying toric data of the blow-up $\hat{Z}_3$. 
From the GKZ-system we derive a complete Picard-Fuchs-system that we solve to obtain the periods of $\hat{\Omega}$ in section \ref{toricbranesP11111-W} at various loci in the moduli space. These are the large radius point of $Z_3$ and the five-brane and selected discriminant loci of the Picard-Fuchs system for $\hat{Z}_3$. 
Since the periods of $\hat{\Omega}$ are also understood as periods on $\mathcal{M}(\Sigma)$, cf.~section \ref{unificationofdeformationsII}, we finally obtain the brane superpotential $W_{\rm brane}$ for the line $\tilde{\P}^1$ as a linear combination of periods fixed by an appropriate flux on $\hat{Z}_3$ that encodes the information about the five-brane. In particular we use open mirror symmetry to obtain predictions for the disk invariants at large volume, that match and extend independent results in the literature. Similarly, the disk instantons are obtained for a different brane phase in section \ref{toricbranesP11111-W-LV}. These are the first available results for branes with \textit{two} open string deformations.

In section \ref{generaltoricstructure} we discuss the toric structure of $\hat{Z}_3$ obtained by blowing up toric curves $\Sigma$. We give a general recipe in section \ref{generaltoricstructure} to obtain a toric polyhedron $\Delta_7^{\hat Z}$ from the charge vectors of the geometry $Z_3$ and the curve $\Sigma$, that can be efficiently used to obtain the open-closed GKZ-system for the brane deformation problem associated to arbitrary toric curves $\Sigma$. We comment on a connection to Calabi-Yau fourfolds which applies to special choices of toric branes, but \textit{not} in general.

Finally in section \ref{ToricBraneBlowupII} we consider another two parameter example of the elliptically fibered Calabi-Yau threefold in $\P^6(1,1,1,6,9)$ with a five-brane supported on a rational curve. There we make similar use of the toric GKZ-system on $\hat{Z}_3$ to obtain the periods of $\hat{\Omega}$ from which we construct the five-brane superpotential. Under toric mirror symmetry we obtain the disk invariants of the dual A-model geometry. In addition, we comment on the connection to heterotic/F-theory duality and to our earlier works \cite{Grimm:2009ef}, \cite{Grimm:2009sy}.

\subsection{Toric Calabi-Yau Hypersurfaces, Toric Branes \& GKZ-Systems}
\label{ToricBranes}

In this section we briefly introduce a basic account on toric geometry used in the remainder of this section. We start by reviewing the construction of toric Calabi-Yau hypersurfaces and toric branes, where we emphasize aspects of closed and open mirror symmetry. Then we introduce the toric GKZ-system which allows a convenient construction of the Picard-Fuchs system for the complex structures on $Z_3$, that is discussed in general in section \ref{fivebranesuperpotential}.

\subsubsection{Toric Mirror Symmetry}

The starting point is a mirror pair of toric Calabi-Yau hypersurfaces denoted $(\tilde{Z}_3, Z_3)$ in type IIA/IIB to which we
want to add a mirror pair of branes. 
The toric ambient variety $\P_{\tilde\Delta}$ of the hypersurface $\tilde{Z}_3$ on the type IIA side is encoded by a set of vectors $\ell^{(i)}$ forming a basis of relations
among the points $\tilde{v}_j$ specifying a polyhedron $\Delta_4^{\tilde Z}$. $\P_{\tilde\Delta}$ can be represented as the quotient $(\mathbb{C}^{k+4}-\text{SR})//\Gamma$ defined by dividing out the isometry or gauge group $\Gamma=(U(1))^k$ 
and by imposing vanishing D-terms or moment maps, 
\begin{equation} \label{eq:ToricVariety}
 	\tilde{x}_j\mapsto e^{i\ell^{(i)}_j\phi_i}\tilde{x}_j\,,\qquad \sum_{j=1}^{k+4} \ell^{(i)}_j |\tilde x_j|^2= r^i\,.
\end{equation}
Here, the $r^i$ denote the real volumes of distinguished, effective curves in $\P_{\tilde\Delta}$. They form the basis of the Mori cone of curves $C_i$ each of which is
associated to one charge vector $\ell^{(i)}$. The basis of charge vectors, thus the generators of the Mori cone, are determined by the triangulation of $\Delta_4^{\tilde Z}$ that correspond to the different phases of $\P_{\tilde\Delta}$ \cite{Witten:1993yc}. This then fixes the form of the D-term constraints in \eqref{eq:ToricVariety} from which the Stanley-Reissner ideal SR can be read off \cite{Denef:2008wq}. The action of $G$ on the coordinates is generated infinitesimally by $k$ vector fields
\begin{equation}
 	V^{(i)}=\sum_j\ell^{i}_j\tilde{x}_j\frac{\partial}{\partial \tilde{x}_j}\,.
\end{equation}
Using these vector fields it is straight forward to construct forms on $\P_{\tilde\Delta}$ from invariant forms of $\C^{k+4}$. Of particular importance is the holomorphic top-form $\Delta_{\P_{\tilde\Delta}}$ on $\P_{\tilde\Delta}$ that is constructed from the holomorphic top-form $\Delta_{\C}=d\tilde{x}_1\wedge\ldots\wedge d\tilde{x}_{k+4}$ on $\C^{k+4}$ as
\begin{equation} \label{eq:toricMeasure}
 	\Delta_{\P_{\tilde\Delta}}=V^{(1)}\lrcorner \ldots\lrcorner V^{(k)}\lrcorner (\Delta_{\C})\,,
\end{equation}
where $\lrcorner$ denotes the interior product defined by contracting a form with a vector. For the example of projective space $\P^n$ this yield the unique holomorphic section of $\Omega_{\P^n}^n(n+1):=\Omega_{\P^n}^n\otimes\mathcal{O}(n+1)$ given by
\begin{equation}\label{eq:MeasurePn}
 	\Delta_{\P^n}=\sum_{i=1,n+1}(-1)^{i-1}x_i dx_1\wedge\ldots \widehat{dx_i}\ldots \wedge dx_n\,,
\end{equation}
where the $x_i$ denote the homogeneous coordinates on $\P^n$ and $\widehat{dx_i}$ indicates, that $dx_i$ is omitted.

In the type IIA theory supersymmetric branes wrap special Lagrangian cycles $L$. In the toric ambient space $\P_{\tilde\Delta}$ one describes 
such a three-cycle $L$ by $r$ additional, so-called brane charge vectors $\hat \ell^{(a)}$ restricting the $|\tilde{x}_j|^2$ and 
their angles $\theta_i$ as \cite{Aganagic:2000gs}
\begin{equation} \label{ABranes}
 	\sum_{j=1}^{k+4} \hat\ell^{(a)}_j |\tilde x_j|^2=c^a\ ,\qquad \theta_i=\sum_{a=1}^r\hat\ell^{(a)}_i\phi_a\ ,
\end{equation}
for angular parameters $\phi_a$. To fulfill the `special' condition of $L$, which is 
equivalent to $\sum_i\theta_i=0$, one demands $\sum_j\hat\ell_j^{(a)}=0$. 

Wrapping a brane on such a cycle we obtain a so-called toric brane, in this case a toric D6-brane, that is a specific brane type well studied in the context of toric mirror symmetry in non-compact Calabi-Yau manifolds \cite{Aganagic:2000gs, Aganagic:2001nx}. 
A toric brane admits an efficient description in terms of toric data similar to the toric charge vectors $\hat{\ell}^{(i)}$ associated to a mirror pair $(\tilde{Z}_3,Z_3)$ of Calabi-Yau
threefolds. In this description the mirror brane can be constructed as
a holomorphic cycle inside $Z_3$ wrapped by the mirror D$p$-brane as the zero set of a number of constraints.

Indeed, the mirror Type IIB description \cite{Aganagic:2000gs,Hori} is be obtained as follows.
First the mirror Calabi-Yau $Z_{3}$ is determined as a hypersurface $P=0$ in a dual toric ambient space $\P_{\Delta}$ with additional $k$ constraints, 
\begin{equation} \label{eqn:HVmirror}
 	P=\sum_{j=0}^m a_jy_j\ ,\qquad \prod_{j=0}^m (a_jy_j)^{\ell^{(i)}_j}=z^i\ ,\qquad i=1,\ldots, k\ ,
\end{equation}
Here the $z^i$ denote the complex structure moduli of $Z_3$ that are related to the complex numbers $a_j$ by the second relation in \eqref{eqn:HVmirror}.
We note that we introduced a further 
coordinate $y_0$ for which we also have to include a zeroth component of the charge vectors as $\ell^{(i)}_0=-\sum^m_{j=1} \ell^{(i)}_j$. 
The mirror toric variety $\P_{\Delta}$ admits also a description as a quotient \eqref{eq:ToricVariety} of homogeneous coordinates $x_i$ associated to a polyhedron $\Delta_4^{Z}$ obtained as the dual of the polyhedron $\Delta_4^{\tilde Z}$. The $y_i$ are then identified with specific monomials $m_i(x_j)$ in the $x_j$, $y_i\mapsto m_i(x_j)$, which are associated to the integral points $\tilde{v}_j$ in $\Delta_4^{\tilde Z}$ by the mirror construction of Batyrev \cite{Hosono:1993qy,Batyrev}. In this description the constraint $P$ reads
\begin{equation} \label{eq:Batyrev}
 	P(\underline{x};\underline{a})=\sum_{\tilde{v}_j\in\Delta_4^{\tilde Z}}a_j\prod_{i}x_i^{\langle \tilde{v}_j,v_i\rangle+1}\,,
\end{equation}
where $v_i$ labels the vertices of $\Delta_4^{Z}$ with associated coordinate $x_i$.
In general there can be a discrete orbifold symmetry group $G$ such that $Z_3$ is an orbifolded hypersurface.

Analogously toric holomorphic submanifolds $\Sigma$ in $\P_{\Delta}$, that can support calibrated B-branes called toric branes, are specified by the constraints 
\begin{equation} \label{eq:BBrane}
 	\prod_{j=0}^m (a_jy_j)^{\hat\ell^{(a)}_j}=u^a,\quad a=1,\ldots, r\ . 
\end{equation}
This can also be re-expressed in terms of the coordinates $x_i$ using the above map $y_i\mapsto m_i(x_j)$. Intersected with the Calabi-Yau constraint $P=0$ this
describes families of submanifolds of codimension $r$ in $Z_3$.
For the configuration $r=2$, $\Sigma$ is a curve in $Z_3$ that we call a toric curve and which is precisely the geometry we are interested in for the study of five-branes.

\subsubsection{The Toric GKZ-System}

In the context of toric hypersurface the problem of complex structure variations of $Z_3$ can be studied efficiently. In the algebraic representation $P(\underline{x};\underline{a})=0$ as a family of hypersurfaces, the residue integrals \eqref{eq:residueZ3} can be used to find Picard-Fuchs equations governing the complex structure dependence of $\Omega$ and its periods. In the toric context it is often very convenient to use instead of the coordinates $\underline{z}$ first the redundant parameterization of the complex structure variables of $Z_3$ by the coefficients $\underline{a}$ on which the toric symmetries of $\P_{\Delta}$ act canonically.  The infinitesimal version of theses symmetries acting on the $\underline{a}$ give rise to very simple differential operators ${\cal Z}_k(\underline{a})$ and  ${\cal L}_l(\underline{a})$  called the Gelfand-Kapranov-Zelevinski or short GKZ differential system of the toric complete intersection. The operators either annihilate $\Omega(\underline{a})$ or, if the toric symmetries are broken in a specific way, annihilate $\Omega(\underline{a})$ up to an exact form as in (\ref{uptoexact}). The latter operators, the ${\cal Z}_k(\underline{a})$, express the fact that $\Omega(\underline{a})$ depends only on specific combinations of the $\underline{a}$, which are the genuine complex structure deformations $\underline{z}$ defined in \eqref{eqn:HVmirror}. From the operators ${\cal L}_k(\underline{a})$ it is possible in  simple situations to obtain a complete set of Picard-Fuchs operators ${\cal D}_a(\underline{z})$~\cite{Hosono:1993qy}. Using monodromy information and knowledge of the classical terms, their solution can be associated to integrals over an integral basis of cycles in $H_3(Z_3,\mathbb{Z})$ and given the flux quanta explicit superpotentials can be written down.  

These operators ${\cal L}_l(\underline{a})$ and ${\cal Z}_k(\underline{a})$ are completely determined by the toric data encoded in the points $\tilde{v}_j$ and relations $\ell^{(a)}$ of $\Delta_4^{\tilde{Z}}$.
The first set of operators is given by
\begin{equation} \label{eq:pfo}
 	\mathcal{L}_{i}=\prod_{\ell^{(i)}_j>0}\left(\frac{\partial}{\partial a_j}\right)^{\ell^{(i)}_j}-\prod_{\ell^{(i)}_j<0}\left(\frac{\partial}{\partial a_j}\right)^{-\ell^{(i)}_j}\,,\qquad i=1,\ldots, k\,.
\end{equation}
These operators annihilate $\Omega(\underline{a})$ in \eqref{eq:residueZ3} and its periods $\Pi^k(\underline{a})$ identically as can be checked as a simple consequence of the second relation in \eqref{eqn:HVmirror}. In other words, the differential operators ${\cal L}_k$ express the trivial algebraic relations between the monomials entering $P(\underline{x};\underline{a})$. 
The second differential system of operators encoding the automorphisms of $\P_{\Delta}$ is given by
\begin{equation} \label{eq:Zs}
 	\mathcal{Z}_{i}=\sum_j(\bar{v}_j)^i\vartheta_j-\beta_i\,,\qquad i=0,\ldots,4\,.
\end{equation}
Here $\beta=(-1,0,0,0,0)$ is the so-called exponent of the GKZ-system and $\vartheta_j=a_j\frac{\partial}{\partial a_j}$ denote the logarithmic derivative. We have embedded the points $\tilde{v}_j$ into a hypersurface at distance $1$ away from the origin by defining $\bar{v}_j=(1,\tilde{v}_j)$ so that all zeroth components are $(\bar{v}_j)^0=1$. The operators $\mathcal{Z}_j$ simply represent the torus symmetries of $\P_{\Delta}$ on the periods $\Pi^k(\underline{a})$ that are functions of the parameters $\underline{a}$. In detail, the operator $\mathcal{Z}_0$ expresses the effect of a rescaling of the constraint $P\mapsto \lambda P$ by the homogeneity property $\Pi^k(\lambda \underline{a})=\lambda \Pi^k(\underline{a})$. It is often convenient to perform the redefinition $\Omega(\underline{a})\mapsto a_0\Omega(\underline{a})$ and $\Pi^k(\underline{a})\mapsto a_0\Pi^k(\underline{a})$ so that the periods become invariant functions under the overall rescaling $\underline{a}\mapsto \lambda \underline{a}$. Accordingly we obtain a shift $\vartheta_0\mapsto \vartheta_0-1$ in all the above operators, such that the $\mathcal{Z}_0=\sum_{i}\vartheta_i$ in particular. The operators $\mathcal{Z}_i$, $i\neq 0$, express rescalings of the coordinates like e.~g.~$x_i\mapsto \lambda x_i$, $x_j\mapsto \lambda^{-1}x_j$ that can be compensated by rescalings of the monomial coefficients $\underline{a}$. In particular these coordinate rescalings leave the measure $\Delta_{\P_{\Delta}}$ on $\P_{\Delta}$ invariant. Examples for the operators $\mathcal{Z}_i$ and the corresponding action on the coordinates $x_j$ are given in section \ref{ToricBraneBlowup}.

The coordinates $z^i$ introduced in \eqref{eqn:HVmirror} are then solutions to the $\mathcal{Z}_i$-system and generally read
\begin{equation} \label{eq:algCoords}
 	z^i=(-)^{\ell^{(i)}_0}\prod_{j=0}^{k+4} a_j^{\ell^{(i)}_j}\,,\quad i=1,\ldots k.
\end{equation}
From monodromy arguments they indeed turn out to be appropriate coordinates at the point of maximally unipotent monodromy in the complex structure moduli space of $Z_3$. The periods $\Pi^l(\underline{z})$ as solutions to the operators $\mathcal{D}_a$ in the coordinates $z^i$ admit the well-known log-grading in terms of powers of $\log(z^a)$ which in particular leads to the well-known monodromy $t^a\mapsto t^a+1$. 

It is the aim of the next sections \ref{ToricBraneBlowup}, \ref{generaltoricstructure} and \ref{ToricBraneBlowupII} to extend the use of the GKZ-system, the coordinates at large volume as well as the periods to the open string sector. This is achieved by formulating a GKZ-system for the blow-up $\hat{Z}_3$ of the toric curve \eqref{eq:BBrane} and by analyzing its solutions.

\subsection{Open-Closed Picard-Fuchs Systems: Branes on the Quintic}
\label{ToricBraneBlowup}

We now apply the blow-up proposal of section \ref{5braneblowupsanddefs} to the case of toric curves\footnote{We emphasize that the five-brane does \textit{not} wrap $\Sigma$ for generic values of the closed and open moduli.} $\Sigma$ on the one parameter quintic. As explained in section \ref{unificationofdeformationsII} the moduli space ${\cal M}(\Sigma)$ of $\Sigma$ is identified with the obstructed deformation space of rational curves $\tilde{\P}^1$, on which we wrap a five-brane. The holomorphic $\P^1$ is directly visible from the complete intersection description of $\Sigma$ at a special sublocus ${\cal M}(\P^1)$ of  ${\cal M}(\Sigma)$ where $\Sigma$ degenerates appropriately modulo the action of the quintic orbifold $G=(\mathds{Z}_5)^3$. This way, we understand the five-brane on a rational curve $\P^1$ as a special case of toric brane and consequently apply open mirror symmetry along the lines of section \ref{ToricBranes}. Thus, we start in section \ref{toricbranesP11111} from the toric curve $\Sigma$, then determine the sublocus ${\cal M}(\P^1)$ and identify the wrapped rational curve $\P^1$ that we represent via the standard Veronese embedding in $\P^4$. After this definition of the deformation problem, we determine the Picard-Fuchs system on the deformation space of the rational curves $\tilde{\P}^1$ as a GKZ-system on the complex structure moduli space of the blow-up $\hat{Z}_3$ along $\Sigma$ in section \ref{toricbranesP11111-GKZ}. The solutions of this system including the large volume expressions for the flux and brane superpotential, the open-closed mirror map and the disk instantons are summarized in section \ref{toricbranesP11111-W}. Furthermore, we find the solutions of the Picard-Fuchs system at its discriminant loci, namely in the vicinity of the sublocus ${\cal M}(\P^1)$ and at the involution brane, where we identify the superpotential $W_{\rm brane}$. Finally in section \ref{toricbranesP11111-W-LV} we obtain the disk invariants for a different brane geometry at large volume.

\subsubsection{Branes on Lines in the Quintic \& the Blow-Up}
\label{toricbranesP11111}

The quintic Calabi-Yau $\tilde{Z}_3$ is given as the general quintic hypersurface $\tilde{P}$ in $\P^4$. It has 101 complex structure moduli corresponding to the independent coefficients of the monomials entering $\tilde{P}$. Its K\"ahler moduli space is one-dimensional generated by the unique K\"ahler class of the ambient $\P^4$. Toric Lagrangian submanifolds of this geometry were discussed in \cite{Aganagic:2000gs} along the lines of section \ref{ToricBranes}.

The mirror quintic threefold $Z_3$ is given as the hypersurface
\begin{equation}
 Z_3\,:\quad P=x_1^5+x_2^5+x_3^5+x_4^5+x_5^5-5\Psi x_1x_2x_3x_4x_5\,,
\label{eq:mirrorquintic}
\end{equation}
where $\Psi$ denotes its complex structure modulus. It is obtained via \eqref{eqn:HVmirror} from the toric data
\begin{equation} \label{eq:quinticTD}
 	\begin{array}{rlllll}
 	 	 \ell^{(1)}=(-5, & 1, & 1,& 1,& 1, &1) \\ \hline \hline
		y_0 & y_1 & y_2 & y_3 & y_4 & y_5\\ 
		x_1x_2x_3x_4x_5 &x_1^5 & x_2^5 & x_3^5  & x_4^5 & x_5^5 
 	\end{array}
\end{equation}
where the $y_i$ corresponding to the entries $\ell^{(1)}_i$ of the charge vector are given as monomials in the projective coordinates $x_i$ constructed using the formula \eqref{eq:Batyrev}. 
In addition we divide by
an orbifold group $G=(\mathds{Z}_5)^3$ that acts on the coordinates so that $x_1x_2x_3x_4x_5$ is invariant. A convenient basis of
generators 
$g^{(i)}$ is given by $v^{(i)}=(1,-1,0,0,0) \mod 5$ and all permutations of its entries where we use
\begin{equation}
	g^{(i)}:\,\,x_k\mapsto e^{2\pi i v^{(i)}_k/5} x_k\,.
\label{eq:quinticorbifold}
\end{equation}
We note that the Fermat point $\Psi=0$ is a point of enhanced symmetry where $G$ enhances to $(\mathds{Z}_5)^4$. As required by mirror symmetry we have $h^{(2,1)}(Z_3)=1$, $h^{(1,1)}(Z_3)=101$. 
 
Next we introduce an open string sector by putting a five-brane on a line $\P^1$ in the quintic $Z_3$. Following the above logic we first construct a toric curve $\Sigma$ to define the deformation space $\tilde{\P}^1$ of the rational curve. The holomorphic $\P^1$ is then obtained at the sublocus ${\cal M}(\P^1)$ where $\Sigma$ degenerates accordingly. Up to a relabeling of the projective coordinates $x_i$ of $\P^4$ we consider the toric curves $\Sigma$ given by
\begin{eqnarray} \label{eq:RSquintic}
	\Sigma&:& P=0\,,\quad h_1\equiv \beta^5x_3^5-\alpha^5x_4^5=0\,,\quad h_2\equiv\gamma^5x_3^5-\alpha^5x_5^5=0\,, \\
	&&\hat{\ell}^{(1)}=(0,0,0,1,-1,0)\,,\quad\hat{\ell}^{(2)}=(0,0,0,1,0,-1)\,,\nonumber
\end{eqnarray} 
where the brane charge vectors $\hat{\ell}^{(i)}$ correspond to the constraints $h_i$ using \eqref{eq:BBrane} and the toric data \eqref{eq:quinticTD}. The complete intersection \eqref{eq:RSquintic} describes for all values of the parameters $\alpha$, $\beta$ and $\gamma$, that take values in $\P^2$, an analytic family of holomorphic curves in the quintic. Consequently, $\alpha$, $\beta$ and $\gamma$ parameterize the unobstructed moduli space of $\Sigma$ on which we introduce coordinates $u^1=\frac{\beta^5}{\alpha^5}$ and $u^2=\frac{\gamma^5}{\alpha^5}$. 

The obstructed deformation problem is defined by the definition of a non-holomorphic family $\tilde{\P}^1$ and the identification of the locus ${\cal M}(\P^1)$. As discussed before in section \ref{unificationofdeformationsII} the obstructed deformation space of lines is identified with the moduli space ${\cal M}(\Sigma)$. For generic values of the moduli in \eqref{eq:RSquintic} the curve $\Sigma$ is an irreducible higher genus Riemann surface. However, we can always linearize \eqref{eq:RSquintic} for generic values of the moduli,
\begin{equation} \label{eq:anholoC}
	\tilde{\P}^1:\quad \eta_1 x_1+\sqrt[5]{x_2^5+x_3^3 m(x_1,x_2,x_3)}=0\,,\quad  \eta_2\beta x_3-\alpha x_4=0\,,\quad \eta_3\gamma x_3-\alpha x_5=0\,,
\end{equation} 
Here we inserted $h_1$ and $h_2$ into $P$, introduced fifths roots of unity $\eta_i^5=1$ and the polynomial
\begin{equation}
	m(x_1,x_2,x_3)=\frac{\alpha^5+\beta^5+\gamma^5}{\alpha^{5}}x_3^2-5\Psi\frac{\beta\gamma}{\alpha^2} x_1x_2\,.  
\label{eq:quinticDiv}
\end{equation} 
This equation \eqref{eq:anholoC} is evidently non-holomorphic due to the non-trivial branching of the fifth root, in other word $\tilde{\P}^1$ is a non-holomorphic family of rational curves in the quintic. 
However at special loci $\Sigma$ degenerates as follows. We rewrite \eqref{eq:RSquintic} as 
\begin{equation}
 \Sigma\,:\quad x_1^5+x_2^5+x_3^3 m(x_1,x_2,x_3)=0 \,,\quad \alpha^5 x_4^5-\beta^5 x_3^5=0\,,\quad \alpha^5 x_5^5- \gamma^5 x_3^5=0\,.
\end{equation}
Whereas $h_1$, $h_2$ can be linearized for generic values of the moduli, $m(x_1,x_2,x_3)$ forbids a holomorphic linearization of \eqref{eq:RSquintic} and accordingly to take the fifths root in \eqref{eq:anholoC}. However, at the sublocus
\begin{equation}
	\mathcal{M}_{\P^1}(\Sigma)\,:\quad \alpha^5+\beta^5+\gamma^5=0\,,\quad \Psi\alpha\beta\gamma=0\,
\label{eq:modulicond}
\end{equation}
the polynomial $m$ vanishes identically and the Riemann surface $\Sigma$ in \eqref{eq:RSquintic} degenerates to
\begin{equation}
	\Sigma:\quad h_0\equiv x_1^5+x_2^5\,,\quad  h_2=\beta^5x_3^5-\alpha^5x_4^5=0\,,\quad h_2=\gamma^5x_3^5-\alpha^5x_5^5=0\,.
\label{eq:tbquintic_h0}
\end{equation}
This can be trivially factorized as in the general discussion \eqref{eq:sublocus} in linear factors that differ only by fifths roots of unity $\eta_i$, that are the 125 solutions to \eqref{eq:anholoC}. In other words, at the locus ${\cal M}(\P^1)$ the curve $\Sigma$ degenerates to $125$ lines corresponding to each choice of $\eta_i$ in the three constraints $h_i$. However, in contrast to \eqref{eq:RSquintic} which is invariant under the orbifold $G$, the linearized equations do transform under $G$. In fact, all the $125$ different lines are identified modulo the action of $G=(\mathds{Z}_5)^3$ so that \eqref{eq:tbquintic_h0} describes a single line on the quotient by $G$,
\begin{equation}
	 {\cal M}(\P^1):\quad\eta x_1+x_2=0\,,\quad \alpha x_4-\beta x_3=0\,,\quad \alpha x_5-\gamma x_3=0\,.
\label{eq:constlines}	
\end{equation}
Equivalently, these lines are given parametrically in $\P^4$ in terms of homogeneous coordinates $U$, $V$ on $\P^1$ as the Veronese mapping
\begin{equation}
 	\quad(U,V)\,\,\mapsto\,\, (U,-\eta U,\alpha V,\beta V,\gamma V)\,,\quad \eta^5=1\,.
\label{eq:paramlines}
\end{equation}
This way, the family $\Sigma$ contains the holomorphic lines \eqref{eq:paramlines} at the sublocus ${\cal M}(\P^1)$ of \eqref{eq:modulicond} defined by the vanishing of $m(x_1,x_2,x_3)$. In summary this shows that a five-brane wrapping the line \eqref{eq:paramlines} falls in the class of toric branes at the sublocus ${\cal M}(\P^1)$ of their moduli space. 
We emphasize again that \eqref{eq:paramlines} is not invariant under the orbifold group $G$ and that the identification of the 125 distinct solutions to \eqref{eq:anholoC} under $G$ is essential to match an in general higher genus Riemann surface $\Sigma$ with a rational curve of genus $g=0$.

 This picture is further confirmed from the perspective of the rational curve \eqref{eq:paramlines} since the constraint \eqref{eq:modulicond} defining ${\cal M}(\P^1)$ is precisely the condition for the line \eqref{eq:paramlines} to lie holomorphically in the quintic constraint $P$. Thus, the sublocus ${\cal M}(\P^1)$ defined in \eqref{eq:modulicond} is precisely the moduli space of the five-brane wrapping the holomorphic lines in the quintic. For generic $\Psi\neq 0$ this moduli space is only a number of discrete points whereas at the Fermat point $\Psi=0$ there is a one-dimensional moduli space of lines in the quintic parametrized by a Riemann surface\footnote{The first constraint in \eqref{eq:modulicond} is a quintic constraint in $\P^2$ describing a Riemann surface of genus $g=6$.} of genus $g=6$, cf.~figure \ref{fig:moduli}. In the language of superpotentials, we understand ${\cal M}(\P^1)$ as the critical locus of $W_{\rm brane}$ at which the five-brane on the rational curve is supersymmetric.
Conversely, deforming away from the critical locus ${\cal M}(\P^1)$ in ${\cal M}(\Sigma)$ is obstructed, inducing a non-trivial superpotential.
Thus, we consider in the following the deformation space defined by $\tilde{\P}^1$, more precisely by the coefficients of $m$,
\begin{equation}
 	\hat{z}^1=\frac{\beta\gamma}{\alpha^2}=(u^1u^2)^{\frac{1}{5}}\,,\quad \hat{z}^2=\frac{\alpha^5+\beta^5+\gamma^5}{\alpha^{5}}=1+u^1+u^2\,, 
\end{equation}
which agrees with the choice of variables used in figure \ref{fig:moduli}. As noted before, we can canonically identify this deformation space with the moduli space ${\cal M}(\Sigma)$ of $\Sigma$ in $Z_3$ by dividing out the orbifold group $G$ and working with the holomorphic constraint \eqref{eq:RSquintic} instead of \eqref{eq:anholoC}.

Most importantly for the blow-up procedure, the description of the toric curve $\Sigma$ of \eqref{eq:RSquintic} is precisely in the form used in section \ref{geometricblowups} to construct the blow-up geometry $\hat{Z}_3$. In particular, we can easily read off the normal bundle $N_{Z_3}\Sigma$ of $\Sigma$ in the quintic which is  $N_{Z_3}\Sigma=\mathcal{O}(5)\oplus\mathcal{O}(5)$ by simply noting the degree of the divisors $h_1=0$, $h_2=0$. Then the blow-up $\hat{Z}_3$ is given by the complete intersection \eqref{eq:blowup}, which in the case at hand reads
\begin{equation}
	\hat{Z}_3\,:\quad P=0\,,\quad Q=l_1(u^2 x_3^5-x_5^5)-l_2(u^1 x_3^5-x_4^5)=0\,.
\label{eq:BUquintic}
\end{equation}   
Since both the closed modulus $\Psi$ as well as the open moduli $u^1$, $u^2$ enter \eqref{eq:BUquintic}, we formally obtain the embedding of the open-closed moduli space of $(Z_3,\Sigma)$, and equivalently the obstructed deformation space of $(Z_3,\tilde{\P}^1)$, into the complex structure moduli space of $\hat{Z}_3$. In particular this trivially embeds the moduli space of the rational curves \eqref{eq:paramlines} by restricting to the critical locus \eqref{eq:modulicond}. 

On the blow-up $\hat{Z}_3$ this embedding as well as the obstructions can be understood purely geometrically. First of all we note that the action of the quintic orbifold $G$ directly lifts to $\hat{Z}_3$. Then by deforming away from the critical values \eqref{eq:modulicond} we change the topology of the blow-up divisor $E$ from a ruled surface over $\P^1$ to a ruled surface $E$ over a Riemann-surface $\Sigma$ of higher genus. The one-cycles of the Riemann-surface in the base lift to new three-cycles on the blow-up $\hat{Z}_3$ that correspond to new non-algebraic complex structure deformations\footnote{Although related these new complex structure deformations should not be confused with the parameters entering $Q$ since these are algebraic by definition.}, compare to the similar discussion of \cite{Kachru:2000ih,Kachru:2000an}. Upon switching on flux on these three-cycles turns on higher order obstructions for the complex structure of $\hat{Z}_3$ destroying the ruled surface $E$ and thus driving us back to the critical locus where $\Sigma$ degenerates to $\P^1$. This way the flux obstructs the complex structure in \eqref{eq:BUquintic} which is expressed by a flux superpotential on $\hat{Z}_3$ that is the sought for superpotential $W_{\rm brane}$. 

In the following we will use the complete intersection \eqref{eq:BUquintic} to analyze the open-closed deformation space $(Z_3,\tilde{\P}^1)$.
The crucial point is that we are working with a well-defined complex moduli space of $(Z_3,\Sigma)$ respectively of complex structures on $\hat{Z}_3$ to describe the space of \textit{anholomorphic} deformations $\tilde{\P}^1$.
In this context this is another reason for the effectiveness of the blow-up $\hat{Z}_3$ for the description of the obstructed brane deformations $\tilde{\P}^1$.

\subsubsection{Toric Branes on the Quintic: GKZ-Systems from Blow-Up Threefolds}
\label{toricbranesP11111-GKZ}

In the following we analyze the open-closed deformation space of $(\tilde{\P}^1,Z_3)$ embedded in the complex structure moduli of $\hat{Z}_3$ augmented by appropriate flux data. We perform this analysis by toric means, i.e.~the GKZ-system. Thus we supplement the polyhedron $\Delta^{\tilde Z}_4$ and the charge vectors $\ell^{(1)}$, $\hat{\ell}^{(1)}$, $\hat{\ell}^{(2)}$ of the quintic Calabi-Yau \eqref{eq:mirrorquintic} and the toric brane \eqref{eq:RSquintic},
\begin{equation}\label{quinticpoly}
	\begin{pmatrix}[c|cccc|c|l||cc]
	    	&   &  \Delta_4^{\tilde Z} &   &    &  \ell^{(1)} &   &\hat{\ell}^{(1)}&\hat{\ell}^{(2)}\\ \hline
		\tilde{v}_0 & 0 & 0 & 0 & 0 	&  -5    & y_0 = x_1x_2x_3x_4x_5 & 0&0\\
		\tilde{v}_1 &-1 &-1 &-1 & -1 	&  1     & y_1 = x_1^5 & 0&0\\
		\tilde{v}_2 & 1 & 0 & 0 & 0 	&  1     & y_2 = x_2^5 & 0&0\\
		\tilde{v}_3 & 0 & 1 & 0 & 0 	&  1     & y_3 = x_3^5 & 1&1\\
		\tilde{v}_4 & 0 & 0 & 1 & 0 	&  1     & y_4 = x_4^5 & -1&0\\
	  \tilde{v}_5 & 0 & 0 & 0 & 1 	&  1     & y_5 = x_5^5 & 0&-1
	\end{pmatrix}.
\end{equation}
The points of the dual polyhedron $\Delta^{ Z}_4$ are given by $v_1=(-1,-1,-1,-1)$, $v_2=(4,-1,-1,-1)$, $v_3=(-1,4,-1,-1)$, $v_4=(-1,-1,4,-1)$ and $v_5=(-1,-1,-1,4)$. These monomials both enter the constraints $P$ and $h_i$ according to \eqref{eq:Batyrev} and \eqref{eq:BBrane} yielding
\begin{equation}
	Z_3\,:\quad P=\sum_{i=1}^5a_ix_i^5+a_0 x_1x_2x_3x_4x_5\,,\quad \Sigma\,:\quad h_1=a_6 x_3^5+a_7x_4^5\,,\quad h_2=a_8 x_3^5+a_9x_5^5\,,
\label{eq:toricquintic}
\end{equation}
where we introduced free complex-valued coefficients $\underline{a}$.\footnote{Conversely to the conventions in \eqref{eq:Zslift}, we denote the parameters $b_i$ by $a_{5+i}$ for convenience.} 
From the polyhedron \eqref{quinticpoly} we readily obtain the standard toric GKZ-system for $Z_3$ along the lines of eqs. \eqref{eq:pfo} and \eqref{eq:Zs},
\begin{eqnarray}\label{eq:GKZquinticclosed}
 	&\mathcal{Z}_0=\sum_{i=0}^5\vartheta_i+1\,,\quad \mathcal{Z}_i=\vartheta_{i+1}-\vartheta_1\,\,(i=1,\ldots,4)\,,\nonumber&\\
	&\displaystyle\mathcal{L}_{1}=\prod_{i=1}^5\frac{\partial}{\partial a_i}-\frac{\partial^5}{\partial a_0^5}\,,\displaystyle&
\end{eqnarray}
where we use the logarithmic derivative $\vartheta_i=a_i\frac{\partial}{\partial a_i}$. The $\mathcal{Z}_i$ express the coordinate rescalings leaving the measure $\Delta$ and the monomial $x_1x_2x_3x_4x_5$ in \eqref{eq:residueZ3} invariant. They express infinitesimal rescalings of the parameters $\underline{a}$ and the coordinates $\underline{x}$ entering $P$. For example the rescaling $(x_1,x_2)\mapsto (\lambda^{1/5} x_1,\lambda^{-1/5}x_2)$ combined with $(a_1,a_2)\mapsto (\lambda^{-1} a_1,\lambda a_2)$ leaves $P$ invariant and consequently the periods have the symmetry $\Pi^k(a_0,\lambda^{-1}a_1,\lambda a_2,a_3,a_4,a_5)=\Pi^k(\underline{a})$. The corresponding generator of this symmetry is $\mathcal{Z}_1$.
These homogeneity properties of the $\Pi^k(a_i)$ imply that they are functions of only a specific combination of the $\underline{a}$, which in the case of the quintic takes the form
\begin{equation}
	z^1=-\frac{a_1a_2a_3a_4a_5}{a_0^5}\,.
\end{equation}
This is perfectly consistent with \eqref{eq:algCoords} and the charge vector $\ell^{(1)}$.

The analysis of the combined system of the quintic and the curve \eqref{eq:toricquintic} is performed by replacing $(Z_3,\Sigma)$ by the blow-up $(\hat{Z}_3,E)$ given by the family of complete intersections in $\mathcal{W}=\P(\mathcal{O}(5)\oplus \mathcal{O}(5))\cong \P(\mathcal{O}\oplus \mathcal{O})$,
\begin{equation}
	\hat{Z}_3\,:\quad P=0\,,\quad Q=\ell_1(a_8 x_3^5+a_9x_5^5)-\ell_2(a_6x_3^5+a_7x_4^5)\,.
\end{equation}
Then the holomorphic three-form $\hat{\Omega}$ is constructed using the residue \eqref{eq:ResZhat}. Using this explicit residue integral expression, it is straightforward to find the Picard-Fuchs system on the blow-up $\hat{Z}_3$ that encodes the complex structure dependence of $\hat{\Omega}$. As it can be directly checked the GKZ-system is given by $\mathcal{L}_{1}$ as in \eqref{eq:GKZquinticclosed} complemented to the system 
\begin{eqnarray}\label{eq:GKZquinticopen}
 	&\mathcal{Z}_0=\sum_{i=0}^5\vartheta_i+1\,,\quad \mathcal{Z}_1=\sum_{i=6}^9\vartheta_i\,,\quad \mathcal{Z}_2=\vartheta_2-\vartheta_1\,,\quad \mathcal{Z}_3=\vartheta_3-\vartheta_1+\vartheta_6+\vartheta_8\,,&\nonumber\\
 	& \mathcal{Z}_4=\vartheta_4-\vartheta_1+\vartheta_7\,,\quad \mathcal{Z}_5=\vartheta_5-\vartheta_1+\vartheta_9\,,\quad \mathcal{Z}_6=\vartheta_8+\vartheta_9-\vartheta_6-\vartheta_7\,,\displaystyle&\nonumber\\
	&\displaystyle\mathcal{L}_{1}=\prod_{i=1}^5\frac{\partial}{\partial a_i}-\frac{\partial^5}{\partial a_0^5}\,,\quad\mathcal{L}_{2}=\frac{\partial^2}{\partial a_3\partial a_7}-\frac{\partial^2}{\partial a_4\partial a_6}\,,\quad \mathcal{L}_{3}=\frac{\partial^2}{\partial a_3\partial a_9}-\frac{\partial^2}{\partial a_5\partial a_8}\,.\displaystyle
\end{eqnarray} 
We emphasize that there are two new second order differential operators $\mathcal{L}_{2}$, $\mathcal{L}_{3}$ that annihilate $\hat{\Omega}$ identically and that incorporate the deformations $a_i$, $i=6,7,8,9$ associated to the curve $\Sigma$. It is clear from the appearance of the $\P^1$-coordinates $(l_1,l_2)$ in the constraint $Q$ that there are no further operators $\mathcal{L}_{a}$ on $\hat{Z}_3$ of minimal degree.
Let us briefly explain the origin of the operators $\mathcal{Z}_k$. The first two are simply associated to an overall rescaling of the two constraints $P\mapsto \lambda P$, $Q\mapsto \lambda' Q$ which acts on $\hat{\Omega}(\underline{a})$ as
$\hat{\Omega}(\lambda a_0,\ldots, \lambda a_6,a_7,\ldots, a_{10})=\lambda\hat{\Omega}(\underline{a})$ and $\hat{\Omega}(a_0,\ldots, a_6,\lambda'a_7,\ldots, \lambda'a_{10})=\hat{\Omega}(\underline{a})$. For the rescaling of $Q$ the factor $\lambda'$ is compensated by the non-trivial prefactor $h_i/\ell_i$ in \eqref{eq:ResZhat}. The third to sixth operators are associated to the torus symmetries of the $\P^4$ as before, $(x_1,x_j)\mapsto(\lambda_j x_1,\lambda_j^{-1}x_j)$, $j=2,\ldots,5$, and the last operator $\mathcal{Z}_6$ is related to the torus symmetry $(l_1,l_2)\mapsto (\lambda l_1,\lambda^{-1}l_2)$ of the exceptional $\P^1$. It is important to note that the operators $\mathcal{Z}_i$ of the $\P^4$ are altered due to the blow-up $\hat{Z}_3$, i.~e.~due to the presence of the five-brane, as compared to the closed string case of \eqref{eq:GKZquinticclosed}. 

Before delving into the determination of the solutions to this differential system let us reconsider the operators we just found from a slightly different perspective. This will in particular allow for a straightforward systematization of the constructions of GKZ-system.

Comparing \eqref{eq:GKZquinticopen} to the closed GKZ-system \eqref{eq:GKZquinticclosed} associated to $Z_3$ we recover a very similar structure. Indeed the above differential system governing 
the complex structure on $\hat{Z}_3$ defines a new GKZ-system with exponent $\beta$. To obtain the set of integral points $\hat{v}_i$ associated to this GKZ-system we apply the general definition of the $\mathcal{Z}_i$ in \eqref{eq:Zs} backwards to obtain
\begin{equation}\label{blowupPolyquintic}
	\begin{pmatrix}[c|ccccccc|ccc|l]
	    		     &  &   &   & \Delta_7^{\hat Z}&&&   		  &\hat{\ell}^{(1)} &\hat{\ell}^{(2)} &\hat{\ell}^{(3)}  &          \\ \hline
		\hat{v}_0    &1 & 0 & 0 & 0 & 0 & 0 & 0   &-5   & 0  		  &	0	&	 \hat{y}_0 = x_1x_2x_3x_4x_5 \\
		\hat{v}_1    &1 & 0 &-1 &-1 &-1 &-1 & 0	  & 1   & 0  		  & 0	&	 \hat{y}_1 = x_1^5           \\
		\hat{v}_2    &1 & 0 & 1 & 0 & 0 & 0 & 0	  & 1   & 0  		  & 0	&	 \hat{y}_2 = x_2^5           \\
		\hat{v}_3    &1 & 0 & 0 & 1 & 0 & 0 & 0	  & 3   & -1  		  & -1	&	 \hat{y}_3 = x_3^5           \\
		\hat{v}_4    &1 & 0 & 0 & 0 & 1 & 0 & 0   & 0   &1  		  & 0	&  \hat{y}_4 = x_4^5            \\
	  \hat{v}_5    &1 & 0 & 0 & 0 & 0 & 1 & 0	  & 0   & 0  		  & 1	&	 \hat{y}_5 = x_5^5            \\
		\hat{v}_6    &0 & 1 & 0 & 1 & 0 & 0 &-1	  & -1   &1  		  & 0	&	 \hat{y}_6 =l_1 \hat{y}_3        \\
		\hat{v}_7    &0 & 1 & 0 & 0 & 1 & 0 &-1	  & 1   & -1  		  & 0	&	 \hat{y}_7 =l_1 \hat{y}_4        \\
		\hat{v}_8    &0 & 1 & 0 & 1 & 0 & 0 & 1	  & -1   & 0  		  & 1	&	 \hat{y}_8 =l_2 \hat{y}_3         \\
		\hat{v}_9    &0 & 1 & 0 & 0 & 0 & 1 & 1	  & 1   & 0  		  & -1	&  \hat{y}_9 =l_2 \hat{y}_5                   
	\end{pmatrix}.                                                                        
\end{equation}
Here we have displayed the points $\hat{v}_i$, the corresponding monomials $\hat{y}_i$ and a basis of relations $\hat{\ell}^{(i)}$, that we obtain as a Mori cone of a triangulation of the polyhedron $\Delta_7^{\hat Z}$. 
We emphasize that besides the closed string charge vectors of $Z_3$ embedded as $\ell^{(1)}=\hat{\ell}^{(1)}+\hat\ell^{(2)}+\hat\ell^{(3)}$, the brane charge vectors $\hat{\ell}^{(a)}$ are among the charge vectors $\hat{\ell}^{(j)}$ of $\Delta_7^{\hat{Z}}$ as well.
Furthermore, for the above triangulation of $\Delta_7^{\hat{Z}}$ we immediately obtain the full GKZ differential system $\mathcal{L}_{a}$, $\mathcal{Z}_i$ of \eqref{eq:GKZquinticopen} by the standard formulas for the \textit{standard} GKZ-system in \eqref{eq:pfo}, \eqref{eq:Zs} using the points $\hat{v}_i$ and relations $\hat{\ell}^{(j)}$ from $\Delta^{\hat{Z}}_7$ with exponent $\beta=(-1,0,0,0,0,0,0)$.

This GKZ-system defines coordinates $\hat{z}^a$ on the complex structure moduli space of $\hat{Z}_3$ as before. We apply the closed string formula \eqref{eq:algCoords} for the charge vectors $\hat{\ell}^{(a)}$ of $\Delta^{\hat{Z}}_7$ to obtain the three coordinates
\begin{equation} \label{eq:zOpenquintic}
 	\hat{z}^1=-\frac{a_1a_2a_3^3a_7a_9}{a_0^5a_6a_8}\,,\quad \hat{z}^2=\frac{a_4a_6}{a_3a_7}\,,\quad \hat{z}^3=\frac{a_5a_8}{a_3a_9}\,.
\end{equation}
We obtain a complete system of differential operators $\mathcal{D}_a$, the Picard-Fuchs operators, by adding to \eqref{eq:GKZquinticopen} further operators $\mathcal{L}_a$ associated to scaling symmetries specified by integer positive linear combinations of the charge vectors $\hat{\ell}^{(a)}$ in (\ref{blowupPolyquintic}). By factorizing these operators $\mathcal{L}_a$ expressed in the coordinates \eqref{eq:zOpenquintic} we obtain the differential system generated by 
\begin{eqnarray}	
\mathcal{D}_1&=&\theta _1 \theta_2 \theta_3 \left(3 \theta _1\!-\!\theta _2\!-\!\theta _3\!\right)-
5 \prod_{i=1}^4 \left(5 \theta _1\!-\!i\right)\hat{z}^1 \hat{z}^2 \hat{z}^3\,,\nonumber\\
\mathcal{D}_i&=&\left(\theta _1-\theta _i\right) \theta _i+ \left(1+\theta _1-\theta _2\right) \left(1+3 \theta _1-\theta _2-\theta _3\right)\hat{z}^i\,,\quad i=2,3\,,
\label{eq:GKZquinticz}
\end{eqnarray}
where we introduced $\theta_i=\hat{z}^i\frac{\partial}{\partial \hat{z}^i}$ and further rescaled the holomorphic three-form $\hat{\Omega}$ by $a_0$. Each of these three operator $\mathcal{D}_a$ corresponds to a linear combination of the charge vectors $\hat{\ell}^{(i)}$, whose integer coefficients can be read off from the powers of $\hat{z}^i$ in the last term of $\mathcal{D}_a$.
Obviously the deformation problem is symmetric under exchange of $\hat{z}^2$ and $\hat{z}^3$. While $\mathcal{D}_1$ is symmetric under that symmetry, $\mathcal{D}_2$ and $\mathcal{D}_3$ map onto each other under $\hat{z}^2\leftrightarrow \hat{z}^3$. 

This Picard-Fuchs system is perfectly consistent with the expected structure from section \eqref{eq:LInZHat}, that in particular implies that the periods of $\Omega$ directly lift to the blow-up $\hat{Z}_3$. Upon the identification of the coordinate  $z^1=\hat{z}^1\hat{z}^2\hat{z}^3$ on the complex structure moduli space of the quintic\footnote{This is perfectly consistent with the embedding of the quintic charge vector as $\ell^{(1)}=\hat{\ell}^{(1)}+\hat\ell^{(2)}+\hat\ell^{(3)}$.}, and keeping $\hat{z}^2$, $\hat{z}^3$ unchanged, we rewrite the operators \eqref{eq:GKZquinticz} as
\begin{eqnarray}
 	\mathcal{D}_1&=&\mathcal{D}_1^{Z_3}+[\theta_1(\theta_1+\theta_3)(\theta_1-\theta_2-\theta_3)\theta_2+(\theta_2\leftrightarrow \theta_3)]\,,\nonumber\\
	\mathcal{D}_i&=&-[\theta_1+\theta_i-z_i(\theta_1-\theta_2-\theta_3)]\cdot\theta_i\,,
\end{eqnarray}
where we write $\theta_1=z^1\frac{\partial}{\partial z^1}$ by abuse of notation.
The first operator $\mathcal{D}_{1}$ splits into the well-known fourth order quintic operator $\mathcal{D}_1^{Z_3}=\theta _1^4-
5 \prod_{i=1}^4 \left(5 \theta _1\!-\!i\right)z^1$ and a term linear in the derivatives $\theta_2$, $\theta_3$. The other operators $\mathcal{D}_{2}$, $\mathcal{D}_{3}$ are proportional to $\theta_2$, $\theta_3$. Consequently, it is ensured that the solutions to \eqref{eq:GKZquinticz} contain the closed string periods $\Pi^k(z^1)$ of the quintic as the unique solutions independent of the open string parameters $\hat{z}^2$, $\hat{z}^3$.

Thus, we summarize by emphasizing that the complete information for the study of complex structure variations in the family $\hat{Z}_3$ of complete intersection threefolds $P=Q=0$ just reduces to the determination of the toric data $\Delta_7^{\hat{Z}}$ and the associated GKZ-system.

\subsubsection{Branes on the Quintic: Superpotentials from Blow-Up Threefolds}
\label{toricbranesP11111-W}

The complex structure moduli space of the blown up quintic orbifold $\hat{Z}_3$ described above is 
the model for our open/closed deformation space and (\ref{eq:GKZquinticz}) is 
the Picard-Fuchs system annihilating its periods.  We will analyze and interpret
the global properties of the deformations space and the solutions at special 
points  in the deformation space. First we analyze the solutions at the locus $\hat z_i=0$.
Different than for systems that can be embedded in a Calabi-Yau fourfold, as the 
one in sections \ref{ToricBraneBlowupII}, we find at $\hat z_i=0$ no maximal unipotent monodromy. Rather the indicial equations of 
the system (\ref{eq:GKZquinticz}) have  the solutions $(0,0,0)^{12},(\frac{1}{3},0,0),(\frac{2}{3},0,0,0)$, 
$(\frac{1}{2},\frac{1}{2},0), (\frac{1}{2},0,\frac{1}{2})$. So in total we find  $16$ 
solutions. The twelve-times degenerate solution $((0,0,0)^{12}$ gives rise to one power 
series  solution 
\begin{equation} 
X^{(0)}_1=1+120 z+113400 z^2+ 168168000 z^3 + {\cal O}(z^4)\ , 
\end{equation}
where $z=\hat{z}_1\hat{z}_2\hat{z}_3$ is the quintic complex structure 
parameter near the point of maximal unipotent monodromy in its moduli space. This solution 
is identified with the fundamental period $X_0$ of the quintic. Denoting 
$\hat l_i:=\log(\hat z_i)$ we get additional eleven logarithmic solutions 
\begin{eqnarray} \label{eq:leadinglogsquintic_b1}
		X^{(1)}_i\,:\,& \hat{l}_1\,,\,\,\hat{l}_2\,,\,\,\hat{l}_3\,,&\\[0.3Em]
		X^{(2)}_{\alpha}\,:\,& \frac{1}{2}\hat{l}_1^2\,,\,\, \hat{l}_2(\frac12\hat{l}_2+ \hat{l}_1)\,,\,\,\hat{l}_3(\frac12\hat{l}_3 +\hat{l}_2 )\,,\,\,\hat{l}_2\hat{l}_3&\nonumber\\[0.3Em]
		X^{(3)}_{\beta}\,:\,& \frac16 \hat{l}_1^3\,,\,\,\frac16 \hat{l}_2^3+\frac12 \hat{l}_1^2 \hat{l}_2  +\frac12 \hat{l}_2^2 \hat{l}_1\,,\,\,\frac16 \hat{l}_3^3+\frac12 \hat{l}_1^2 \hat{l}_3  +\frac12 \hat{l}_3^2 \hat{l}_1\,,\,\,\frac12 \hat{l}_2^2 \hat{l}_3 +\frac12 \hat{l}_3 \hat{l}_3^2+ \hat{l}_1 \hat{l}_2\hat{l}_3\, . &\nonumber
\end{eqnarray}
The  single logarithmic solutions are 
\begin{eqnarray}
		X^{(1)}_1&=& X^{(0)}\log(\hat{z}_1)-60 \hat{z}_1(\hat{z}_2 +\hat{z}_3)+770 z+ 9450 {\hat z}_1^2({\hat z}_2^2+{\hat z_3}^2)
                    +60 {\hat z}_1 ({\hat z}_2^2{\hat z}_3 +{\hat z}_2 {\hat z}_3^2)+\mathcal{O}(\underline{\hat{z}}^5)\,,\nonumber\\
		X^{(1)}_2&=& X^{(0)}\log(\hat{z}_2)+60 \hat{z}_1 \hat{z}_3-9450 {\hat z}_1^2 {\hat z_3}^2
                    -60 {\hat z}_1 {\hat z}_2^2{\hat z}_3 +  \mathcal{O}(\underline{\hat{z}}^5)   \,,\\
		X^{(1)}_3&=& X^{(0)}\log(\hat{z}_3) +60 \hat{z}_1 \hat{z}_2-9450 {\hat z}_1^2 {\hat z_2}^2
                    -60 {\hat z}_1 {\hat z}_3^2{\hat z}_2 +\mathcal{O}(\underline{\hat{z}}^5)\,.\nonumber
\end{eqnarray}
It is easy to check that the single logarithmic period of the mirror quintic is obtained as $\sum_i X^{(1)}_i$. Similarly 
we have chosen the normalization of (\ref{eq:leadinglogsquintic_b1}) so that $\sum_\alpha X^{(2)}_{\alpha}$  and 
$\sum_\beta X^{(3)}_{\beta}$ are double and triple logarithmic solutions of the Picard-Fuchs equation 
of the mirror quintic $Z_3$. Using the information about the classical terms of the mirror 
quintic~\cite{Candelas:1990rm,Hosono:1994ax} one can identify the precise combination 
of periods corresponding to a basis of $H_3(Z_3,\mathbb{Z})$.

Notable are the four fractional power series solutions to the remaining indices, 
\begin{eqnarray}
X^{(0)}_2&=&\hat{z}_1^\frac{1}{3}+\hat{z}_1^\frac{1}{3}(\frac{1}{2}\hat{z}_2+\frac{1}{2}\hat{z}_3+\frac{6545}{2592}\hat{z}_1)+{\cal O}(\underline{\hat{z}}^\frac{7}{3})\,,  \\
X^{(0)}_3&=&\hat{z}_1^\frac{2}{3}+\hat{z}_1^\frac{2}{3}(4\hat{z}_2+4 \hat{z}_3+\frac{86944}{10125}\hat{z}_1)+{\cal O}(\underline{\hat{z}}^\frac{7}{3})\,,  \nonumber\\
X^{(0)}_4&=&\sqrt{\hat{z}_1 \hat{z}_2} +\sqrt{\hat{z}_1 \hat{z}_2} \hat{z}_3-\frac{5005}{72} (\hat{z}_1\hat{z}_2)^\frac{3}{2}+ {\cal O}(\underline{\hat{z}}^4)\,, \nonumber\\ 
X^{(0)}_5&=&\sqrt{\hat{z}_1 \hat{z}_3} +\sqrt{\hat{z}_1 \hat{z}_3} \hat{z}_2-\frac{5005}{72} (\hat{z}_1\hat{z}_3)^\frac{3}{2}+ {\cal O}(\underline{\hat{z}}^4)\,.\nonumber 
\end{eqnarray}

Let us discuss now the global properties of the moduli space of the branes 
on the quintic orbifold defined by~(\ref{eq:RSquintic}). As discussed in section \ref{toricbranesP11111}
there are critical points, where the unobstructed deformation problem of the complete 
intersection (\ref{eq:RSquintic}) gives rise to superpotentials associated to obstructed deformation 
problems such as the lines in the quintic orbifold. Clearly these loci must occur at the 
discriminant of the Picard-Fuchs equation determined by $\mathcal{D}_1,\mathcal{D}_2,\mathcal{D}_3$ 
described  in the last section. We find
\begin{eqnarray}
\Delta&=&(1 + \hat{z}_2) (1 - \hat{z}_2) (1+\hat{z}_3) (1-\hat{z}_3) (1-\hat{z}_2-\hat{z}_3) (1+2 \hat{z}_2-\hat{z}_3)(1+2 \hat{z}_3-\hat{z}_2) 
 \,  \\
& &  \times(4+ 5^5\hat{z}_1\hat{z}_2 (1-\hat{z}_3)^2) (4+5^5 \hat{z}_1 \hat{z}_3 (1-\hat{z}_2)^2) (1 - 5^5 \hat{z}_1 \hat{z}_2\hat{z}_3) (27 + 5^5 \hat{z}_1 (1 - \hat{z}_2 - \hat{z}_3)^3)\ .\nonumber
\end{eqnarray}
We expect to get a degeneration of the holomorphic curve $\Sigma$ of (\ref{eq:RSquintic}) at the discriminate locus and 
thus obstructed deformation problems that can be characterized by appropriate 
flux quantum numbers. Let us consider two discriminant loci of particular interest.

At the locus $\hat{z}_2=-1$ and $\hat{z}_3=-1$\footnote{We note that $\hat{z}_2=\hat{z}_3=-1$ agrees with $u^1=u^2=1$ in the notation of \eqref{eq:BBrane} since $\hat{z}_a=-u^a$.} the complete intersection becomes holomorphic 
in the quintic and in fact the toric A-brane, which is mirror to the holomorphic 
constraint, becomes compatible with the involution brane, i.e.  the fixpoint 
locus of the involution 
\begin{equation}
 (x_1,x_2,x_3,x_4,x_5)\rightarrow (\bar x_1,\bar x_2 ,\bar x_3, \bar x_4, \bar x_5)\ . 
\end{equation}
More precisely the toric $A$-branes is given by the constraints (\ref{ABranes}) 
defined by the charge vectors $\hat{\ell}^{(a)}$ in (\ref{eq:toricquintic}) with vanishing relative K\"ahler/Wilson 
line parameters $c^a=0$. Comparing the solutions at that locus we obtain a 
two open parameter deformation of the brane discussed in~\cite{Walcher}. 
The relevant periods at the involution brane point are trivially obtained 
from the solutions at the large complex structure point by analytic continuation. In particular the solutions 
at large complex structure, which are at most linear in the logarithms 
of the $\hat{z}_a$ converge in the variables $(v_1=z_1,v_2=(1+\hat{z}_2),v_3=(1+\hat{z}_3))$.
The solutions with the square root cuts $X^{(0)}_{4}$ and $X^{(0)}_{5}$ are expected to 
specialize to the superpotential for the involution brane, if the open moduli $v_2$ 
and $v_3$  are set to zero. Indeed, if we symmetrize in the two square root solutions,
we find up to a normalization worked out in~\cite{Walcher} the series
\begin{equation}  \label{eq:involutionW}
\begin{array}{rl} 
W^{quant}=&\displaystyle{\frac{30}{4 \pi^2}\biggl(v_1^{1/2}+\frac{5005}{9} v_1^{3/2}+\frac{52055003}{75}v_1^{5/2}+
v_1^{1/2}( \frac{1}{2}(v_2+ v_3)- \frac{1}{16}(v_2^2-4v_1 v_2 + v_3^2))}\\ 
&\displaystyle{+\frac{5005}{6} v_1^{3/2} (v_2+v_3)+{\cal O}(\underline{v}^{7/2})\biggl)}\ .
\end{array}
\end{equation} 
In particular we note that for $v_2=v_3=0$ this superpotential is exactly the 
one for the involution brane obtained in~\cite{Walcher}. Using the mirror map of the quintic
it is possible to obtain from \eqref{eq:involutionW} at $v_2=v_3=0$ the disk instantons for the involution brane. 
We expect that the scalar potential induced by \eqref{eq:involutionW} has a minimum along the $v_2=v_3=0$ direction. However to see 
this   minimalization explicitly requires  construction of the K\"ahler potential, 
a choice of flat coordinates and a choice of the gauging of the superpotential 
in the K\"ahler line bundel. 
We note that the above discussion of the involution brane is similar to the one of~\cite{Alim:2009rf} in the context of a one open parameter 
family of a toric brane on the quintic.  

A similarly interesting locus is the $(1-\hat{z}_2-\hat{z}_3)=0$ and $\frac{1}{\hat{z}_1}=0$. 
According to the discussion in section \ref{toricbranesP11111} this is 
the locus $\mathcal{M}_{\P^1}(\Sigma)$ of \eqref{eq:modulicond}, 
$\frac{(\alpha^5+\beta^5+\gamma^5)}{\alpha^5}=0$ and $\frac{\psi \beta \gamma}{\alpha^2}=0$, where 
the constraints \eqref{eq:RSquintic} factorize and the holomorphic lines occur. 
We expect the superpotential to vanish at this locus. Indeed if we expand in 
$(w_1=\frac{1}{\hat{z}_1}, w_2=(1-\hat{z}_2-\hat{z}_3),w_3=\hat{z}_2-\hat{z}_3)$ we find 16 solutions having 
the indicials $(\frac{k}{5},i,j)$, where $k=1,\ldots,4$ and 
$(i,j)=(0,0),(1,0),(0,1),(1,1)$. Thus, the solutions vanish with 
$\hat{z}_1^{-\frac15},\,\hat{z}_1^{-\frac25},\,\hat{z}_1^{-\frac35},\,\hat{z}_1^{-\frac45}$ for $\hat{z}_1^{-\frac15}=\frac{\psi\beta\gamma}{\alpha^2}$. 
This is compatible with the vanishing of the superpotential at the locus of the holomorphic lines. 
Again one would  need  the flat coordinates and the gauge choice in order to 
to perform a detailed local analysis of the orbifold 
superpotential. 

In summary, from the two examples above it is clear that the descriminant of 
the Picard-Fuchs equation contains the expected information about the 
degeneration of the two open parameter brane system ar special loci, 
where the problem can be related to obstructed deformation problems. 
We expect this also to be true at the other loci of the descriminant, where a brane interpretation
is not yet available.

\subsubsection{Brane Superpotential at Large Volume: Disk Instantons}
\label{toricbranesP11111-W-LV}

In this section we apply the blow-up $\hat{Z}_3$ to a different five-brane on the quintic. The following analysis is focused on the determination of the disk instanton invariants at large radius of the A-model and thus brief at several points for the sake of brevity.

The Calabi-Yau geometry of the B-model is given by the one parameter mirror quintic with the constraint $P$ as in \eqref{eq:toricquintic}. We add an open string sector of a five-brane, that we describe as before under the identification of the moduli space $\mathcal{M}(\Sigma)$ with the deformation space $\tilde{P}^1$, by the toric curve $\Sigma$ specified by brane charge vectors $\hat{\ell}^{(a)}$ as
\begin{eqnarray} \label{eq:toricquintic_b2}
 \Sigma&:& P=0\,,\quad h_1\equiv a_6x_1x_2x_3x_4x_5+a_7x_1^5=0\,,\quad h_2\equiv a_8x_1^5+a_9x_2^5=0\,, \nonumber\\
& & \hat{\ell}^{(1)}=(-1,1,0,0,0,0)\,,\quad\hat{\ell}^{(2)}=(0,-1,1,0,0,0)\,.
\end{eqnarray}

For this geometry we readily construct the blow-up $\hat{Z}_3$ as the complete intersection in the toric variety $\mathcal{W}=\mathds{P}(\mathcal{O}(5)\oplus\mathcal{O}(5))\cong \P(\mathcal{O}\oplus\mathcal{O})$,
\begin{equation}
	\hat{Z}_3\,:\, \quad P=0\,,\quad Q=l_1(a_8x_1^5+a_9x_2^5)-l_2(a_6x_1x_2x_3x_4x_5+a_7x_1^5)\,.	
\end{equation}
From these constraints we construct the holomorphic three-form $\hat{\Omega}$ as the residue \eqref{eq:ResZhat} from which we read off the GKZ-system for $\hat{Z}_3$ as
\begin{eqnarray}\label{eq:GKZquinticopen_b2}
 	&\mathcal{Z}_0=\sum_{i=0}^5\vartheta_i+1\,,\qquad \mathcal{Z}_1=\sum_{i=6}^9\vartheta_i\,,\qquad \mathcal{Z}_2=\vartheta_2-\vartheta_1-\vartheta_7-\vartheta_8+\vartheta_9\,,\nonumber\\ &\mathcal{Z}_i=\vartheta_i-\vartheta_1-\vartheta_7-\vartheta_8\,,\,\, i=3,4,5\,,\qquad \mathcal{Z}_6=\vartheta_8+\vartheta_9-\vartheta_6-\vartheta_7\,,\displaystyle&\nonumber\\
	&\displaystyle\mathcal{L}_{1}=\prod_{i=1}^5\frac{\partial}{\partial a_i}-\frac{\partial^5}{\partial a_0^5}\,,\qquad\mathcal{L}_{2}=\frac{\partial^2}{\partial a_1\partial a_6}-\frac{\partial^2}{\partial a_0\partial a_7}\,,\qquad \mathcal{L}_{3}=\frac{\partial^2}{\partial a_2\partial a_8}-\frac{\partial^2}{\partial a_1\partial a_9}\,,\displaystyle
\end{eqnarray} 
for the logarithmic derivative $\vartheta_i=a_i\frac{\partial}{\partial a_i}$.
Again there are two second order differential operators $\mathcal{L}_2$, $\mathcal{L}_3$ that include the curve moduli $a_i$, $i=6,7,8,9$, and one fifth order operator $\mathcal{L}_1$ which is lifted from the quintic Calabi-Yau to the blow-up. There are no further operators of minimal degree.
We obtain the GKZ-system \eqref{eq:GKZquinticopen_b2} from the following toric data of $\hat{Z}_3$
\begin{equation}\label{blowupPolyquintic_b2}
	\begin{pmatrix}[c|ccccccc|ccc|l]
	    		     &  &   &   & \Delta_7^{\hat Z}&&&   		  &\hat{\ell}^{(1)} &\hat{\ell}^{(2)} &\hat{\ell}^{(3)}  &          \\ \hline
		\hat{v}_0    &1 & 0 & 0 & 0 & 0 & 0 & 0   &-3   & -1  		  &	0	&	 \hat{y}_0 = x_1x_2x_3x_4x_5 \\
		\hat{v}_1    &1 & 0 &-1 &-1 &-1 &-1 & 0	  & 0   & 1  		  & -1	&	 \hat{y}_1 = x_1^5           \\
		\hat{v}_2    &1 & 0 & 1 & 0 & 0 & 0 & 0	  & 0   & 0  		  & 1	&	 \hat{y}_2 = x_2^5           \\
		\hat{v}_3    &1 & 0 & 0 & 1 & 0 & 0 & 0	  & 1   & 0  		  & 0	&	 \hat{y}_3 = x_3^5           \\
		\hat{v}_4    &1 & 0 & 0 & 0 & 1 & 0 & 0   & 1   &0  		  & 0	&  \hat{y}_4 = x_4^5            \\
	  \hat{v}_5    &1 & 0 & 0 & 0 & 0 & 1 & 0	  & 1   & 0  		  & 0	&	 \hat{y}_5 = x_5^5            \\
		\hat{v}_6    &0 & 1 & 0 & 1 & 0 & 0 &-1	  & -2   &1  		  & 0	&	 \hat{y}_6 =l_1 \hat{y}_0        \\
		\hat{v}_7    &0 & 1 & 0 & 0 & 1 & 0 &-1	  & 2   & -1  		  & 0	&	 \hat{y}_7 =l_1 \hat{y}_1        \\
		\hat{v}_8    &0 & 1 & 0 & 1 & 0 & 0 & 1	  & -1   & 0  		  & 1	&	 \hat{y}_8 =l_2 \hat{y}_1         \\
		\hat{v}_9    &0 & 1 & 0 & 0 & 0 & 1 & 1	  & 1   & 0  		  & -1	&  \hat{y}_9 =l_2 \hat{y}_2                   
	\end{pmatrix}.                                                                        
\end{equation}
We note that the second and third charge vector realize the brane charge vectors \eqref{eq:toricquintic_b2} and the closed string charge vector of the quintic is embedded as $\ell^{(1)}=\hat{\ell}^{(1)}+2\hat{\ell}^{(2)}+\hat{\ell}^{(3)}$. Here the generators of the Mori cone $\hat{\ell}^{(a)}$ are obtained as a triangulation of the polyhedron $\Delta_7^{\hat{Z}}$.

The GKZ-system \eqref{eq:GKZquinticopen_b2} defines three local coordinates $\hat{z}^a$ on the complex structure moduli space of $\hat{Z}_3$, that are chosen according to the basis of charge vectors in \eqref{blowupPolyquintic_b2},
\begin{equation} \label{eq:zOpenquintic_b2}
 	\hat{z}^1=-\frac{a_3a_4a_5a_7^2a_9}{a_0^3a_6^2a_8}\,,\quad \hat{z}^2=\frac{a_1a_6}{a_0a_7}\,,\quad \hat{z}^3=\frac{a_2a_8}{a_1a_9}\,.
\end{equation}
The complete system of differential operators $\mathcal{D}_a$ constituting the Picard-Fuchs system are found by linear combinations of the charge vectors $\hat{\ell}^{(a)}$ in \eqref{blowupPolyquintic_b2}. They are obtained by factorizing the corresponding operators $\mathcal{L}_a$, that are directly associated to the scaling symmetries of the charge vectors. We obtain the system
\begin{eqnarray}
\mathcal{D}_1&=&\theta _1^3 \left(2 \theta _1-\theta _2\right) \left(\theta _1-\theta _3\right)+\left(2 \theta _1-\theta _2-2\right) \left(\theta _1-\theta_3-1\right)\prod_{i=0}^{i=2} \left(3 \theta _1+\theta _2-i\right) \hat{z}^1\,,\nonumber\\
\mathcal{D}_2&=&\left(2 \theta _1-\theta _2\right) \left(\theta _2-\theta _3\right)+\left(2 \theta _1-\theta _2+1\right) \left(3 \theta _1+\theta _2\right) \hat{z}^2\,,\nonumber\\
\mathcal{D}_3&=&\left(\theta _1-\theta _3\right) \theta _3-\left(\theta _1-\theta _3+1\right) \left(-\theta _2+\theta _3-1\right) \hat{z}^3\,,\nonumber\\
\mathcal{D}_4&=&-\theta _1^3 \left(2 \theta _1-\theta _2\right) \theta _3-\left(2 \theta _1-\theta _2-1\right) \prod_{i=0}^3\left(3 \theta _1+\theta _2-i\right) \hat{z}^1\hat{z}^2\hat{z}^3\,,\nonumber\\
\mathcal{D}_5&=&\theta _1^3 \left(\theta _1-\theta _3\right) \left(-\theta _2+\theta _3\right) \left(1-\theta _2+\theta _3\right)+\left(-1+\theta _1-\theta _3\right)\prod_{i=0}^4\left(3 \theta _1+\theta _2-i\right) \hat{z}^1(\hat{z}^2)^2\,,\nonumber\\
\end{eqnarray}
\begin{eqnarray}
\mathcal{D}_6&=&\theta _1^3 \left(-\theta _2+\theta _3\right) \theta _3+\prod_{i=0}^4\left(3 \theta _1+\theta _2-i\right) \hat{z}^1(\hat{z}^2)^2\hat{z}^3\,,
\end{eqnarray}
where the corresponding linear combination of the $\hat{\ell}^{(a)}$ can be read off from the powers of the $\hat{z}^a$. We note that this system has the structure advertised in eq. \eqref{eq:LInZHat} and thus the periods $\Pi^k(z^1)$ of the quintic $Z_3$ with $z^1=\hat{z}^1(\hat{z}^2)^2\hat{z}^3$ are solutions to it.

Indeed, we identify 12 solutions of the following form at $\hat{z}^i\rightarrow 0$. There is one solution $X^{(0)}$ with a power series expansion, three single logarithmic solutions $X^{(1)}_i$, four double logarithmic solutions $X^{(2)}_{\alpha}$ and four triple logarithmic solutions $X^{(3)}_{\beta}$. The unique power series solution starts with a constant, that we normalize to $1$,
\begin{equation}
	X^{(0)}=1+120 z^1+113400 (z^1)^2+168168000 (z^1)^3+305540235000 (z^1)^4+\mathcal{O}((z^1)^5)\,,
\end{equation}
where we set $z^1=\hat{z}^1 (\hat{z}^2)^2 \hat{z}^3$. Thus, we identify this as the fundamental periode $\Pi^0(z^1)$ of the quintic. We recover the three other quintic periods by first noting that the leading logarithms of the solutions are given by
\begin{eqnarray} \label{eq:leadinglogsquintic_b2}
		X^{(1)}_i\,:\,& \hat{l}_1\,,\,\,\hat{l}_2\,,\,\,\hat{l}_3\,,&\\[0.3Em]
		X^{(2)}_{\alpha}\,:\,& \frac{1}{2}\hat{l}_1^2\,,\,\, \hat{l}_2(\hat{l}_1-2 \hat{l}_3)\,,\,\,\hat{l}_3(\hat{l}_1 +2\hat{l}_2 +\frac{1}{2}\hat{l}_3)\,,\,\,\hat{l}_2(\frac{1}{2}\hat{l}_2+\hat{l}_3)&\nonumber\\[0.3Em]
		X^{(3)}_{\beta}\,:\,& \frac16 \hat{l}_1^3\,,\,\,\frac12 \hat{l}_1^2 \hat{l}_2-\frac13 \hat{l}_2^3-2 \hat{l}_1 \hat{l}_2 \hat{l}_3-3 \hat{l}_2^2 \hat{l}_3- \hat{l}_2 \hat{l}_3^2\,,& \nonumber\\
		&\,\,\frac12 \hat{l}_1^2 \hat{l}_3+2 \hat{l}_1 \hat{l}_2 \hat{l}_3+2 \hat{l}_2^2 \hat{l}_3+\frac12 \hat{l}_1 \hat{l}_3^2+ \hat{l}_2 \hat{l}_3^2+\frac16 \hat{l}_3^3\,,\,\,\frac12 \hat{l}_1 \hat{l}_2^2+\frac12 \hat{l}_2^3+\hat{l}_1 \hat{l}_2 \hat{l}_3+\frac32 \hat{l}_2^2 \hat{l}_3+\frac12 \hat{l}_2 \hat{l}_3^2\,,&\nonumber
\end{eqnarray}
where we used the abbreviation $\log(\hat{z}^i)=\hat{l}_i$. We immediately observe that all quintic periods $\Pi^k(z^1)$ with leading logarithms $l_1$, $\frac12 l_1^2$ and $\frac16l_1^3$ for $l_1=\hat{l}_1+2\hat{l}_2+\hat{l}_3$ are indeed contained in the leading logarithms \eqref{eq:leadinglogsquintic_b2} of the solutions on $\hat{Z}_3$. We readily check that the complete $z^1$-series of the quintic periods $\Pi^k(z^1)$ are reproduced as well on the blow-up.

The remaining six logarithmic solutions are related to the open string sector. In particular, we can cross-check this statement by finding the brane superpotential $W_{\rm brane}$ by its A-model interpretation at large volume as a generating functional for disk instantons. First we interpret the single logarithms in \eqref{eq:leadinglogsquintic_b2} as the mirror map of the open-closed system at $z\rightarrow 0$ defining the flat coordinates via $\hat{t}_i=X^{(1)}_i/X^{(0)}$,
\begin{eqnarray}
		X^{(1)}_1&=& X^{(0)}\log(\hat{z}_1)+2 \hat{z}_2-\hat{z}_2^2-60 \hat{z}_1 \hat{z}_2^2+\frac{2\hat{z}_2^3 }{3}-\frac{\hat{z}_2^4}{2}+\frac{2\hat{z}_2^5}{5}-48 \hat{z}_1 \hat{z}_2 \hat{z}_3+462 z_1+\mathcal{O}(\underline{\hat{z}}^6)\,,
		\nonumber\\
		X^{(1)}_2&=& X^{(0)}\log(\hat{z}_2)-\hat{z}_2+\frac{\hat{z}_2^2}{2}-\frac{\hat{z}_2^3}{3}+\frac{\hat{z}_2^4}{4}-\frac{\hat{z}_2^5}{5}+24 \hat{z}_1 \hat{z}_2 \hat{z}_3+154 z_1-360 \hat{z}_1 \hat{z}_2^3 \hat{z}_3+\mathcal{O}(\underline{\hat{z}}^6)\,,\nonumber\\
		X^{(1)}_3&=& X^{(0)}\log(\hat{z}_3)+60 \hat{z}_1 \hat{z}_2^2-9450 \hat{z}_1^2 \hat{z}_2^4+75600 \hat{z}_1^2 \hat{z}_2^4 \hat{z}_3-60 \hat{z}_1 \hat{z}_2^2 \hat{z}_3^2+\mathcal{O}(\underline{\hat{z}}^8)\,.
\end{eqnarray}
Here we omit a factor of $\frac{1}{2\pi i}$ in front of the logarithms for brevity\footnote{We also label the variables $\hat{z}_i$ by a subscript instead of a superscript in order to shorten the expressions.}.
This is perfectly consistent with the mirror map of the quintic that is obtained as $t=\hat{t}_1+2\hat{t}_2+\hat{t}_3$ or as $\Pi^{1}(z_1)=X^{(1)}_1+2X^{(1)}_2+X^{(1)}_3=X^{(0)}\log(z_1)+770 z_1+\ldots$ as required by the charge vectors $\hat{\ell}^{(a)}$ in \eqref{blowupPolyquintic_b2}. Upon inversion of the mirror map, we obtain the $\hat{z}^i$ as a series of $q_a=e^{2\pi i \hat{t}_a}$, that we readily insert into the double logarithmic solutions in \eqref{eq:leadinglogsquintic_b2}. 
 Then we construct a linear combination of the double logarithmic solutions in \eqref{eq:leadinglogsquintic_b2} as 
\begin{equation}
 	W_{\rm brane}=(2X^{(2)}_1+4X^{(2)}_2+a X^{(2)}_3+4X^{(2)}_4)/X^{(0)}
\end{equation}
in which we insert the inverse mirror map to obtain 
\begin{equation}
	W_{\rm brane}=2 t^2+2 \hat{t}_2^2+\frac{1}{2} (4-a) \hat{t}_3^2-t\hat{t}_2-(4-a) t\hat{t}_3-\frac{1}{4\pi}\sum_{n_i}n_{d_1,d_2,d_3}\text{Li}_2(q_1^{d_1} q_2^{d_2}q_2^{d_3})\,,
\end{equation}
where $a$ denotes a free complex parameter.
This has the expected integrality properties of the Ooguri-Vafa Li$_2$-double cover formula, such that we obtain the disk instantons $n_{d_1,d_2,d_3}$. Selected invariants $n_{j,i+j,j}$ are summarized in table \ref{tab:instantonsLVquintic_b2}, where the rows and columns are labelled by $i$ and $j$, respectively. 
\begin{table}[!ht]
\centering
$
\scriptstyle
 \begin{array}{|c|rrrrrr|}
\hline
\rule[-0.2cm]{0cm}{0.6cm}  i&j=0&j=1&j=2&j=3&j=4&j=5\\
\hline
 0 & 0 & -320 & 13280 & -1088960 & 119783040 & -15440622400 \\
 1 & 20 & 1600 & -116560 & 12805120 & -1766329640 & 274446919680 \\
 2 & 0 & 2040 & 679600 & -85115360 & 13829775520 & -2525156504560 \\
 3 & 0 & -1460 & 1064180 & 530848000 & -83363259240 & 16655092486480 \\
 4 & 0 & 520 & -1497840 & 887761280 & 541074408000 & -95968626498800 \\
 5 & 0 & -80 & 1561100 & -1582620980 & 931836819440 & 639660032468000 \\
 6 & 0 & 0 & -1152600 & 2396807000 & -1864913831600 & 1118938442641400 \\
 7 & 0 & 0 & 580500 & -2923203580 & 3412016521660 & -2393966418927980 \\
 8 & 0 & 0 & -190760 & 2799233200 & -5381605498560 & 4899971282565360 \\
 9 & 0 & 0 & 37180 & -2078012020 & 7127102031000 & -9026682030832180 \\
 10& 0 & 0 & -3280 & 1179935280 & -7837064629760 & 14557931269209000 \\
 11& 0 & 0 & 0 & -502743680 & 7104809591780 & -20307910970428360 \\
 12& 0 & 0 & 0 & 155860160 & -5277064316000 & 24340277955510560 \\
 13& 0 & 0 & 0 & -33298600 & 3187587322380 & -24957649473175420 \\
 14& 0 & 0 & 0 & 4400680 & -1549998228000 & 21814546476229120 \\
 15& 0 & 0 & 0 & -272240 & 597782974040 & -16191876966658500 \\
 16& 0 & 0 & 0 & 0 & -178806134240 & 10157784412551120 \\
 17& 0 & 0 & 0 & 0 & 40049955420 & -5351974901676280 \\
 18& 0 & 0 & 0 & 0 & -6332490480 & 2348019778753280\\
\hline
\end{array}
$
\caption{Disk instanton invariants $n_{j, i+j, j}$ on the quintic at large volume. These results agree with \cite{Alim:2009bx}.}
\label{tab:instantonsLVquintic_b2}
\end{table}
We note that the parameter $a$ does not affect these instantons, however, it does affect the classical terms\footnote{The classical term $4t^2-2t_2^2$ of \cite{Alim:2009bx} can not be reproduced by tuning the parameter $a$. The ``closest'' match is $2 (t-\hat{t}_2)^2$ for $a=4$, for which the only non-vanishing disk instantons are those in table \ref{tab:instantonsLVquintic_b2}.}. It should be fixed by the determination of the symplectic basis on the blow-up $\hat{Z}_3$.

\subsection{Open-Closed GKZ-Systems from Blow-Up Threefolds}
\label{generaltoricstructure}

In the following section we present a general recipe to easily obtain the toric GKZ-system of an arbitrary toric brane $\Sigma$ in an arbitrary toric Calabi-Yau hypersurface $Z_3$. 

Motivated by the above example it is possible in a simple manner to construct the toric data $\Delta_7^{\hat{Z}}$ right from the original polyhedron $\Delta^{\tilde{Z}}_3$ and the toric curve $\Sigma$ as specified by the charge vectors $\hat{\ell}^{(1)}$, $\hat{\ell}^{(2)}$. We denote the vertices of $\Delta_3^{\tilde{Z}}$ by $\tilde{v}_i$, $i=1,\ldots, n$, with $\tilde{v}_0$ the origin, its charge vectors by $\ell^{(i)}$, $i=1,\ldots,n-4$ and the two brane vectors by $\hat{\ell}^{(1)}$, $\hat{\ell}^{(2)}$ as before. We define $n+5$ points $\hat{v}_i$ spanning a seven-dimensional polyhedron $\Delta_7^{\hat{Z}}$ as
\begin{eqnarray}\label{eq:genConstruction}
 	Z_3:& \quad\,\ \,\,\hat{v}_i=&(1,0,\tilde{v}_i,0)\,,\qquad\quad i=0,\ldots,n\,,\nonumber\\ \hat{\ell}^{(1)}:&\quad \hat{v}_{n+1}=&(0,1,v^{(-)}_1,-1)\,,\,\,\,\hat{v}_{n+2}=(0,1,v^{(+)}_1,-1)\,,\nonumber\\
	\hat{\ell}^{(2)}:&\quad\hat{v}_{n+3}=&(0,1,v^{(-)}_2,1)\,,\,\,\,\hat{v}_{n+4}=(0,1,v^{(+)}_2,1)\,,
\end{eqnarray}
where we use the abbreviations
\begin{equation}
 	v^{(+)}_1=\sum_{\hat{\ell}^{(1)}_i>0}\hat{\ell}^{(1)}_i\tilde{v}_i\,,\,\, v^{(-)}_1=-\sum_{\hat{\ell}^{(1)}_i<0}\hat{\ell}^{(1)}_i\tilde{v}_i\,,\,\,
	v^{(+)}_2=\sum_{\hat{\ell}^{(2)}_i>0}\hat{\ell}^{(2)}_i\tilde{v}_i\,,\,\,\quad v^{(-)}_2=-\sum_{\hat{\ell}^{(2)}_i<0}\hat{\ell}^{(2)}_i\tilde{v}_i\,.
\end{equation}
The first line of \eqref{eq:genConstruction} simply embeds the original toric data associated to $Z_3$ into $\hat{Z}_3$, whereas the second and third line translate the brane data 
into geometric data of $\hat{Z}_3$. The structure of the points $\hat{v}_i$ is quite generic for the description of a toric complete intersection, cf.~\cite{Hosono:1994ax} for the Calabi-Yau case.

In our context this structure in addition reflects the distinction between the closed and open string sector. It is encoded by the two canonical hyperplanes in the first and second row of the $\hat{v}_i$. Points in the hyperplane $H_1=\{(1,0,w_1,w_2,w_3,w_4,w_5)\}$ correspond to the closed string sector, i.e.~the geometry of the Calabi-Yau encoded in the constraint $P$, whereas points in the hyperplane $H_2=\{(0,1,w_1,w_2,w_3,w_4,w_5)\}$ contribute to the open string sector as encoded in the constraints $h_1$, $h_2$ of $\Sigma$. On the blow-up $\hat{Z}_3$, that we construct as a complete intersection \eqref{eq:blowup}, this translates to the rule, which monomial $y_i(\underline{x})$ contributes to which of the constraints $P$, $Q$. Points $\hat{v}_i$ in $H_1$ contribute to $P$, whereas those in $H_2$ contribute to $Q$. We summarize this in the following table
\begin{equation}\label{blowupDelta7}
	\begin{pmatrix}[c|cccc||ccc|cc|cc]
	    		        &        &       &   \Delta_7^{\hat Z}     &   & \hat{\ell}^{(1)}&\ldots&\hat{\ell}^{(n-4)}&\hat{\ell}^{(n-3)}&\hat{\ell}^{(n-2)} & &\text{monomials}\\ \hline
		\hat{v}_0       & 1      & 0     & \tilde{v}_0    & 0      & \vert    & \ldots & \vert        & \vert           &\vert           & &\hat{y}_0=y_0 \\
	        \ldots          & \vdots & \vdots& \vdots & \vdots & \ell^{(1)}& \ldots & \ell^{(n-4)}& \hat{\ell}^{(1)}&\hat{\ell}^{(2)}& P:&\ldots \\
		\hat{v}_n       & 1      & 0     & \tilde{v}_n    &0       & \vert     & \ldots & \vert       & \vert           &\vert           & &\hat{y}_{n-4}=y_{n-4}\\ \hline
		\hat{v}_{n+1}   &0 & 1   & v^{(-)}_1  & -1         & 0         & \ldots & 0           & 1               & 0              & &\hat{y}_{n-3}=l_1\prod_{\hat{\ell}^{(1)}_i<0}y_i^{-\hat{\ell}^{(1)}_i}\\
		\hat{v}_{n+2}   &0 & 1   & v^{(+)}_1  & -1	   & 0         & \ldots & 0           &-1               & 0              & Q:&\hat{y}_{n-2}=l_1\prod_{\hat{\ell}^{(1)}_i>0}y_i^{\hat{\ell}^{(1)}_i}    \\
		\hat{v}_{n+3}   &0 & 1   & v^{(-)}_2  & 1          & 0         & \ldots & 0           & 0               & 1              & &\hat{y}_{n-1}=l_2\prod_{\hat{\ell}^{(1)}_i<0}y_i^{-\hat{\ell}^{(2)}_i}      \\
		\hat{v}_{n+4}   &0 & 1   & v^{(+)}_2  & 1          & 0         & \ldots & 0           & 0               & -1             & &\hat{y}_{n}=l_2\prod_{\hat{\ell}^{(1)}_i>0}y_i^{\hat{\ell}^{(2)}_i}     \\
	\end{pmatrix}.                                                                        
\end{equation}
Here we displayed besides the points $\hat{v}_i$ of \eqref{eq:genConstruction} also a natural choice of basis $\hat{\ell}^{(a)}$ of the lattice of relations of $\Delta_7^{\hat{Z}}$. In this basis the first $n-4$ charge vectors are identical to those of $\Delta_4^{\tilde{Z}}$ up to the extension by four further entries $0$. More importantly the last two charge vectors naturally contain the two original brane vectors $\hat{\ell}^{(a)}$ extended by four further entries with $\pm1$, $0$. As before their entries sum up to zero. In the last row we associated monomials $\hat{y}_i$ to the points $\hat{v}_i$ where the $y_i(x_j)$ merely denote the polynomials on the original geometry of $\Delta_4^{\tilde{Z}}$ computed by the Batyrev formula \eqref{eq:Batyrev}. The coordinates $l_1$, $l_2$ denote the homogenous coordinates on $\P^1$. We note that the form of the polynomials $\hat{y}_i$ associated to the four new points $\hat{v}_{n},\ldots,\hat{v}_{n+4}$ reflects the definition of the brane constraints $h_1$, $h_2$ defined via \eqref{eq:BBrane}. As mentioned before and indicated in \eqref{blowupDelta7} the constraints $P$ and $Q$ are given by
\beq \label{eq:blowupConstrains}
	\hat{Z}_3\,:\,\quad P=\sum_{i=0}^{n-4}a_i\hat{y_i}\equiv \sum_{i=0}^{n-4}a_iy_i\,,\qquad Q=\sum_{i=n-3}^n a_i\hat{y}_i\,,
\eeq 
where the $\underline{a}$ denote free complex parameters.
As can be easily checked the general toric data in \eqref{blowupDelta7} immediately reproduce the toric data of the blow-up \eqref{blowupPolyquintic} of the curve $\Sigma$ in the quintic $Z_3$. Similarly we obtain the toric data \eqref{blowupPolyCand} of our second example in section \ref{ToricBraneBlowupII}

To the general form of the toric data \eqref{blowupDelta7} of $\hat{Z}_3$ we associate a GKZ-system on the complex structure moduli space of $\hat{Z}_3$ by the standard formulae
\begin{eqnarray} \label{eq:GKZoopen}
 	\mathcal{L}_{a}&=&\prod_{\hat{\ell}^{(a)}_i>0}\left(\frac{\partial}{\partial a_i}\right)^{\hat{\ell}^{(a)}_i}-\prod_{\hat{\ell}^{(a)}_i<0}\left(\frac{\partial}{\partial a_i}\right)^{-\hat{\ell}^{(a)}_i}\,,\qquad a=1,\ldots n-2\,,\\
	\mathcal{Z}_{j}&=&\sum_i(\hat{v}_i)^j\vartheta_i-\beta_j\,,\qquad j=1,\ldots,7\,.\nonumber
\end{eqnarray}
This immediately yields a natural choice of complex coordinates given by
\begin{equation}
	\hat{z}^a=(-)^{\hat{\ell}^{(a)}_0}\prod_{i=0}^{n+4} a_i^{\hat{\ell}^{(a)}_i}\,,\quad a=1,\ldots n-2.
\end{equation} 
We readily apply these formulae to reproduce the GKZ-system \eqref{eq:GKZquinticopen} and the coordinates \eqref{eq:zOpenquintic} for the choice of charge vectors $\hat{\ell}^{(a)}$ as in \eqref{blowupPolyquintic}. The same applies for the GKZ-system in the second example discussed in section \ref{ToricBraneBlowupII}.

Let us conclude by mentioning some remarkable properties of the geometry of $\hat{Z}_3$ as encoded by $\Delta^{\hat Z}_7$. First, the last row in \eqref{blowupDelta7} is associated to the toric symmetries of the exceptional $\P^1$ in the blow-up divisor $E$. In fact, this $\P^1$ can be made directly visible in $\Delta_7^{\hat{Z}}$ by projection on the ray $(0,0,0,0,0,0,w_1)$.
Second, one might be tempted to map the toric data $\Delta_7^{\hat{Z}}$ of the complete intersection \eqref{eq:blowupConstrains} to toric data of a hypersurface defined by six-dimensional vectors obtained by adding the first and second row, $((\hat{v}_i)^1,(\hat{v}_i)^2,\ldots)\,\mapsto\,((\hat{v}_i)^1+(\hat{v}_i)^2,\ldots)\equiv (1,v_i')$, where we note that $(\hat{v}_i)^1+(\hat{v}_i)^2=1$. This defines a five-dimensional polyhedron $\Delta_5$ with points $v_i'$. Clearly, this polyhedron has one further charge vector $\hat{\ell}^{(n-1)}$ so that the dimension of a corresponding toric variety is five-dimensional. Furthermore, for special choices of the brane charge vectors $\hat{\ell}^{(a)}$, but by far not for all choices\footnote{The polyhedron $\Delta_5$ defined by the $v_l'$ is not generically reflexive. This is the case in section \ref{toricbranesP11111-W-LV}.}, this toric data defines a mirror pair of \textit{compact} Calabi-Yau fourfolds $\tilde{X}_4$, $X_4$ with the hypersurface constraint given by the standard Batyrev formalism \eqref{eq:Batyrev}. In combination with the first observation about the universal presence of the $\P^1$ in $\Delta_7^{\hat{Z}}$, the geometry of $\tilde{X}_4$ will contain this very $\P^1$ as the basis of a Calabi-Yau threefold fibration with generic fiber $\tilde{Z}_3$. This is precisely the geometric structure we encountered in the Calabi-Yau geometries used in \cite{Grimm:2009ef} and in the context of heterotic/F-theory duality in \cite{Grimm:2009sy}. In fact, the Calabi-Yau fourfold $X_4$ we obtain from $\Delta_7^{\hat{Z}}$ for the example of the next section \ref{ToricBraneBlowupII} precisely agrees with the F-theory dual fourfold of a heterotic setup with horizontal five-branes as predicted by heterotic/F-theory. We suspect that heterotic/F-theory duality for horizontal five-branes is in general the reason for the occurrence of a Calabi-Yau fourfold geometry associated to some blow-up threefolds $\hat{Z}_3$. This nicely completes the discussion of the application of the blow-up proposal for heterotic five-branes and heterotic/F-theory duality in \cite{Grimm:2009sy}. Let us conclude by emphasizing that only parts of the fourfold
geometry $X_4$, if present, are intrinsically related to the original five-brane as e.g.~signalled by the additional charge vector $\hat{\ell}^{(n-1)}$.\footnote{On the heterotic side, the additional data of $X_4$ is related to the heterotic bundle.} However, the toric data $\Delta_7^{\hat{Z}}$ of the blow-up $\hat{Z}_3$ should by construction always carry the minimal amount
of information in order to study the open-closed system of the five-brane in $Z_3$.

A technical similar but differently motivated method to obtain toric data and a GKZ-system governing the deformations of toric D5-branes was presented in \cite{Mayr:2001xk,Lerche:2001cw,Jockers,Alim:2009rf,Alim:2009bx} and formulated mathematically rigorously in \cite{Li:2009dz}.

\subsection{Open-Closed Picard-Fuchs Systems: Branes on $Z_3(1,1,1,6,9)$}
\label{ToricBraneBlowupII}

As a second demonstration of the application of the blow-up proposal we consider a two-parameter Calabi-Yau threefold $Z_3$. The discussion will be similar to the case of the quintic, thus, we keep it as brief as possible.

\subsubsection{Branes on Lines in $Z_3(1,1,1,6,9)$ and the Blow-Up}
\label{toricbranesP11169}

The Calabi-Yau threefold $Z_3(1,1,1,6,9)$\footnote{We will abbreviate this simply by $Z_3$.} is defined as the mirror of the Calabi-Yau hypersurface $\tilde Z_3$ in $\mathds{P}^4(1,1,1,6,9)$ which admits $h^{2,1}(Z_3)=2$ complex structure moduli and is an elliptic fibration over $\P^2$. In the conventions of \cite{Candelas:1994hw} the two complex structures denoted $\Psi_1$, $\Psi_2$ enter the constraint $P$ as
\begin{equation}
	P=y^2+x^3+u_1^{18}+u_2^{18}+u_3^{18}-3\Psi_1 (u_1u_2u_3)^6-18\Psi_2 xyu_1u_2u_3\,,
\label{eq:Cand2param}
\end{equation}
where we introduce the homogeneous coordinates $(u_i,x,y,z)$, $i=1,2,3$, for the $\P^2$-base and the elliptic fiber $\P^{2}(1,2,3)$, respectively. Note however that we are working in an affine patch $z=1$ of the elliptic fiber\footnote{Strictly speaking one has to include the divisor $z$ to resolve a curve of $\mathds{Z}_3$-singularity in $\P^4(1,1,1,6,9)$ \cite{Candelas:1994hw}.}. 
This is reflected in the toric data used to obtain $P$,
\begin{equation} \label{eq:P11169TD}
 	\begin{array}{cclllll}
 	 	 \ell^{(1)}=(\phantom{-}0, & -3,\phantom{i} & 1,& 1,& 1, &0 &0) \\ 
		\ell^{(2)}=(-6, & 1, & 0,& 0,& 0, &2 &3) \\\hline \hline
		y_0 & y_1 & y_2 & y_3 & y_4 & y_5& y_6\\ 
		zxyu_1u_2u_3 &(zu_1u_2u_3)^6 & z^6u_1^{18} & z^6u_2^{18}  & z^6u_3^{18} & x^3 & y^2 
 	\end{array}
\end{equation}
where the $y_i$ corresponding to the entries $\ell^{(j)}_i$ of the charge vectors are monomials in the homogeneous coordinates on $\mathds{P}^4(1,1,1,6,9)$.
This hypersurface data is augmented by the action of a discrete orbifold group $G$ which is $G=\mathds{Z}_{6}\times\mathds{Z}_{18}$ generated by $v^{(i)}=(0,1,3,2,0) \text{ mod } 6$, $v'^{(j)}=(1,-1,0,0,0)\text{ mod } 18$ acting on the coordinates as 
\begin{equation}
	g^{(i)}:\,\,x_k\mapsto e^{2\pi i v^{(i)}_k/6} x_k\,,\quad g'^{(j)}:\,\,x_k\mapsto e^{2\pi i v'^{(j)}_k/18} x_k\,.
\label{eq:GCand}
\end{equation}

To this setup we add five-brane wrapping a rational curve $\P^1$ in $Z_3$ that will be of similar type as the lines \eqref{eq:paramlines} considered in the quintic. As before we use the moduli space $\mathcal{M}(\Sigma)$ as a model for the deformation space $\tilde{P}^1$ of this line, where we specify the analytic family of curves $\Sigma$ in the form of toric branes 
\begin{eqnarray} \label{eq:constlinesCand}
	\Sigma&:& P=0\,,\quad h_1\equiv \beta^{12}(u_1u_2u_3)^6-\alpha^6\gamma^6 u_2^{18}=0\,,\quad h_2\equiv\gamma^{12}(u_1u_2u_3)^6-\alpha^6\beta^6u_3^{18}=0\,,\nonumber\\
	&&\hat{\ell}^{(1)}=(0,-1,0,1,0,0,0)\,,\quad\quad\hat{\ell}^{(2)}=(0,-1,0,0,1,0,0)\,.
\end{eqnarray}
The brane charge vectors $\hat{\ell}^{(i)}$ are used to construct the constraint $h_i$ via \eqref{eq:BBrane}.
This basis of constraints and parameters might look inconvenient, is however justified by noting the convenient, equivalent form 
\begin{equation}
	\Sigma\,:\quad P=0\,,\quad\gamma^{18}u_2^{18}-\beta^{18}u_3^{18}=0\,,\quad \gamma^{18}u_1^{18}-\alpha^{18}u_3^{18}=0\,
\label{eq:constlinesCandNF}
\end{equation}
upon a trivial algebraic manipulation.  
We introduce affine coordinates parameterizing this analytic family of curves in $Z_3$ that we choose to be $u^1=\frac{\beta^{18}}{\gamma^{18}}$ and $u^2=\frac{\alpha^{18}}{\gamma^{18}}$.
 
Next we proceed by linearizing \eqref{eq:constlinesCandNF} to describe a non-holomorphic deformation $\tilde{P}^1$ of rational curves given by
\begin{equation} \label{eq:anholoCCand}
	\tilde{\P}^1:\quad \eta_1 x+\sqrt[3]{y^2+ m(u_3,x,y)}=0\,,\quad  \eta_2\gamma u_2-\beta u_3=0\,,\quad \eta_3\gamma u_1-\alpha u_3=0\,.
\end{equation} 
Here we inserted $h_1$ and $h_2$ into $P$ and introduced the third and eighteenth roots of unity $\eta_1$, respectively $\eta_2$, $\eta_3$ as well as the polynomial
\begin{equation}
 	m(u_3,x,y)=\frac{\alpha^{18}+\beta^{18}+\gamma^{18}}{\gamma^{18}}-3\Psi_1\left(\frac{\alpha\beta}{\gamma^2}\right)^6u_3^{18}-18\Psi_2\frac{\beta\alpha}{\gamma^2}xyu_3^{18}\,.
\label{eq:CandDiv}
\end{equation}
At the critical locus of the parameter space $\alpha$, $\beta$, $\gamma$ where the polynomial $m$ vanishes identically, the generically higher genus Riemann surface $\Sigma$ degenerates. This locus reads
\begin{equation}
	\mathcal{M}_{\P^1}(\Sigma)\,:\quad\alpha^{18}+\beta^{18}+\gamma^{18}-3\Psi_1\alpha^6\beta^6\gamma^6=0\,,\quad \Psi_2\alpha\beta\gamma=0\,.
\label{eq:modulicondCand}
\end{equation}
At this locus the Riemann surface $\Sigma$ in \eqref{eq:constlinesCandNF} degenerates to
\begin{equation}
	\Sigma\,:\quad h_0\equiv y^2+x^3\,,\quad\gamma^{18}u_2^{18}-\beta^{18}u_3^{18}=0\,,\quad \gamma^{18}u_1^{18}-\alpha^{18}u_3^{18}=0\,.
\end{equation}
Modulo the action of $G$ identifying the different solutions in \eqref{eq:anholoCCand} we can solve \eqref{eq:anholoCCand} holomorphically and consistent with the weights of $\P^2(1,2,3)$ at the locus $\mathcal{M}_{\P^1}(\Sigma)$ by the Veronese embedding of a line in $\P^{4}(1,1,1,6,9)$,
\begin{equation}
	(U,V)\,\mapsto\,(\alpha U,\beta U,\gamma U,-{\eta_1}^2 V^6,V^9)\,,\quad \eta_1^3=1.
\label{eq:paramlinesCand}
\end{equation}
From the perspective of this line, the constraint \eqref{eq:modulicondCand} on the parameters $(\alpha,\beta,\gamma)$ is precisely the condition for it to lie holomorphically in the Calabi-Yau constraint $P$ of $\hat{Z}_3$.
This implies that at the point $\Psi_2=0$ there is an analytic family of lines in $Z_3$ and otherwise, for $\Psi_2\neq 0$, only a discrete number of lines.

To study the open-closed system defined by the five-brane in $Z_3$ we construct the blow-up threefold $\hat{Z}_3$. As explained above, cf.~section \ref{unificationofdeformationsII}, we use the holomorphic description by the toric curve $\Sigma$ of the anholomorphic brane deformations $\P^1$ in order to construct the blow-up. Before we construct $\hat{Z}_3$, we switch to a full toric description.

\subsubsection{Toric Branes on $Z_3(1,1,1,6,9)$: the GKZ-System}

We begin the analysis of the open-closed moduli space using the toric means. First of all let us recall the toric construction of the Calabi-Yau $Z_3$ by giving its constraint as well as the curve $\Sigma$,
\begin{eqnarray} \label{eq:constraintsCand}
	Z_3:& P=a_6y^2+a_5x^3+a_1u_1^6u_2^6u_3^6+a_2u_1^{18}+a_3u_2^{18}+a_4u_3^{18}+a_0xyu_1u_2u_3\,,\\
	\Sigma:& h_1=a_7u_1^6u_2^6u_3^6+a_8u_2^{18}\,,\quad h_2=a_9u_1^6u_2^6u_3^6+a_{10}u_3^{18}\,. \nonumber
\end{eqnarray}
The coefficients $\underline{a}$ redundantly parameterize the complex structure of $Z_3$ respectively the moduli of the curve $\Sigma$ in $Z_3$. The information in \eqref{eq:constraintsCand} is directly encoded in the toric data specifying $(Z_3,\Sigma)$ via the polyhedron $\Delta^{\tilde{Z}}_4$ and the two brane-vectors $\hat{\ell^{(1)}}$, $\hat\ell^{(2)}$ reading
\begin{equation}\label{3foldellp2}
	\begin{pmatrix}[c|cccc|cc|l||cc]
	    	&   &  \Delta_4^{\tilde Z} &   &    &  \ell^{(1)} & \ell^{(2)} &   &\hat{\ell}^{(1)}&\hat{\ell}^{(2)}\\ \hline
		\tilde{v}_0 & 0 & 0 & 0 & 0 	  &  0  &  -6 & y_0 = zxyu_1u_2u_3 & 0&0\\
		\tilde{v}_1 & 0 & 0 & 2 & 3 	  &  -3   &1  & y_1 = z^6 u_1^6 u_2^6 u_3^6 & -1&-1\\
		\tilde{v}_2 & 1 & 1 & 2 & 3 	&  1   & 0  & y_2 = z^6 u_1^{18} & 0&0\\
		\tilde{v}_3 &-1 & 0 & 2 & 3 	&  1   & 0  & y_3 = z^6 u_2^{18} & 1&0\\
		\tilde{v}_4 & 0 &-1 & 2 & 3 	&  1   & 0  & y_4 = z^6 u_3^{18} & 0&1\\
	        \tilde{v}_5 & 0 & 0 &-1 & 0 	&  0   & 2  & y_5 = x^3 & 0&0\\
		\tilde{v}_6 & 0 & 0 & 0 &-1 	&  0   & 3  & y_6 = y^2 & 0&0 
	\end{pmatrix}.
\end{equation}
The points of the dual polyhedron $\Delta_4^Z$ are given by $v_1=(-12,6,1,1)$, $v_2=(6,-12,1,1)$, $v_3=(6,6,1,1)$, $v_4=(0,0,-2,1)$ and $v_5=(0,0,1,-1)$, where the point $(0,0,1,1)$ corresponds to the $z$-coordinate on the elliptic fiber that we set to $1$.
The Calabi-Yau as well as the two brane constraints of \eqref{eq:constraintsCand} are then associated to this data via \eqref{eqn:HVmirror} and \eqref{eq:BBrane}, respectively.

Accordingly, the variational problem of complex structures on $Z_3$ can be studied by exploiting the existence of the GKZ-system \eqref{eq:pfo}, \eqref{eq:Zs} associated to $\Delta_4^{\tilde Z}$. For the example at hand it reads
\begin{eqnarray}\label{eq:GKZclosed}
 	&\mathcal{Z}_0=\sum_{i=0}^6\vartheta_i+1\,,\quad \mathcal{Z}_i=\vartheta_2-\vartheta_{i+2}\,\, (i=1,2)\quad \mathcal{Z}_3=2\displaystyle\sum_{i=1}^4\vartheta_i-\vartheta_5\,,\displaystyle\quad
	\displaystyle\mathcal{Z}_4=3\sum_{i=1}^4\vartheta_i-\vartheta_6\,,\displaystyle\nonumber\\
	\displaystyle&\displaystyle\mathcal{L}_{1}=\prod_{i=2}^4\frac{\partial}{\partial a_i}-\frac{\partial^3}{\partial a_1^3}\,,\qquad \mathcal{L}_{2}=\frac{\partial^6}{\partial a_1\partial a_5^2\partial a_6^3}-\frac{\partial^6}{\partial a_0^6}\,,\displaystyle&\displaystyle
\end{eqnarray}
for $\vartheta_i=a_i\frac{\partial}{\partial a_i}$ as before. The differential system $\mathcal{Z}_i$ then determines two algebraic coordinates $z^1$, $z^2$ on the complex structure moduli space that are given in terms of the Mori generators $\ell^{(i)}$ according to \eqref{eq:algCoords} as 
\begin{equation} 
		z^1=\frac{a_2a_3a_4}{a_1^3}\,,\quad z^2=\frac{a_1a_5^2a_6^3}{a_0^6}\,.
\end{equation}

To analyze the open-closed system $(Z_3,\Sigma)$ described by \eqref{eq:constraintsCand} we now apply the blow-up proposal, i.e.~construct the geometry $(\hat{Z}_3,E)$ given as the family of complete intersections \eqref{eq:blowup} in $\mathcal{W}=\P(\mathcal{O}(18H)\oplus \mathcal{O}(18H))\cong \P(\mathcal{O}\oplus\mathcal{O})$ which now reads
\begin{eqnarray}
 	\hat{Z}_3:& P=0\,,\quad Q=l_1(a_7u_1^6u_2^6u_3^6+a_8u_2^{18})-l_2(a_9u_1^6u_2^6u_3^6+a_{10}u_3^{18})\,.
\end{eqnarray}
We define the three-form $\hat{\Omega}$ by the residue expression \eqref{eq:ResZhat} and determine a system of differential operators, the Picard-Fuchs operators, for the family $\hat{Z}_3$.
First we determine the GKZ-system on the complex structure moduli space of the blow-up $Z_3$. We check that $\hat{\Omega}$ is identically annihilated by the two differential operators $\mathcal{L}_{1}$ and $\mathcal{L}_{2}$ of \eqref{eq:GKZclosed} that are complemented to the system 
\begin{eqnarray} \label{eq:GKZopenCand}
	&\displaystyle\mathcal{Z}_0=\sum_{i=0}^6\vartheta_i+1\,,\quad \mathcal{Z}_1=\sum_{i=7}^{10}\vartheta_i\,,\quad \mathcal{Z}_2=\vartheta_2-\vartheta_3-\vartheta_8\,,\quad \mathcal{Z}_3=\vartheta_2-\vartheta_4-\vartheta_{10}\,,&\nonumber\\
	&\displaystyle\mathcal{Z}_4=2\Big(\sum_{i=1}^4+\sum_{i=7}^{10}\Big)\vartheta_i-\vartheta_5\,,\quad  \mathcal{Z}_5=3\Big(\sum_{i=1}^4+\sum_{i=7}^{10}\Big)\vartheta_i-\vartheta_6\,,\quad \mathcal{Z}_6=\vartheta_9+\vartheta_{10}-\vartheta_7-\vartheta_8\,.\nonumber \displaystyle&\\
\displaystyle&\displaystyle\mathcal{L}_{1}\,,\qquad\mathcal{L}_2\,,\qquad\mathcal{L}_{3}=\frac{\partial^2}{\partial a_3\partial a_7}-\frac{\partial^2}{\partial a_1\partial a_8}\,,\qquad \mathcal{L}_{4}=\frac{\partial^2}{\partial a_4\partial a_9}-\frac{\partial^2}{\partial a_1\partial a_{10}}\displaystyle&\displaystyle
\end{eqnarray}
There are two additional second order differential operators $\mathcal{L}_{3}$ and $\mathcal{L}_{4}$ that annihilate $\hat{\Omega}$ and incorporate the parameters $a_7,\ldots a_{10}$ that are associated to the moduli of $\Sigma$. Clearly, there are no further operators of low minimal degree.
The operators $\mathcal{Z}_k$ are related to the symmetries of $\mathcal{W}$.
The first two operators are associated to an overall rescaling of the two constraints $P\mapsto \lambda P$, $Q\mapsto \lambda' Q$. The third and fourth operator describe the torus symmetries of the $\P^2$-base, $(u_1,u_j)\mapsto(\lambda_j u_1,\lambda_j^{-1}u_j)$, $j=2,3$, the fifth and sixth operator are due to the $\P^2(1,2,3)$-fiber symmetries, $(x,y,z)\mapsto(\lambda'_1x,y,{\lambda'_1}^{-1}z)$, $(x,y,z)\mapsto(x,\lambda'_2y,{\lambda'_2}^{-1}z)$ and the last operator $\mathcal{Z}_6$ is related to the torus symmetry $(l_1,l_2)\mapsto (\lambda l_1,\lambda^{-1}l_2)$ of the exceptional $\P^1$. All operators $\mathcal{Z}_i$ of the original system \eqref{eq:GKZclosed} are altered due to the lift to the blow-up $\hat{Z}_3$. 

As discussed in detail in section \ref{generaltoricstructure} the constructed GKZ-system can be obtained as a GKZ-system associated to toric data of the blow-up $\hat{Z}_3$.
The set of integral points $\hat{v}_i$ reads
\begin{equation}\label{blowupPolyCand}
	\begin{pmatrix}[c|ccccccc|cccc|l]
	    		     &  &   &   & \Delta_7^{\hat Z}&&&   		  &\hat{\ell}^{(1)} &\hat{\ell}^{(2)} &\hat{\ell}^{(3)}  &\hat{\ell}^{(4)}   &          \\ \hline
		\hat{v}_0    &1 & 0 & 0 & 0 & 0 & 0 & 0   & 0   &-6  		  &	0	&	0	& \hat{y}_0 = zxyu_1u_2u_3          \\
		\hat{v}_1    &1 & 0 & 0 & 0 & 2 & 3 & 0	  &-2   & 0  		  &    -1	&	1	& \hat{y}_1 = z^6 u_1^6 u_2^6 u_3^6 \\
		\hat{v}_2    &1 & 0 & 1 & 1 & 2 & 3 & 0	  & 1   & 0  		  &	0	&	0	& \hat{y}_2 = z^6 u_1^{18}          \\
		\hat{v}_3    &1 & 0 &-1 & 0 & 2 & 3 & 0	  & 0   & 0  		  &	1	&	0	& \hat{y}_3 = z^6 u_2^{18}          \\
		\hat{v}_4    &1 & 0 & 0 &-1 & 2 & 3 & 0   & 1   & 1  		  &	0	&      -1	& \hat{y}_4 = z^6 u_3^{18}          \\
	        \hat{v}_5    &1 & 0 & 0 & 0 &-1 & 0 & 0	  & 0   & 2  		  &	0	&	0	& \hat{y}_5 = x^3            	      \\
		\hat{v}_6    &1 & 0 & 0 & 0 & 0 &-1 & 0	  & 0   & 3  		  &	0	&	0	& \hat{y}_6 = y^2                   \\
		\hat{v}_7    &0 & 1 & 0 & 0 & 2 & 3 &-1	  &-1   & 0  		  &	1	&	0	& \hat{y}_7 =l_1 \hat{y}_1                   \\
		\hat{v}_8    &0 & 1 &-1 & 0 & 2 & 3 &-1	  & 1   & 0  		  &    -1	&	0	& \hat{y}_8 =l_1 \hat{y}_3                   \\
		\hat{v}_9    &0 & 1 & 0 & 0 & 2 & 3 & 1	  & 0   & 1  		  &	0	&      -1	& \hat{y}_9 = l_2\hat{y}_1                   \\
		\hat{v}_{10} &0 & 1 & 0 &-1 & 2 & 3 & 1	  & 0   &-1  		  &	0	&	1	& \hat{y}_{10} = l_2\hat{y}_4                \\
	\end{pmatrix}.                                                                        
\end{equation}
Here we have displayed the points $\hat{v}_i$, the basis of relations $\hat{\ell}^{(i)}$ and the corresponding monomials $\hat{y}_i$. We emphasize that besides the closed string charge vectors of $Z_3$ embedded as $\ell^{(1)}=\hat{\ell}^{(1)}+\hat{\ell}^{(3)}$, $\ell^{(2)}=\hat{\ell}^{(2)}+\hat{\ell}^{(4)}$ the brane charge vectors $\hat{\ell}^{(a)}$ are among the $\hat{\ell}^{(i)}$ of $\Delta_7^{\hat{Z}}$ as well.
We note that this toric data is completely consistent with the general formula \eqref{eq:genConstruction} to obtain $\Delta_7^{\hat{Z}}$. Similarly the associated GKZ-system precisely reproduces \eqref{eq:GKZopenCand} upon using the general formula of the GKZ-system \eqref{eq:GKZoopen}. We confirm as mentioned in section \ref{generaltoricstructure} that this polyhedron can be mapped to the five-dimensional polyhedron with an associate compact Calabi-Yau fourfold by adding the first and second row. This agrees with the heterotic/F-theory dual fourfold when considering the elliptic $Z_3$ as a heterotic compactification \cite{Grimm:2009sy}. Furthermore, the GKZ-system \eqref{eq:GKZopenCand} is a closed and more restrictive subsystem of GKZ-system for the Calabi-Yau fourfold.

The GKZ-system \eqref{eq:GKZopenCand} defines four coordinates $\hat{z}^a$ on the complex structure moduli space of $\hat{Z}_3$ that we calculate, according to the triangulation \eqref{blowupPolyCand} of $\Delta_7^{\hat{Z}}$, as
\begin{equation} \label{eq:zOpenCand}
 	\hat{z}^1=\frac{a_2a_4a_8}{a_1^2a_7}\,,\quad \hat{z}^2=\frac{a_4a_5^2a_6^3a_9}{a_0^6a_{10}}\,,\quad \hat{z}^3=\frac{a_3a_7}{a_1a_8}\,,\quad\hat{z}^4=\frac{a_1a_{10}}{a_4a_9}\,.
\end{equation}
We obtain a complete system of differential operators $\mathcal{D}_a$, the Picard-Fuchs system, by considering operators $\mathcal{L}_a$ associated to linear combinations of the charge vectors $\hat{\ell}^{(a)}$ in $\Delta_7^{\hat{Z}}$. By factorizing these operators as expressed in the coordinates \eqref{eq:zOpenCand} we obtain
\begin{eqnarray}
\mathcal{D}_{1}&=&\theta _1 \left(\theta _1-\theta _3\right) \left(\theta _1+\theta _2-\theta _4\right)+\left(\theta _1-\theta _3-1\right) \prod_{i=1}^2\left(2 \theta _1+\theta _3-\theta _4-i\right) \hat{z}^1\,,\\
\mathcal{D}_{2}&=&\theta _2 \left(\theta _2-\theta _4\right) \left(\theta _1+\theta _2-\theta _4\right)+12 \left(6 \theta _2-5\right) \left(6 \theta _2-1\right) \left(\theta _2-\theta _4-1\right) \hat{z}^2\,,\nonumber\\
\mathcal{D}_{3}&=&\left(\theta _1-\theta _3\right) \theta _3-\left(1+\theta _1-\theta _3\right) \left(2 \theta _1+\theta _3-\theta _4-1\right) \hat{z}^3\,,\nonumber\\
\mathcal{D}_{4}&=&\left(\theta _2-\theta _4\right) \left(-2 \theta _1-\theta _3+\theta _4\right)+\left(1+\theta _2-\theta _4\right) \left(1+\theta _1+\theta _2-\theta _4\right) \hat{z}^4\,,\nonumber\\
\mathcal{D}_{5}&=&\theta _1 \theta _3 \left(\theta _1+\theta _2-\theta _4\right)+\prod_{i=1}^3\left(2 \theta _1+\theta _3-\theta _4-i\right) \hat{z}^1\hat{z}^3\,,\nonumber\\
\mathcal{D}_{6}&=&\theta _1 \left(\theta _1-\theta _3\right) \left(\theta _2-\theta _4\right)+\left(\theta _1-\theta _3-1\right) \left(1+\theta _2-\theta _4\right) \left(2 \theta _1+\theta _3-\theta _4-1\right) \hat{z}^1\hat{z}^4\,,\nonumber\\
\mathcal{D}_{7}&=&\theta _2 \left(2 \theta _1+\theta _3-\theta _4\right)+12 \left(6 \theta _2-5\right) \left(6 \theta _2-1\right) \hat{z}^2\hat{z}^4\,,\nonumber\\
\mathcal{D}_{8}&=&\theta _1 \theta _3 \left(\theta _2-\theta _4\right)- \left(-\theta _2+\theta _4-1\right)\prod_{i=1}^2\left(2 \theta _1+\theta _3-\theta _4-i\right)  \hat{z}^1\hat{z}^3\hat{z}^4\,.\nonumber
\label{eq:GKZCandz}
\end{eqnarray}
The linear combination of charge vectors corresponding to each of these operators can be read off from the powers of the $\hat{z}^a$ in the last term of the $\mathcal{D}_a$.
We note that this system has the expected structure advertised in eq. \eqref{eq:LInZHat} of section \ref{hatomega}. Consequently the periods $\Pi^k(z^1,z^2)$ lift to the blow-up $\hat{Z}_3$ upon the identification $z^1=\hat{z}^1\hat{z}^3$, $z^2=\hat{z}^2\hat{z}^4$.

\subsubsection{Brane Superpotential at Large Volume: Disk Instantons}
\label{toricbranesP11169-W-LV}

In this section we solve the Picard-Fuchs system \eqref{eq:GKZCandz} at the point of maximal unipotent monodromy $\hat{z}^{a}\rightarrow 0$ in the complex structure moduli space of $\hat{Z}_3$. In addition we exploit the local limit $K_{\P^2}=\mathcal{O}_{\P^2}(-3)$ of $Z_3$, which is given by a decompactification of the elliptic fiber $t_2\rightarrow i\infty$, to determine the compact brane superpotential $W_{\rm brane}$ and the compact disk instanton invariants.

We find 16 solutions at large volume, that split into one power series solution $X^{(0)}$, four single logarithmic solutions $X^{(1)}_i$, seven double logarithmic solutions $X^{(2)}_\alpha$ and four triple logarithmic solutions $X^{(3)}_\beta$. As already expected from the observation made below \eqref{blowupPolyCand}, that $\hat{Z}_3$ connects to a compact Calabi-Yau fourfold, these are the only solutions. In particular, there are now square root and third root at large volume. The unique power series solution reads in the chosen normalization as
\begin{equation}
 	X^{(0)}=1+60 z_2+13860 z_2^2+4084080 z_2^3+24504480 z_1 z_2^3 +1338557220 z_2^4+\mathcal{O}(\underline{\hat{z}}^{10})\,,
\end{equation}
where we identify $z_1=\hat{z}_1\hat{z}_3$ and $z_2=\hat{z}_2\hat{z}_4$ as the complex structure moduli of $Z_3$ corresponding to $\ell^{(1)}$, $\ell^{(2)}$ in \eqref{3foldellp2}. Thus, $X^{(0)}$ is precisely the fundamental period $\Pi^0(z_1,z_2)$ of $Z_3$, cf. \cite{Candelas:1994hw,Hosono:1993qy}. In addition we recover the other five periods of $Z_3(1,1,1,6,9)$ as linear combinations of the solutions of the GKZ-system on $\hat{Z}_3$ with leading logarithms
\begin{eqnarray} \label{eq:leadinglogs11169}
		X^{(1)}_i\,:\,& \hat{l}_1\,,\,\,\hat{l}_2\,,\,\,\hat{l}_3\,,\,\,\hat{l}_4&\\[0.3Em]
		X^{(2)}_{\alpha}\,:\,& \frac{\hat{l}_1^2}{2}\,,\,\,  \hat{l}_2 (\hat{l}_1 +2 \hat{l}_4) \,,\,\,
\hat{l}_3(\hat{l}_1 +\frac{\hat{l}_3}{2})\,,\,\,\hat{l}_4(\hat{l}_1 +\hat{l}_4)\,,\,\,\frac{\hat{l}_2^2}{2}\,,\,\,\hat{l}_2 (\hat{l}_3+ \hat{l}_4)\,,\,\, \hat{l}_4(\hat{l}_3 +\frac{\hat{l}_4}{2} )&\nonumber\\[0.3Em]
		X^{(3)}_{\beta}\,:\,& \frac{1}{2} \hat{l}_1^2 \hat{l}_2+\frac{1}{2} \hat{l}_2^2 \hat{l}_3+\frac{1}{2} \hat{l}_1^2 \hat{l}_4+2 \hat{l}_1 \hat{l}_2 \hat{l}_4+\frac{1}{2} \hat{l}_2^2 \hat{l}_4+\hat{l}_1 \hat{l}_4^2+2 \hat{l}_2 \hat{l}_4^2+\frac{2 \hat{l}_4^3}{3} \,,\,\, \hat{l}_1 \hat{l}_2^2-\frac{3}{2} \hat{l}_2^2 \hat{l}_3+\frac{1}{2} \hat{l}_2^2 \hat{l}_4\,,& \nonumber\\
		&\!\!\!\!\hat{l}_1 \hat{l}_2 \hat{l}_3+\hat{l}_2^2 \hat{l}_3+\frac{1}{2} \hat{l}_2 \hat{l}_3^2+\hat{l}_1 \hat{l}_2 \hat{l}_4+\hat{l}_2^2 \hat{l}_4+\hat{l}_1 \hat{l}_3 \hat{l}_4+3 \hat{l}_2 \hat{l}_3 \hat{l}_4+\frac{1}{2} \hat{l}_3^2 \hat{l}_4+\frac{1}{2} \hat{l}_1 \hat{l}_4^2+\frac{5}{2} \hat{l}_2 \hat{l}_4^2+\frac{3}{2} \hat{l}_3 \hat{l}_4^2+\frac{5 \hat{l}_4^3}{6}\,,\!\!\!\nonumber\\ &  \frac{\hat{l}_2^3}{3}+\frac{1}{2} \hat{l}_2^2 \hat{l}_3+\frac{1}{2} \hat{l}_2^2 \hat{l}_4 \,,&\nonumber
\end{eqnarray}
where we used the abbreviation $\hat{l}_i=\log(\hat{z}_i)$. Indeed the threefold periods $l_1=\hat{l}_1+\hat{l}_3$, $l_2=\hat{l}_2+\hat{l}_4$, $l_1^2$, $l_1l_2+\frac{3}{2}l_2^2$ and $l_2(\frac12 l_1^2+\frac32 l_1 l_2+\frac32 l_2^2)$ are linear combinations of the solutions in \eqref{eq:leadinglogs11169}. Furthermore, we readily check that the complete $(z_1,z_2)$-series of the periods agree with the solutions on the blow-up.

Next we obtain the disk instanton invariants of the A-model dual to the five-brane on $Z_3$ from the local limit geometry $K_{\P^2}$. First we use the single logarithms of \eqref{eq:leadinglogs11169} as the mirror map $\hat{t}_i=\frac{X_i^{(1)}}{X^{(0)}}$ at large volume, where we have the series expansions
\begin{eqnarray}
		X^{(1)}_1&\!\!\!=&\!\!\! X^{(0)}\log(\hat{z}_1)-4 \hat{z}_1 \hat{z}_3+120 \hat{z}_2 \hat{z}_4+60 \hat{z}_2 \hat{z}_3 \hat{z}_4+30 \left(\hat{z}_1^2 \hat{z}_3^2+4 \hat{z}_1 \hat{z}_2 \hat{z}_3 \hat{z}_4+1386 \hat{z}_2^2 \hat{z}_4^2\right) +\mathcal{O}(\underline{\hat{z}}^5),\!\!
		\nonumber\\
		X^{(1)}_2&\!\!\!=&\!\!\! X^{(0)}\log(\hat{z}_2)
-60 \hat{z}_2-3080 \hat{z}_2^2 \left(442 \hat{z}_2+9 \hat{z}_4\right)+6 \hat{z}_2 \left(1155 \hat{z}_2+62 \hat{z}_4\right)+\mathcal{O}(\underline{\hat{z}}^4)\,,\nonumber\\
		X^{(1)}_3&\!\!\!=&\!\!\! X^{(0)}\log(\hat{z}_3)-2 \hat{z}_1 \hat{z}_3+60 \hat{z}_2 \hat{z}_4-60 \hat{z}_2 \hat{z}_3 \hat{z}_4+15 \left(\hat{z}_1^2 \hat{z}_3^2+4 \hat{z}_1 \hat{z}_2 \hat{z}_3 \hat{z}_4+1386 \hat{z}_2^2 \hat{z}_4^2\right)+\mathcal{O}(\underline{\hat{z}}^5),\!\nonumber\\
		X^{(1)}_4&\!\!\!=&\!\!\! X^{(0)}\log(\hat{z}_4)+60 \hat{z}_2-6930 \hat{z}_2^2+2 \hat{z}_1 \hat{z}_3-60 \hat{z}_2 \hat{z}_4+3080 \hat{z}_2^2 \left(442 \hat{z}_2+9 \hat{z}_4\right)+\mathcal{O}(\underline{\hat{z}}^4)\,.
\end{eqnarray}
Here we omit a factor $2\pi i$ for brevity. This confirms the consistency with the mirror of the threefold $Z_3$, $t_1=\hat{t}_1+\hat{t}_3$ and $t_2=\hat{t}_2+\hat{t}_4$, since the periods agree as $\Pi^{(1)}(\underline{z})=X^{(1)}_1+X^{(1)}_3=X^{(0)}\log(z_1)-6 z_1 + 45 z_1^2+\ldots$ and $\Pi^{(2)}(\underline{z})=X^{(1)}_1+X^{(1)}_3=X^{(0)}\log(z_2)+2 z_1+312 z_2+\ldots$. Upon inversion of the mirror map, we obtain the $\hat{z}^i$ as a series of $\hat{q}_a=e^{2\pi i \hat{t}_a}$, that we readily insert into the double logarithmic solutions in \eqref{eq:leadinglogs11169}. Then we construct a linear combination of these solutions to match the disk instantons in \cite{Aganagic:2000gs,Aganagic:2001nx} of the local geometries. Since $q_2=e^{2\pi it_2}\rightarrow 0$ in the local limit this means that we match, as a first step, only the part of the $q$-series, that is independent of $q_2$. Morally speaking, this procedure fixes part of the flux on $\hat{Z}_3$ specifying the five-brane. Indeed we obtain a perfect match of the disk instantons for both brane phases $I/II$, $III$ considered in \cite{Aganagic:2001nx} for the choices of superpotential
\begin{eqnarray} \label{eq:W11169}
 	W^{I/II}_{\rm brane}&=&\Big((\tfrac12(a_4-a_7)+a_3-\tfrac12)X^{(2)}_1+\sum_{i=2}^7 a_iX^{(2)}_i\Big)/X^{(0)}\,,\\
	W^{III}_{\rm brane}&=&\Big(a_1X^{(2)}_1+\tfrac{1}{3}(1+a_4+a_5-2 a_6+a_7)X^{(2)}_1+\sum_{i=3}^7 a_iX^{(2)}_i\Big)/X^{(0)}\,,
\end{eqnarray}
where we in addition fix the parameters $a_3=a_7=0$ to switch off the contribution of the closed instantons of $Z_3$ through its double logarithmic periods. We note that the two choices corresponding to \eqref{eq:W11169} are compatible with each other so that we can also find a single superpotential matching both phases of \cite{Aganagic:2001nx} simultaneously, corresponding to the parameters $a_1=-1+\tfrac{3}{2}a_2+a_3-\tfrac{1}{2}a_5+a_6-a_7,a_4=-1+3 a_2-a_5+2 a_6-a_7$.
Most notably, this match and the match in \eqref{eq:W11169} of the disk instantons of the local geometry already predicts the parts of the disk instantons in the \textit{compact} threefold $Z_3$. The compact disk instantons according to the Ooguri-Vafa multi-covering formula are listed in the tables in appendix \ref{instantonsLV11169}.

\section{Brane Blow-Ups as $SU(3)$ Structure Manifolds}
\label{su3structur}

In the previous sections we have argued that five-brane deformations
can be equivalently described on $\hat Z_3$ with a two-form flux 
localized on the blow-up divisor $E$. The aim of this section 
is to delocalize the flux further to three-form flux and propose an $SU(3)$
structure on the open manifold $\hat Z_3-E$ and $\hat Z_3$. 
Let us note that this section is independent of any explicit 
toric construction and the reader only interested in the enumerative 
aspects of the superpotential can safely skip this section. 
We hope to provide concrete proposals which should only be viewed 
as first steps to identify a complete back-reacted vacuum.
 
To set the stage we recall in section \ref{SU(3)rev} 
some basic facts about $SU(3)$ structure manifolds 
and the superpotential. 
As an intermediate step, we show how the blow-up space
$\hat Z_3$ can be endowed with a K\"ahler structure 
in section \ref{blow-up_as_Kaehler}. However, it is 
well-known that a supersymmetric vacuum with background 
three-form fluxes requires that the internal space is non-K\"ahler (section \ref{SU(3)rev}).
Furthermore, there exists no globally defined, nowhere vanishing holomorphic three-form on $\hat Z_3$. 
We propose a resolution to these issues in two steps in section \ref{non-Kaehlertwist}. 
In a first step we argue that there is a natural non-K\"ahler structure $\hat J$ on the 
open manifold $\hat Z_3 - E$ which, in a supersymmetric vacuum, 
matches the flux via $i(\bar \partial - \partial)\hat H_3 = d\hat J$.
While $(\hat J, \hat \Omega, \hat H_3)$ are well-defined forms 
on the open manifold, they have poles (as it is the case for $\hat J$, $\hat H_3$) and 
zeros (as we have seen for $\hat \Omega$) when extended to all of $\hat Z_3$. 
Hence, in a second step, we argue that the poles and zeros can be removed by an appropriate
local logarithmic transformation yielding new differential forms $(J',\Omega',H_3')$ on $\hat Z_3$ . 
In fact, the new global forms are defined 
such that the zeros and poles precisely cancel in the superpotential which 
can now be evaluated on $\hat Z_3$.

\subsection{Brief Review on $SU(3)$ Structures and the Superpotential} \label{SU(3)rev} 

To begin with, let us recall some basic facts about compactifications 
on non-Calabi-Yau manifolds $\hat Z_3$. In order that the four-dimensional effective theory 
obtained in such compactifications has $\cN=1$ supersymmetry one 
demands that $\hat Z_3$ has $SU(3)$ structure \cite{Koerber:2010bx}.
$SU(3)$ structure manifolds can be characterized by the
existence of two no-where vanishing forms, a real two-form $J'$ and a real three-form $\rho'$. 
Following \cite{Hitchin:2000jd} one demands that $J'$ and $\rho'$ are stable forms, 
i.e.~are elements of open orbits under the action of general linear transformations $GL(6,\bbR)$ at every point of the 
tangent bundle $T\hat Z_3$. Then one can set $\hat\rho'=*\rho'$ and show that $\hat \rho'$ is 
a function of $\rho'$ only \cite{Hitchin:2000jd}.
These forms define a reduction of the structure group from $GL(6,\bbR)$ to
$SU(3)$ if they satisfy $J \wedge \Omega' = 0$,
with a nowhere vanishing three-form $\Omega'=\rho'+i\hat\rho'$ and $J'$.

By setting $I_m^{\ n} = J'_{mp}g^{pn}$ one defines 
an almost complex structure with respect to which the metric $g_{mn}$ is hermitian.
The almost complex structure allows to introduce a $(p,q)$ grading of forms. Within this decomposition the form  
$J'$ is of type $(1,1)$ while $\Omega'$ is of type $(3,0)$. In general, 
neither $J'$ nor $\Omega'$ are closed. The non-closedness is parameterized 
by five torsion classes $\cW_i$ which transform as 
$SU(3)$ irreducible representations \cite{Chiossi:2002tw,Cardoso:2002hd}. One has 
\bea\label{dJ}
    dJ'&=&\tfrac{3}{2}\I (\bar \WV_1\Omega')+\WV_4\wedge J'+\WV_3\nn\\
     d\Omega'
   &=&\WV_1 J' \wedge J'+\WV_2\wedge J'+\overline\WV_5\wedge\Omega' \ ,
\eea
with constraints $J'\wedge J'\wedge\WV_2=J'\wedge\WV_3=\Omega' \wedge\WV_3=0$.
The pattern of vanishing torsion classes defines the properties of 
the manifold $\hat Z_3$. In a supersymmetric vacuum the 
pattern of torsion classes is constraint by the superpotential.  

Let us first discuss the pure flux superpotential for heterotic and Type IIB orientifolds with O5-planes. 
Recall from section \ref{fivebranesuperpotential} that the pure flux superpotential of these theories 
is of the form $W_{\rm flux} = \int \Omega \wedge H_3$ and $W_{\rm flux} = \int \Omega \wedge F_3$.
It is easy to check that there are no supersymmetric flux vacua for Calabi-Yau compactifications.
In fact, in the absence of branes $W_{\rm flux}$ is the only perturbative superpotential for a 
Calabi-Yau background. The supersymmetry conditions are 
\beq \label{DW=0}
  D_{z^k} W_{\rm flux} = 0 \ , \qquad W_{\rm flux}= 0 \ ,
\eeq
where the latter condition arises from the fact that $D_S W_{\rm flux} = K_S W_{\rm flux} = 0$,
for other moduli $S$ which do not appear in $W_{\rm flux}$. One easily checks that the 
first condition in \eqref{DW=0} implies that $H_3$ cannot be of type $(2,1)+(1,2)$, while 
the second condition implies that it cannot be of type $(3,0)+(0,3)$. This implies that 
$H_3$ has to vanish and there are no flux vacua in a Calabi-Yau compactification.

The situation changes for non-Calabi-Yau compactifications since the superpotential in 
this case is of more general form. 
More precisely, denoting by $\hat Z_3$ a generic $SU(3)$ structure manifold it takes
the form \cite{Behrndt:2000zh,LopesCardoso:2003af,Benmachiche:2008ma}
\beq
   W = \int_{\hat Z_3} \Omega' \wedge (H'_3+idJ')\ .
\eeq
It is straightforward to evaluate the supersymmetry conditions for 
this superpotential. Firstly, we note that in a supersymmetric background 
the compact manifold $\hat{Z}_3$ is complex, thus the torsion classes vanish, $\cW_1 = \cW_2=0$.
Second the superpotential $W$
is independent of the dilaton superfield and hence one 
evaluates in the vacuum that $W=0$, which implies that the $(0,3)$
part of $H'_3 + idJ'$ has to vanish.
However, since the $(3,0)+(0,3)$ component of $dJ'$ vanishes 
for a vanishing $\cW_1$, one concludes that also $H_3'+i dJ'$ has no 
$(3,0)$ component, and hence can be non-zero in the $(2,1)$ and $(1,2)$ directions.
The K\"ahler covariant derivative of $\Omega'$ yields 
the condition that $H_3'+idJ'$ has only components along the
$(2,1)$ direction. Hence, using the fact that $J'$ is of type $(1,1)$ 
one finally concludes  
\beq \label{dJ=H}
   H'_3 = i (\bar \partial - \partial) J'\ .
\eeq
This matches the long-known relation found in \cite{Strominger:1986uh} for
a supersymmetric vacuum of the heterotic string with background 
three-form flux. It should be stressed that 
there will be additional conditions which have to be respected 
by the heterotic vacuum. These involve a non-constant dilaton 
and cannot be captured by a superpotential. 

\subsection{The Blow-Up Space as a K\"ahler Manifold} \label{blow-up_as_Kaehler}

Before introducing an $SU(3)$ structure on $\hat Z_3$, it will be necessary to 
recall that each $\hat Z_3$ obtained by blowing up a holomorphic curve 
naturally admits a K\"ahler structure \cite{Griffiths,Voisin}. We will `twist' this K\"ahler 
structure to obtain a non-K\"ahler $SU(3)$ structure in 
subsection \ref{non-Kaehlertwist}. It will be crucial to look at the geometry of 
$\hat Z_3$ near $E$ more closely, and introduce the K\"ahler structure very explicitly. 

\subsubsection{K\"ahler Geometry on the Blow-Up: Warm-Up in two complex Dimensions} 
\label{Kahleronpointblowup}

To warm up for the more general discussion, let us first consider a simpler 
example and blow up a point in a complex surface. In a small patch $U_\epsilon$ 
around this point this looks like blowing up the 
the origin in $\bbC^2$ into an exceptional divisor $E = \bbP^1$. 
Let us denote by $B_{\bbC^2}$
the space obtained after blowing up as in section \ref{geometricblowups}. 
Our aim is to explicitly define a K\"ahler form $\tilde J$ on $B_{\bbC^2}$ 
following ref.~\cite{Griffiths}.

To define $\tilde J$ the key object we will study is the line bundle $\cL \equiv \cO_{B_{\bbC^2}}(E)$, 
or rather $\cL^{-1} \equiv \cO_{B_{\bbC^2}}(-E)$. To get a clearer picture of this bundle, we 
give a representation of $\cL$ near $E$. As in subsection \ref{geometricblowups} we first introduce the patch
$\hat U_\epsilon = \pi^{-1}(U_\epsilon)$.  
One can embed 
the fibers of $\cL$ into $\hat U_{2\epsilon} = \{U_{2\epsilon} \times \bbP^1: y_2 l_1 - y_1 l_2=0\}$ as
\beq \label{expl_cL}
   \cL_{(y,l)} = \{\lambda\cdot (l_1 ,l_2), \, \lambda \in \bbC \}\ ,
\eeq
where $(l_1,l_2)$ are the projective coordinates of $\bbP^1$ and $y$ collectively 
denote the coordinates on $U_\epsilon$. This implies that 
holomorphic sections $\sigma$ of $\cL$ are locally specified by $\sigma \equiv \lambda(\underline{y},\underline{l})$.
To explicitly display the expressions we introduce local coordinates on patches 
$\hat U_{2 \epsilon}^{(1)}$ and $\hat U_{2 \epsilon}^{(2)}$, which cover the 
$\bbP^1$ such that $l_i \neq 0 $ on $\hat U_{2 \epsilon}^{(i)}$.
We set 
\beq \label{local_coords_uell}
   \hat U_{2 \epsilon}^{(1)}: \quad u_1 = y_1,\quad \ell_1 = \frac{l_2}{l_1} \ ,\qquad \quad  
    \hat U_{2 \epsilon}^{(2)}: \quad u_2 = y_2,\quad \ell_2 = \frac{l_1}{l_2} \ ,
\eeq
Using the blow-up relation one finds the following 
coordinate transformation on the overlap
\beq \label{coord_trans}
  \hat U_{2 \epsilon}^{(1)} \cap \hat U_{2 \epsilon}^{(2)}:\qquad \quad (\ell_2,u_2) = (1/\ell_1 , \ell_1 u_1)\ ,
\eeq
which shows that $B_{\bbC^2}$ is identified with $\cO_{\bbP^1}(-1)$. Let us point out that 
this matches the local description presented in appendix \ref{App:Local} if we interpret $B_{\bbC^2}$ as 
a local model of $(N_{\hat Z_3} \Sigma)_p$ at a point $p$ on $\Sigma$.\footnote{To 
avoid cluttering of indices we introduce the new notation for $z_1^{(1)}\equiv u_1,$ $z_2^{(1)} \equiv \ell_1$ and
$z_2^{(2)} \equiv u_2$, $z_2^{(2)}\equiv \ell_2$.}

One now can introduce a metric $|| \simga ||$ for sections $\sigma$ of 
the line bundle $\cL$ as follows. Since $\cL$ is non-trivial
one cannot simply specify $|| \simga ||$ by using a single global holomorphic 
section $\sigma$. Each such global holomorphic section will have either poles of zeros.
However, we can specify $||\cdot||$ on local holomorphic sections patchwise and 
glue these local expressions together.
Let us define the local expression on $B_{\bbC^2}-E$ by evaluation on a global holomorphic section $\sigma_{(0)}$ with 
zeros along $E$. One defines
\beq \label{patch1}
  B_{\bbC^2} - E:\qquad ||\sigma_{(0)}||_1 := 1 \ .
\eeq
On the patches $\hat U_{2\epsilon}$ covering $E$ one needs to 
use other sections which are non-vanishing also along $E$.  
Using the explicit realization of $\cL$ as in \eqref{expl_cL} 
with sections $\sigma = \lambda$ one can 
specify $||\cdot||_2$ in $\hat U_{2\epsilon}$ 
setting
\beq\label{patch2}
  \hat U_{2\epsilon}:\qquad ||\sigma||_2:= |\lambda|\, (|l_1|^2 + |l_2|^2)^{1/2} \ .  
\eeq
Note that the section $\sigma_{(0)}$ is also defined in $\hat U_{\epsilon}-E$ and can 
be given in 
the representation \eqref{expl_cL} of $\cL$. In the same representation we 
can also define local sections $\sigma_{(i)}$ near $E$, such that 
\beq \label{loc_sections_near_D}
  \hat U_{2\epsilon}: \quad  \sigma_{(0)} = \frac{y_1}{l_1} = \frac{y_2}{l_2}\ , 
     \qquad  \qquad \hat U_{2\epsilon}^{(i)}: \quad \sigma_{(i)} = \frac{1}{l_i}\ ,   
\eeq
Recall that the $y_i$, as also introduced in subsection \ref{geometricblowups}, 
specify the point which is blown up as $y_1 =y_2 =0$. The metric 
\eqref{patch2} for these sections is simply given by 
\beq  \label{expl_metric_2}
   \hat U_{\epsilon}^{(i)}:\quad || \sigma_{(i)}||_2 = (1 + |\ell_i|^2)^{1/2}\ ,\qquad \quad
    || \sigma_{(0)}||_2 = |u_i|\cdot (1 + |\ell_i|^2)^{1/2}  \ ,\\
\eeq 
where we have used the local coordinates \eqref{local_coords_uell}.
To give the global metric one next splits 
$B_{\bbC^2}$ into patches $(B_{\bbC^2} - \hat U_\epsilon, \hat U_{2\epsilon})$,
and introduces a partition of unity $(\rho_1,\rho_2)$. The local expressions 
\eqref{patch1} and \eqref{patch2} are glued together as 
\beq \label{patched_metric}
   ||\cdot ||:=\rho_1 ||\cdot||_1 + \rho_2 ||\cdot||_2\ .
\eeq

Using this metric one can now determine the Chern curvature form $\frac{i}{2}\Theta $ of the 
line bundle $\cL$.
Locally one has to evaluate
\beq 
  \frac{i}{2}\Theta = - \frac{1}{4\pi}\partial \bar \partial \log || \sigma||^2\ ,
\eeq 
for holomorphic sections $\sigma$ which have no poles or zeros in the considered patch. 
Using \eqref{patch1} and \eqref{expl_metric_2} one finds that $\cL$ is trivial on $B_{\bbC^2}-\hat U_{2\epsilon}$, 
but non-trivial in the patches $\hat U_\epsilon^{(i)}$:
\bea \label{Theta_patches}
  B_{\bbC^2} - \hat U_{2\epsilon}: \quad \Theta &=& 0 \ , \\
 \hat U_\epsilon^{(i)}: \quad \Theta &=& - \frac{i}{2\pi} \partial \bar \partial  \log \big(1+|\ell_i|^2 \big) = - \omega_{FS} \ , \nn
\eea
where $\omega_{FS}$ is the Fubini-Study metric. 
One notes that the form $\Theta$ is strictly negative on $E$, since
it is given by minus the pull-back of the Fubini-Study metric under the restriction map to $\bbP^1$. 
Hence, one finds that $-\Theta > 0$ on $E$, and $-\Theta \geq 0$ on $\hat U_\epsilon$.
In the region $\hat U_{2\epsilon}-\hat U_\epsilon$ the form $\Theta$ interpolates
in a continuous way. Finally, we can give the K\"ahler form $\tilde J$  on $B_{\bbC^2}$.
Since $-\Theta \geq 0$ on $\hat U_\epsilon$ and $\Theta=0$ on 
$B_{\bbC^2} - \hat U_{2\epsilon}$ continuity implies that $-\Theta$ is bounded from below. 
This fact can be used to define a K\"ahler form on $B_{\bbC^2}$ by setting
\beq \label{Kahleronblowup}
    \tilde J =\pi^* J - v_{\rm bu} \Theta\ ,
\eeq
where $\pi^* J$ is the pull-back of the K\"ahler form on $\bbC^2$.
$\tilde J$ is a closed $(1,1)$-form, and positive for a sufficiently small blow-up volume $v_{\rm bu}$. 
In other words, one finds that the manifold $B_{\bbC^2}$ is naturally endowed with a 
K\"ahler structure. Since the blow-up is a local operation, one uses
this construction to blow up a point in any K\"ahler surface identifying $J$ in 
\eqref{Kahleronblowup} with the K\"ahler form before the blow-up. 

\subsubsection{Blow-Up Spaces as K\"ahler Manifolds}

Having discussed the K\"ahler structure on the blow-up of a point in a K\"ahler surface, we 
can now generalize this construction to blowing up curves in $Z_3$. 
This is again textbook knowledge \cite{Voisin} 
and we can be brief. 

As in subsection \ref{Kahleronpointblowup} we 
study the line bundles $\cL \equiv \cO_{\hat Z_3}(E)$ and 
$\cL^{-1} \equiv \cO_{B_{\hat Z_3}}(-E)$. One proceeds as in 
the two-dimensional example and first examines the bundle 
near $E$. One shows that restricted to $E$ one obtains  
\beq
   \cL^{-1}|_{E} \ \cong\ N^*_{\hat Z_3} E\ \cong\ \cO_E(1)\ ,
\eeq
where $ N^*_{\hat Z_3} E$ is the co-normal bundle to $E$ in $\hat Z_3$.
The bundle $E=\bbP^1 (N_{Z_3} \Sigma) \rightarrow \Sigma$ 
admits a natural closed $(1,1)$-form which is positive on the fibers. 
This form can be obtained from a hermitian metric induced from 
$N_{Z_3} \Sigma$ and is obtained from the Chern curvature of $\cO_E(1)$.
As in the previous section this curvature can be extended to $\cL^{-1}$ by 
a partition of unity as in \eqref{patched_metric}.
Let us denote the Chern curvature of $\cL$ again by $\frac{i}{2}\Theta$.
The K\"ahler form on $\hat Z_3$ is then given by 
$\tilde J= \pi^* J   - v_{\rm bu} \Theta $, for sufficiently 
small blow-up volume $v_{\rm bu}$.
While the construction of $\Theta$ depends on the 
explicit metric $||\cdot||$ on $\cL$ one can also 
evaluate the corresponding cohomology classes which are topological in nature. Clearly, one has
\beq
  c_1(\cL) = \tfrac{1}{\pi} [\Theta]\ , \qquad \quad c_1 (\cL)|_E = c_1(N_{\hat Z_3} E) = [E]_{E}\ ,
\eeq
where we have also displayed the restriction to the exceptional divisor.

The above discussion identifies $\hat Z_3$ as a K\"ahler manifold.
If one directly uses this K\"ahler structure, however, 
one finds that $\hat Z_3$ cannot arise as an actual 
supersymmetric flux background. Recall that, for example, in heterotic compactifications 
with background fluxes $H_3$ the internal manifold has to be non-K\"ahler to satisfy \eqref{dJ=H} as
shown in \cite{Strominger:1986uh}.
In the following we will show that there actually exists a natural $SU(3)$ 
structure on $\hat Z_3$ which renders it to be non-K\"ahler and allows to 
identify a supersymmetric flux vacuum on $\hat Z_3$.

\subsection{Defining the $SU(3)$ Structure: the Non-K\"ahler Twist}
\label{non-Kaehlertwist}

In the following we propose an $SU(3)$ structure on the open manifold $\hat Z_3 - E$ and the 
blow-up space $\hat Z_3$ in subsections \ref{SU3_on_open} and \ref{SU(3)onblowup}, respectively. 
Before turning to the three-dimensional case we first discuss 
the complex two-dimensional case $B_{\bbC^2}$ in subsection \ref{Hopf}. 

\subsubsection{The Non-K\"ahler Twist: Warm-Up in two complex Dimensions}
\label{Hopf}

To warm up for the more general discussion, let us first introduce a 
non-K\"ahler structure on the simpler
two-dimensional example $B_{\bbC^2}$. Recall from \eqref{coord_trans} that $B_{\bbC^2}$
is the total space of the universal line bundle $\cO_{\bbP^1}(-1)$ over $\bbP^1$.
There are two geometries related to $B_{\bbC^2}$ which admit a non-trivial 
non-K\"ahler structure. Firstly, to render 
$B_{\bbC^2}$ into a compact space $\hat B_{\bbC^2}$ one can replace each fiber $\bbC$ 
of the line bundle $B_{\bbC^2}$ by a two-torus $\bbC^*/\bbZ$.\footnote{More precisely,
parameterizing a fiber by $\lambda = r e^{i\theta}$ one removes the origin $r=0$ 
and identifies $r \cong r +1$.} Secondly, one can consider the open manifold $B_{\bbC^2} - E$,
where one simply removes the origin and the attached exceptional blow-up divisor $E = \bbP^1$. 
On the geometries $\hat B_{\bbC^2}$ and $B_{\bbC^2}$ 
on can introduce coordinates $(\ell_i, u_i)$ as in \eqref{local_coords_uell}. 
The new geometries have been modified from $B_{\bbC^2}$ such that, in particular, one 
has $u_i \neq 0 $. For the compact space $\hat B_{\bbC^2}$ one 
further has a periodicity since $u_i \in \bbC^*/\bbZ$.
Note that one inherits from the blow-up the coordinate transformation \eqref{coord_trans}
on the overlaps $\hat U^{(1)} \cap \hat U^{(2)}$ covering the north and 
south pole of the $\bbP^1$.

One observes that with the transformations \eqref{coord_trans} the surface $\hat B_{\bbC^2}$ can be identified as 
the \textit{Hopf surface} $S^1 \times S^3$. This surface does not admit a K\"ahler structure 
since $h^{1,1}=0$, while $h^{0,0}=h^{1,0}=h^{2,1}=h^{2,2}=1$. 
Note however, that $\hat B_{\bbC^2}$ admits a natural globally defined no-where vanishing 
$(1,1)$-form $\hat J$. The construction of $\hat J$ was given in ref.~\cite{Gualtieri:2010fd}. 
In fact, one can also introduce 
a three-form flux $\hat H_3$ such that $i(\bar \partial - \partial) \hat J = \hat H_3$, just as in 
the supersymmetry conditions \eqref{dJ=H}. 

Keeping this connection with the Hopf surface in mind, we aim 
to introduce $\hat J$ and $\hat H_3$ on $B_{\bbC^2} - E$. 
Firstly, we introduce on $U^{(i)}_{\epsilon}$ the B-field 
\bea \label{hatBi}
  U^{(i)}_{\epsilon}-E:\quad  \hat B_i &=& \tfrac{1}{4 \pi} \big(\partial \log (1+ |\ell_i|^2) \wedge \partial \log u_i + \bar \partial \log (1+ |\ell_i|^2) \wedge \bar \partial \log \bar u_i  \big)\ ,\\
   &=& \tfrac{1}{2 \pi} \text{Re}\big(\partial \log || \sigma_{(i)}||^2 \wedge \partial \log ||\sigma_{(0)}||^2 \big) \ . \nn 
\eea
Here we have inserted the definitions \eqref{loc_sections_near_D} of the sections $\sigma_{(0)},\sigma_{(i)}$ and 
the norm $||\cdot||$ given in \eqref{expl_metric_2}.
One realizes that these $\hat B_i$ are of type $(2,0)+(0,2)$. They extend continuously to
$B_{\bbC^2} - E$ and one checks that $\hat B_i=0$ on $B_{\bbC^2} - \hat U_{2 \epsilon}$, since 
$|| \sigma_{(0)}|| =1 $ outside $\hat U_{2 \epsilon}$. 
However, note that the $ \hat B_i$ do not patch together on the overlap $\hat U^{(1)}_\epsilon \cap \hat U^{(2)}_\epsilon$ as a form, but rather 
satisfy
\beq
  \hat U^{(1)}_\epsilon \cap \hat U^{(2)}_\epsilon: \qquad  \hat B_2- \hat B_1 = F_{21}\ ,  
\eeq
where one identifies 
\beq 
   F_{21} = - \tfrac{1}{2\pi} \text{Re} \big( d \log \ell_1 \wedge d \log u_1\big)\ ,
\eeq
The $(2,0)+(0,2)$ form $F_{21}$ on $\hat  U^{(1)}_\epsilon \cap \hat U^{(2)}_\epsilon$ can be used to define a Hermitian line bundle on this overlap.
In mathematical terms $F_{01}$ and $\hat B_i$ define a gerbe with curvature 
\beq \label{def-H3_surf}
  \hat H_3 = d \hat B_i = - \tfrac{i}{2} \omega_{FS} \wedge (\bar \partial - \partial) \log |u_i|^2\ , 
\eeq
where we have introduced the Fubini-Study metric $\omega_{FS} = \tfrac{i}{2\pi} \partial \bar \partial   \log || \sigma_{(i)}||^2$ on $\bbP^1$.
One can now check that indeed on the Hopf surface $\hat B_{\bbC^2} \cong S^1 \times S^3$ one 
has a non-vanishing integral of $\hat H_3$ over the $S^3$. Similarly, 
one can evaluate the integral on the open manifold $B_{\bbC^2} - E$, where the integral of 
$\hat H_3$ is performed on $\bbP^1$ and the $S^1$ encircling the zero section which has been removed.

It is now straightforward to read off the non-K\"ahler $(1,1)$-form $\hat J$, satisfying the 
supersymmetry condition $i(\bar \partial - \partial) \hat J = \hat H_3$.
There are two immediate choices. One could set
\beq \label{def-J_ex1}
 \hat U_\epsilon^{(1)}-E: \quad   
   \hat J = \tfrac{1}{2 \pi} \text{Im} \big(\partial \log ||\sigma_{(1)}||^2 \wedge \bar \partial \log \bar u_1 \big) \ ,
\eeq
which is the choice used on the Hopf surface in \cite{Gualtieri:2010fd}.
To evaluate $\hat J$ in the second patch $U_1$ one uses the coordinate transformation \eqref{coord_trans} and 
transforms $\hat J$ like a standard differential form. However, the second choice 
\beq \label{def-J_ex2}
   \hat U_\epsilon^{(1)}-E: \quad   
   \hat J =  - \tfrac{1}{2} \log |u_1|^2 \, \omega_{FS}\ ,  
\eeq
is more appropriate to the blow-up case. The reason is that 
\eqref{def-J_ex2} can be extended to $B_{\bbC^2}-E$ by replacing 
$\omega_{FS} \rightarrow - \Theta$, where $\Theta$ is proportional to 
the Chern curvature of the line bundle $\cL = \cO(E)$ introduced in subsection \ref{Kahleronpointblowup}. 
As noted in \eqref{Theta_patches} the Chern curvature of $\cL$ is localized near 
$E$ and vanishes outside $\hat U_{2 \epsilon}$. It will be the 
expression \eqref{def-J_ex2} which we will extend to the three-dimensional 
case in the next subsection.

Note that for the expressions in \eqref{hatBi}-\eqref{def-J_ex2}
to define well-behaved differential forms, one has to exclude the origin $u_i =0$ in $\hat U^{(i)}_\epsilon$.
This is precisely the reason why we considered the restriction of the blow-up space $B_{\bbC^2}$
to the open manifold $B_{\bbC^2}-E$. The extension of this structure 
to $B_{\bbC^2}$ will be discussed in subsection \ref{SU(3)onblowup} below. However, 
one can extend the above constructions to $\bbC^2$ and $B_{\bbC^2}$ 
by including currents as in sections \ref{N=1branes} and \ref{potentialhatZ3minusD}.
Note that on $\bbC^2$ the delta-current $\delta_{\{0\}}=\bar \partial T_\beta$ 
localizing on the origin is defined using the Cauchy kernel 
\beq \label{Cauchykernel}
   \beta = -\tfrac{1}{4\pi}\partial \log (|x_1|^2 + |x_2|^2) \wedge \partial \bar \partial   \log (|x_1|^2 + |x_2|^2) \ ,
\eeq
where $x_i$ are the coordinates on $\bbC^2$. Here one has to use the same formalism as in section \ref{N=1branes}. 
The Cauchy kernel $\beta$ can be lifted to the blow-up space $B_{\bbC^2}$ using the blow-down map $\pi: B_{\bbC^2} \rightarrow \bbC^2$
as 
\beq
  \pi^* \beta =  i \partial \log u_i \wedge \omega_{FS} \ ,
\eeq
where $u_i$ is identified with the coordinate on 
the fiber of $B_{\bbC^2}$ viewed as a line bundle over $\bbP^1$, and 
$\omega_{FS}$ is 
the Fubini-Study metric on $\bbP^1$ defined below \eqref{def-H3_surf}. Hence, we see that 
$\hat H_3$ as defined 
in \eqref{def-H3_surf} is not a form on the full space $B_{\bbC^2}$.
Rather the $(2,1)$-part of $\hat H_3$ is seen to be the pull-back of the Cauchy 
kernel \eqref{Cauchykernel}, $\hat H_3^{(2,1)} = - i \partial \hat J = \pi^* \beta$.
One readily evaluates 
\beq
  d \hat H_3 = \pi^* d \beta = \pi^* \delta_{\{ 0\}} = \omega_{FS} \wedge \delta_{E}\ ,
\eeq
where $\delta_E$ is the delta-current localizing on the exceptional divisor $E$ as
in section \ref{potentialhatZ3minusD}.
Therefore, we can reduce the integrals involving the so-defined $\hat H_3$ to chain integrals over 
a one-chain ending on $\{ 0 \}$ in $\bbC^2$ since 
\beq \label{chain_rewrite}
   \int_{\Gamma} \gamma = \int_{\bbC^2} \gamma \wedge \bar \beta = \int_{B_{\bbC^2}} \pi^* \gamma \wedge \hat H_3^{(1,2)} 
   = \tfrac12 \int_{B_{\bbC^2}} \pi^* \gamma \wedge (\hat H_3 + i d \hat J) \ , 
\eeq
for a compactly supported $(1,0)$-form $\gamma$.
Note that this is the analog of the computation performed in section \ref{potentialhatZ3minusD}, where we have shown how the 
five-brane superpotential is translated to a superpotential on the blow-up space $\hat Z_3$. 
We are now in the position to introduce an $SU(3)$ structure on the open manifold $\hat Z_3 - E$.

\subsubsection{The $SU(3)$ Structure on the Open Manifold $\hat Z_3 - E$}
\label{SU3_on_open}

We are now in the position to discuss the $SU(3)$ structure on the open manifold $\hat Z_3 - E$
by introducing forms $\hat \Omega, \hat J$ of type $(3,0)$ and $(1,1)$, respectively.
We show that these forms satisfy 
\beq \label{dhatJ}
   d \hat \Omega = 0 \ ,\qquad d \hat J = \cW_4 \wedge \hat J + \cW_3 \ ,
\eeq
for non-trivial $\cW_4$ and $\cW_3$.
Note that $\hat \Omega = \pi^* \Omega$ has been already discussed in detail in section \ref{hatomega}.
It was noted that $\hat \Omega$ has a zero along $E$, but is a well-defined, nowhere vanishing form on the 
open manifold $\hat Z_3 - E$. The basic idea to introduce 
$\hat J$ is to extend the definition \eqref{def-J_ex2} to $\hat Z_3 - E$ by 
using the sections and metric on $\hat Z_3$ defined in section \ref{blow-up_as_Kaehler}. It is further illuminating to view the construction of subsection \ref{Hopf} as a local model for $N_p \Sigma$. 

Let us recall that on patches $\hat U_{2 \epsilon}$ one can introduce 
the holomorphic sections $\sigma_{(0)}$ and $\sigma_{(1)},\sigma_{(2)}$ of $\cL= \cO_{\hat Z_3}(E)$ as in \eqref{loc_sections_near_D}. 
One recalls that $\sigma_{(0)}$ is a global section which has zeros along 
$E$ just as the form $\hat \Omega$. We therefore work on $\hat Z_3 - E$. Moreover, 
we first consider a situation where a supersymmetric five-brane curve 
has been blown up, and hence the flux $\hat H_3$ on $\hat Z_3$ is supersymmetric and 
satisfies $i (\bar \partial - \partial) \hat J = \hat H_3$.
Following the steps of subsection \ref{Hopf} we first introduce the 
B-fields
\beq
  \hat U^{(i)}_{2\epsilon} - E: \qquad  \hat B_i =  \tfrac{1}{2\pi} \text{Re} \big( \partial  \log || \sigma_{(i)}||^2 \wedge \partial \log || \sigma_{(0)}||^2 \big) \ ,
\eeq
where $\hat U^{(i)}_{2\epsilon}$
are the patches which cover $\Sigma$  and the blow-up $\bbP^1$'s such that $\sigma_{(i)}$ is well-defined. 
Note that this implies that we have to refine the cover $\hat U^{(i,\alpha)}_{2\epsilon}$ as in section \ref{geometricblowups}. 
For simplicity we will drop 
these additional indices in the following.
Evaluating $\hat H_3 = d \hat B_i$ this implies 
\bea
  \hat U^{(i)}_{2\epsilon} - E:\qquad   
   \hat H_3 & = & - \tfrac{i}{2} \Theta \wedge ( \bar \partial - \partial)  \log \big(|| \sigma_{(0)}||^2 / || \sigma_{(i)}||^2 \big) \qquad  \\
    &=&  - \tfrac{i}{2} \Theta \wedge ( \bar \partial - \partial)  \log | z^{(i)}_i|^2 \ , \nn
\eea
where $\frac{i}{2}\Theta$ is the Chern curvature of $\cL$ as introduced 
in section \ref{blow-up_as_Kaehler}, and  we inserted the sections as given in \eqref{loc_sections_near_D}
using local coordinates $z^{(i)}_i = y_i$ in $\hat U^{(i)}_{2\epsilon}$. Note that $E$ is 
given in $\hat U^{(i)}_{2\epsilon}$ as $z^{(i)}_i=0$ and has been excluded from $\hat U^{(i)}_{2\epsilon}$.
One can further evaluate $\hat B_2 - \hat B_1 = F_{21}$ on the overlap $ \hat U^{(i)}_{\epsilon}\cap \hat U^{(i)}_{\epsilon}$
and show in local coordinates that $dF_{21}=0$.

With this preparation one next defines $\hat J$ using the 
above supersymmetric flux.
One finds the natural extension of \eqref{def-J_ex2} given by 
\beq \label{def-hatJ}
  \hat U^{(1)}_{2\epsilon} - E :\qquad \hat J = \pi^* J - (1- \log |y_1|^2 ) \Theta \ ,   
\eeq
where $J$ is the K\"ahler form on $Z_3$ and $\pi$ is the blow-down map.
In order to transform this $\hat J$ into the other patches  $\hat U^{(2)}_{2\epsilon}$ covering $\bbP^1$ one transforms 
$\hat J$ as a differential form. Note that $\Theta$ vanishes outside 
$\cup \hat  U_{2 \epsilon}$ and one finds
\beq \label{oldJ}
  \hat Z_3 -\cup \hat  U_{2 \epsilon}:\qquad \hat J = \pi^* J\ .
\eeq
In other words, the departure from the original K\"ahler structure only 
arises in a small neighborhood around $E$. 
Using the explicit form of $\hat J$ one checks 
that $d \hat J$ has non-trivial torsion classes $\cW_3,\cW_4$ in \eqref{dhatJ}.
In conclusion one finds that near $E$ we introduced a non-K\"ahler geometry on the open manifold. 

Let us close this section by noting that the constructed structure $(\hat J,\hat \Omega)$ 
and $\hat H_3$ are only well-defined differential forms on the open manifold $\hat Z_3 - E$.
Furthermore, one finds on $\hat Z_3 -E$ that $d \hat H_3= 2i  \partial \bar \partial \hat J =0$, which implies that 
there is no source term for the blown-up five-brane. To include such a source one has to work on the whole manifold $\hat Z_3$
using currents as in section \ref{potentialhatZ3minusD}. One then finds 
\beq
  d \hat H_3 =2 i \partial \bar \partial \hat J = \delta_E \wedge \Theta\ ,   
\eeq
in accord with \eqref{dhatH_3}. Analog to \eqref{chain_rewrite} one also rewrites the 
chain integral on $Z_3$ to $\hat Z_3$ as 
\beq 
   \int_{\Gamma} \chi = \int_{\hat Z_3} \pi^* \chi \wedge (\hat H_3 + i d\hat J)  \ ,
\eeq
where $\Gamma$ is a three-chain ending on $\Sigma$. 
Here we have used $\hat H_3 = i (\bar \partial - \partial) \hat J $ for a supersymmetric 
configuration. Clearly, upon varying the complex structure of $\hat Z_3$ this relation 
will no longer hold, since $(\hat \Omega,\partial\hat J)$ are still forms of type $(3,0)$ and $(2,1)$
in the new complex structure, respectively. The flux $\hat H_3$, however, is fixed as a form but
changes its type under a complex structure variation. Thus in the supersymmetric configuration 
the complex structure on $\hat{Z}_3$ is adjusted so that $\hat{H}_3$ is of type $(2,1)+(1,2)$. 

Note that the described structure is not yet satisfying, since $(\hat \Omega,\hat J,\hat H_3)$
are no well-defined differential forms on all of $\hat Z_3$.
In the next subsection we resolve this issue by proposing a redefinition 
which allows us to work with differential forms on the full space $\hat Z_3$.

\subsubsection{The $SU(3)$ Structure on the Manifold $\hat Z_3$}
\label{SU(3)onblowup}

In this subsection we will finally completely resolve the 
five-brane into a non-K\"ahler geometry. Recall that 
so far we had to work on the open manifold if we wanted to 
use forms, while on $\hat Z_3$ we had to use currents due 
to the singularities of $\hat H_3$ and $\hat J$ along the 
exceptional divisor $E$.
In the following we will introduce an $SU(3)$ structure on $\hat Z_3$
by specifying a new $(1,1)$-form $J'$, a non-singular three-form flux 
$H_3'$ and a non-where vanishing $(3,0)$-form $\Omega'$.\footnote{Recent 
constructions of non-K\"ahler geometries can be found in ref.~\cite{Larfors:2010wb,Chen:2010bn}.
The constructions of \cite{Larfors:2010wb,Chen:2010bn} share sensible similarities with our approach. We hope to come back to this issue in future works.}

To begin with we note that there exists no holomorphic no-where vanishing 
$(3,0)$ form on $\hat Z$ since $K\hat Z_3$ is non-trivial. 
Hence, $\Omega'$ has to be non-holomorphic and 
we will find non-vanishing torsion classes $\cW_3,\cW_4,\cW_5$ such that  
\beq
  d \Omega' = \bar \cW_5 \wedge \Omega'\ , \qquad d J' = \cW_4 \wedge J' + \cW_3\ . 
\eeq 
The basic idea to define $\Omega'$ and $J'$ is rather simple. We first note 
that $\hat H_3$ and $d\hat J$ have a pole of order one along $E$, while $\hat \Omega$ has a first order zero along $E$. Then we want to cancel the zero against the 
pole and introduce $(J',\Omega')$ such that
\beq \label{match_super}
   \int_{\hat Z_3 - E} (\hat H_3 + i d \hat J) \wedge \hat \Omega = \int_{\hat Z_3} (H'_3 +i d J') \wedge \Omega' \ , 
\eeq
which is essential in the matching of the superpotentials.

In order to implement this cancellation we note that $d\hat J$ is proportional to $d|y_1|/|y_1|$
in $\hat U^{(1)}_\epsilon$. We thus define 
\beq  
 \hat U^{(1)}_\epsilon:\quad \Omega' = \frac{1}{|y_1|} \hat \Omega\ ,
\eeq
and transform $\Omega'$ to the other patches like a differential form.
Note that $\Omega'$ no longer admits a zero along $E$, but also is no longer 
holomorphic. Evaluating $d\Omega'$ one finds a non-vanishing real $\cW_5 = - d \log |y_1|$ in $\hat U^{(1)}_\epsilon$. 
Let us stress, however, that by construction of $\hat{Z}_3$,
\beq 
  \hat Z_3 - \cup \hat U_{2 \epsilon}:\qquad \Omega' = \pi^* \Omega, \quad \cW_5 = 0 \ . 
\eeq 
This implies that we need to 
`glue in' the non-holomorphic dependence of $\Omega'$ using a partition 
of unity for $(\hat Z_3 -\cup \hat U_\epsilon,\hat U_{2 \epsilon})$.

Let us comment on a more global approach to define $\Omega'$ on 
the blow-up $\hat Z_3$ realized as in \eqref{eq:blowup}.
Instead of starting with $\hat \Omega$ we 
begin with a globally defined and no-where vanishing $\tilde \Omega$.
However, this $\tilde \Omega$ is not a differential from, but rather  
a section of $\cL^{-1} \otimes K\hat Z_3$ given by the residue expression 
\beq
   \tilde \Omega = \int_{\epsilon_1}\int_{\epsilon_2} \frac{\Delta}{P Q}\ .
\eeq 
Comparing this expression with $\hat \Omega$ given in \eqref{eq:ResZhat} 
one immediately sees that $\tilde \Omega$ does not vanish along $E$. 
This $\tilde \Omega$ captures the complex structure dependence of the 
$\hat Z_3$. It also satisfies the same Picard-Fuchs equations as 
$\hat \Omega$, and hence can be given in terms of the same periods.
By the blow-up construction it depends on the complex structure 
deformations of $Z_3$ and the five-brane deformations.
However, $\tilde \Omega$ transforms under $\cL^{-1}$ since the scaling-weight 
of $\Delta$ is not entirely canceled.  In other words, if one insists on 
holomorphicity in the coordinates of $\hat Z_3$ one can either work with $\hat \Omega$
which has zeros along $E$, or with $\tilde \Omega$ which transforms also
under $\cL^{-1}$ and thus is not a three-form. A natural global definition
of the non-holomorphic $\Omega'$ is then given by the residue integral
\beq \label{globalOmega'}
  \Omega' = \int_{\epsilon_1}\int_{\epsilon_2} \frac{\Delta}{P Q} \frac{h_i}{l_i} \frac{|l_i|}{|h_i|}\ .
\eeq
Note that this expression is invariant under phase transformations of the 
global projective coordinates $(l_1,l_2)$ and $x_k$ entering the constraints $h_i$ as in \eqref{eq:blowup}.
In contrast $\Omega'$ would scale under real scalings of the global coordinates. However, 
these are fixed by the conditions defining the K\"ahler volumes. For example, the blow-up volume $v_{\rm bu}$
fixes the real scalings of the $(l_1,l_2)$ 
\beq
   |l_1|^2 + |l_2|^2 = v_{\rm bu}\ .
\eeq
Similarly, the real scalings of the $h_i(x)$ are fixed by the definitions
of the K\"ahler moduli of $Z_3$. It would be very interesting to check 
if the conjecture \eqref{globalOmega'} for a global $\Omega'$ 
can be used to derive Picard-Fuchs equations for $\hat Z_3$, probably including anti-holomorphic derivatives.

Finally, let us turn to the definition of the non-K\"ahler form 
$J'$. In a patch $\hat U^{(1)}_{\epsilon}$ it is natural to identify  
\beq
  \hat U^{(1)}_{\epsilon}:\quad J' = \pi^* J  - v_{\rm bu} (1-|y_1|) \Theta\ ,
\eeq
which agrees with $\hat J$ given in \eqref{def-hatJ} up to the logarithmic singularity along $E$.
The non-trivial torsion classes from $dJ'$ are again $\cW_3,\cW_4$.
Since we expect the non-K\"ahlerness to be localized near $E$ one has to have 
\beq
  \hat Z_3 - \cup \hat U_{2 \epsilon}:\qquad J' = \pi^* J\ .
\eeq
This condition is indeed satisfied for any ansatz for $J'$ which involves the curvature 
$\Theta$ of $\cL$, since $\Theta$ vanishes outside a patch covering $E$. 
For a blow-up of a curve wrapped by a holomorphic five-brane one infers 
the flux $H_3'$ using $H'_3 = i (\bar \partial - \partial) J'$. $H'_3$ is 
a differential form on all of $\hat Z_3$, and appears in the superpotential 
\beq \label{SU(3)super}
   W = \int_{\hat Z_3} (H_3' + idJ') \wedge \Omega'\ ,
\eeq
which is now valid also for complex structure variation yielding a setup departing from a 
supersymmetric configuration. 
Clearly, by construction one can use \eqref{match_super} to equate the superpotential on the 
open manifold $\hat Z_3 - E$ with the expression \eqref{SU(3)super}.

Let us close this section by noting that the above construction should be considered 
as a first step in finding a fully back-reacted solution of the theory which 
dissolves the five-brane into flux. It will be interesting to extend these considerations 
to include the remaining supersymmetry conditions of \cite{Strominger:1986uh} which are not encoded by a superpotential. 
In particular, this requires a careful treatment of the dilaton $\phi$ and the warp factor already in $Z_3$, 
where $e^{\phi}$ becomes infinite near the five-brane. 

\section{Conclusion and Outlook} 

In this work we studied the dynamics  
of five-brane wrapped on curves in a compact Calabi-Yau threefold $Z_3$. Our focus was on 
NS5-branes and D5-branes in $\cN=1$ heterotic and 
orientifold compactifications. In these setups  
five-branes can be included if global tadpoles are 
canceled by a non-trivial bundle or orientifold five-planes.
Since five-branes are represented by delta-currents, these 
tadpole conditions are studied in cohomology of currents which naturally extends 
the standard cohomology of forms by singular forms.
 
While these global consistency conditions have to be always satisfied,
the five-brane curve can be deformed within its cohomology class. In 
a fixed background geometry there is an infinite set of 
deformations of the embedding curve. Only a finite set of 
fields can be massless, while most fields are massive 
with a leading scalar potential computed in \cite{Simons,McLean}, or 
a higher order scalar potential. We have shown 
that at leading order even on this infinite space one finds 
a superpotential which together with the 
field space metric determines the potential.
The superpotential is given by a chain integral 
with a three-chain ending on the curve wrapped by the 
five-brane. Higher order obstructions can be studied 
by computing the superpotential directly as a function 
of the complex structure deformations of $Z_3$ and of a
finite number of brane deformations which become 
massless at special loci.

The aim of this work was to study the 
deformations of the five-brane curve $\Sigma$ in $Z_3$ by replacing the setup 
with a new geometry $\hat Z_3$ which is no longer Calabi-Yau. 
The new geometry $\hat Z_3$ is obtained by blowing up $Z_3$ along the curve $\Sigma$. 
This space has a negative first Chern class proportional to the 
new exceptional divisor which is a ruled surface over $\Sigma$. 
We have argued that the $\cN=1$ superpotential 
can be computed by studying the geometry of $\hat Z_3$ together 
with a background three-form flux indicating the presence of the 
five-brane. A crucial point was to realize that certain deformations 
of the five-brane become new complex structure deformations of the 
blow-up space $\hat Z_3$. To make this map more precise we 
had to recall some basic facts about deformation spaces
and their obstructions, and to develop a picture how obstructed 
deformations can be included by studying complex curves 
moving with the Calabi-Yau constraint. We considered brane excitations 
that are only generically obstructed but become
massless at some point in the complex structure moduli space of $Z_3$.

While the proposal to study deformations of holomorphic curves
in a complex variety $Z_3$ by blowing them up to a rigid divisor
and by studying the complex structure deformation of the blown
up manifold ${\hat Z}_3$ is very general,
we obtained the corresponding Picard-Fuchs equations for the
deformation problem in some generality for hypersurfaces
in toric varieties in the presence of toric branes.
For the concrete computations we focused on
rational curves in the mirror quintic threefold and
the mirror of the degree-18 hypersurface in $\bbP^4(1,1,1,6,9)$.
As we noted in these examples there exist maps from
the brane on complex higher genus curves, which are
determined by a complete intersection of two holomorphic
constraints in the Calabi-Yau space $P=0$, to generically obstructed
configurations such as branes on rational curves or the involution
brane studied in~\cite{Walcher}. For example one can identify the
special locus in  the deformation space where the complex
higher genus curve degenerates and coincides with the rational
curves wrapped by the five-brane.
Away from this locus one has to identify the rational curves
using a non-holomorphic map which has branch cuts and
the corresponding obstructions are encoded in a superpotential
calculable on $\hat{Z}_3$. In particular we found that the discrimante components
of the Picard-Fuchs system for $\hat{Z}_3$ describe
these special loci. In order to understand the properties
of the superpotential near these loci in more detail
one needs to indentify the flat coordinates in
the open moduli space.

In the final section of this paper we made a proposal 
for an $SU(3)$ structure on the blow-up space $\hat Z_3$.
Firstly, we argued that in order to define a no-where 
vanishing $(3,0)$-form $\Omega'$ one has to cancel the zeros 
of the pull-back form $\hat \Omega$ along the exceptional 
divisor $E$. This could only be done by introducing a 
non-holomorphic $\Omega'$ which is no longer closed 
near $E$. In addition to a local patchwise definition 
we proposed a global residuum expression for $\Omega'$
if $\hat Z_3$ can be realized in a toric ambient space.
Secondly, we recalled that in a local back-reacted geometry 
with a five-brane, the exterior derivative of the three-form 
flux $\hat H_3$ has a delta-pole near $E$.
Using the supersymmetry conditions this implies that the 
Laplacian of the original K\"ahler form is also singular, and 
we have argued that the extension of the K\"ahler form to 
$\hat Z_3$ thus has a logarithmic singularity. To smooth out 
this singularity we introduced a new $(1,1)$ non-K\"ahler-form $J'$ by 
locally removing the logarithmic singularity. Similarly, one 
defines a smoothed out three-form flux $H_3'$. This 
procedure was motivated by asserting a cancellation 
of poles and zeros such that the superpotential takes the 
form $\int_{\hat Z_3}  (H_3' + i dJ') \wedge \Omega' $.  
Clearly, this should be only considered as a first step in finding 
a fully back-reacted supersymmetric vacuum solution. Of crucial importance 
is the careful treatment of the varying dilaton and warp-factor.

Let us close by listing a number of directions which 
appear to be of interest for future investigations:
\begin{itemize}
	\item  It will be interesting to gain a more complete 
	picture of the open-closed field space as analyzed by 
	the blow-up space $\hat Z_3$. In technical terms, it would 
	be desirable to find a general method to fix the 
	symplectic basis of three-forms on $\hat Z_3$ which 
	allows an identification of the open-closed mirror map 
	similar to the Calabi-Yau threefold case.
	 
	\item In addition to the superpotential it is of 
	crucial importance to compute the remaining $\cN=1$ characteristic 
	data. It was shown in \cite{Grimm:2008dq} that the K\"ahler potential mixes 
	the deformations of a D5-brane curve $\Sigma$ with the K\"ahler moduli 
	of $Z_3$. There exists an analogous leading order expression for 
	the K\"ahler potential for $\hat Z_3$. One might hope that higher order 
	corrections enter through the new periods on $\hat Z_3$. For heterotic 
	five-branes, such an assertion can be motivated by using heterotic/F-theory duality. 
	
	\item It is important to stress that so far independent checks 
	using a direct A-model computation for the compact disk instantons with open moduli are still
	lacking. It would be of crucial importance to perform such computations, e.g.~in order to check the compact disk 
	invariants listed in app. \ref{instantonsLV11169}. 
	Moreover, note that the computation of disk instanton numbers in the 
	superpotential is only the 
	first step in exploiting the power of mirror symmetry. 
	One can attempt to compute amplitudes of higher genus 
	and with more boundaries. It would be of interest to investigate 
	if this can be done on the blow-up space $\hat Z_3$. 
	Comparing these results in a local limit with \cite{Bouchard:2007ys} will 
	give further indication for an existence of a duality between 
	the setups. 
  
  \item It is important to note that in the blow-up procedure one finds  
   new fields, which measure the size of the blow-up $\bbP^1$'s. In a 
   physical duality these need to be identified with new quantum degrees of 
   freedom. In the heterotic compactifications it is natural to identify 
   these fields with the positions of the heterotic five-branes in the 
   interval of heterotic M-theory. It would be interesting to make this 
   map explicit and clarify the interpretation of the blow-up mode in the 
   Type II setups.

   \item The construction of a non-K\"ahler structure on blow-up spaces 
   is also of importance in F-theory compactifications. 
   In particular, one expects that one is able to define an $SU(4)$ structure
   on the blow-up of a Calabi-Yau fourfold along a four-cycle. A special class of 
   these non-K\"ahler deformations are of importance when studying the 
   F-theory effective action using an M-theory lift as in \cite{Grimm:2010ez,Grimm:2010ks}.  
   It was recently argued \cite{Grimm:2010ez} that the non-K\"ahler deformations lift 
   to massive $U(1)$ vector fields in the four-dimensional effective theory. 
\end{itemize}

\subsection*{Acknowledgments}

We gratefully acknowledge discussions with B.~Haghighat, H.~Hayashi, S.~Katz, D.~L\"ust, E.~Palti, M.~Poretschkin, C.~Vafa, J. Walcher, T.~Weigand, E.~Witten, E.~Zaslow and especially D.~Huybrechts for many helpful explanations and comments. A.K. and D.K.~thank the Simons Center for Geometry and Physics for hospitality. T.G. and D.K.~would like to thank the KITP for hospitality. This work was supported in parts by the European Union 6th framework program MRTN-CT-2004-503069 ``Quest for unification'', MRTN-CT-2004-005104 ``ForcesUniverse'', MRTN-CT-2006-035863 ``UniverseNet'', SFB-Transregio 33 ``The Dark Universe'' by the DFG.  The work of D.K.\ is supported by the German Excellence Initiative via the graduate school
``Bonn Cologne Graduate School" and a scholarship of the ``Deutsche Telekom Stiftung". 

\begin{appendix}

\section{Calculation of the F-term Potential}
\label{app:potcalc}

This section provides the necessary background to perform the calculation of the F-term potential \eqref{eq:Ftermpot}. The following calculation extends the analysis made in \cite{Grimm:2008dq} for deformations associated to holomorphic sections $H^0(\Sigma,N_{Z_3}\Sigma)$ to the case of the infinite dimensional space of deformations $\mathcal{C}^{\infty}(\Sigma,N_{Z_3}\Sigma)$. In particular, the K\"ahler metric for the open string deformations is accordingly generalized.

First we need the general form of the K\"ahler potential of the D5-brane action that is given by \cite{Grimm:2008dq}
\beq \label{eqn:kaehler-pot}
    K =-\ln\big[ -i\int\Omega\wedge\bar\Omega \Big]+K_q\ ,\qquad  K_q=-2\ln\big[
      \sqrt{2}e^{-2\phi}\mathcal V \Big]\ ,
\eeq
that immediately implies that 
\begin{equation} \label{eq:e^K}
e^K=\frac{ie^{4\phi}}{2\mathcal{V}^2\int\Omega\wedge\bar\Omega}.
\end{equation}
In order to evaluate the K\"ahler metric the potential $K$ has to to be expressed as a function of the $\mathcal{N}=1$ complex coordinates. For the purpose of our discussion in section \ref{N=1onalldef} we only need the part of the K\"ahler metric for the open string deformations $u^a$. 

We straight forwardly extend the K\"ahler metric deduced in \cite{Grimm:2008dq} for the fields associated to $H^0(\Sigma,N_{Z_3}\Sigma)$ to the infinite dimensional space $\mathcal{C}^{\infty}(\Sigma,N_{Z_3}\Sigma)$.  This is possible since the condition of holomorphicity of sections does not enter the calculations of \cite{Grimm:2008dq}. Thus, we obtain the inverse K\"ahler metric
\beq \label{eqn:Kaehlerpart}
       K^{a\bar b}=2\mu_5^{-1}e^{-\phi}\mathcal{G}^{a\bar b}\,
\eeq
cf.~appendix C of \cite{Grimm:2008dq} for the process of inversion of the full K\"ahler metric, where the matrix $\mathcal{G}^{a\bar b}$ is the inverse of 
\beq
	\mathcal{G}_{a\bar b}=\frac{-i}{2\mathcal{V}}\int_{\Sigma}s_a\lrcorner \bar{s}_{\bar b}\lrcorner (J\wedge J).
\eeq 
Here we introduce a basis $s_a$ of $\mathcal{C}^{\infty}(\Sigma,N_{Z_3}\Sigma)$ so that a generic section $s$ enjoys the expansion $s=u^as_a$. 
Next we Taylor expand the superpotential \eqref{eq:chain} in the open string deformations $u^a$ around the supersymmetric vacuum of the holomorphic curve $\Sigma=\Sigma_0$ as
\beq
\displaystyle
W_{\rm brane}=\displaystyle\int_{\Gamma_u}\Omega\equiv\int_{\Sigma_{\rm rev}}^{\Sigma+\delta \Sigma_u}\Omega=\int_{\Gamma_0}\Omega+\int_{\Sigma}u\lrcorner \Omega+\frac12 u^au^b\int_{\Sigma}s_a\lrcorner d s_b\lrcorner\Omega)+\mathcal{O}(u^3)\,.
\eeq
To evaluate the derivatives $\frac{\partial^k}{\partial^k u^a} W_{\rm brane}\vert_{u=0}$ we use that for every $\frac{\partial}{\partial u^a}$ the Lie-derivative $\mathcal{L}_{s_a}=d s_a\lrcorner +s_a\lrcorner d$ acts on the integrand $\Omega$, where we further denote the interior product with a vector $s_a$ by $s_a\lrcorner$. In addition we use that on the holomorphic curve $\Sigma$ there are no $(2,0)$-forms such that the linear term in $u^a$ in the Taylor expansion vanishes identically. Then the derivative with respect to $u^a$ is obtained as
\beq
	\partial_{u_a}W_{\rm brane}=-\mu_5\int_{\Sigma}\bar\partial s\lrcorner s_a\lrcorner \Omega
\eeq
where we perform a partial integration on $\Sigma$. In addition, we rescale the superpotential by the D5-brane charge, $W_{\rm brane}\mapsto \mu_5 W_{\rm brane}$, as in \cite{Grimm:2008dq}. Now we can calculate the F-term potential 
\beq
	V=\frac{2\mu_5}{e^{\phi}}e^K\int_{\Sigma}\bar\partial s\lrcorner s_a\lrcorner\Omega\ \mathcal{G}^{a\bar b}\int_{\Sigma}\partial \bar{s}\lrcorner \bar{s}_{\bar b}\lrcorner\bar{\Omega}\,.
\eeq
To further evaluate this we have to rewrite the matrix $\mathcal{G}_{a\bar b}$ as follows.
Consider the integral 
\begin{equation}
 	I_{a\bar{b}}:=\int_\Sigma \left(s_a\lrcorner\Omega\right)_{ij}\left(\bar{s}_{\bar{b}}\lrcorner\bar{\Omega}\right)^{ij}\iota^\ast\left(J\right).
\end{equation}
The contracted indices $i$, $j$ denote the coordinates on $Z_3$, one of which is tangential and two are normal to $\Sigma$. 
Then using $\bar{s}_{\bar a}\lrcorner\Omega=0$, $\Omega\wedge\bar\Omega=\tfrac{\int\Omega\wedge\bar\Omega}{6\mathcal V}J^3$ where $\mathcal{V}$ denotes the compactification volume and the rule $s_a\lrcorner (\alpha\wedge\beta)=(s_a\lrcorner\alpha)\wedge\beta+(-1)^p\alpha\wedge (s_a\lrcorner \beta)$ for a $p$-form $\alpha$ we can rewrite this as
\begin{eqnarray}
 	I_{a\bar{b}}&=&-\int_\Sigma \left(s_a\lrcorner\bar{s}_{\bar{b}}\lrcorner\left(\Omega\wedge\bar{\Omega}\right)\right)_{ij}^{ij}\iota^\ast\left(J\right)=-\tfrac{\int\Omega\wedge\bar\Omega}{6\mathcal V}\int_\Sigma\left(s_a\lrcorner s_{\bar b}\lrcorner\left(J^3\right)\right)_{ij}^{ij}\iota^\ast\left(J\right)\nonumber\\
&=&-\tfrac{\int\Omega\wedge\bar\Omega}{6\mathcal V}\int_\Sigma\left[3\left( s_a\lrcorner s_{\bar b}\lrcorner J\right)J^2-6\left(\bar{s}_{\bar b}\lrcorner J\right)\wedge \left(s_a\lrcorner J\right)\wedge J\right]_{ij}^{ij}\iota^\ast\left(J\right)\nonumber\\
&=&\tfrac{\int\Omega\wedge\bar\Omega}{4\mathcal V}\int_\Sigma s_a\lrcorner s_{\bar b}\lrcorner J^2=\tfrac{i\int\Omega\wedge\bar\Omega}{2}\mathcal{G}_{a\bar b}.
\end{eqnarray}
Consequently, introducing the abbreviation $\Omega_a=s_a\lrcorner \Omega$ we can write the matrix $\mathcal{G}_{a\bar b}$ as
\beq
	\mathcal{G}_{a\bar b}=\frac{-i}{2\mathcal{V}}\int_{\Sigma}s_a\lrcorner \bar{s}_{\bar b}\lrcorner (J\wedge J)=\frac{-2i}{\int\Omega\wedge \bar\Omega}\int_{\Sigma}(\Omega_a)_{ij}(\bar{\Omega}_{\bar b})^{ij} \iota^\ast(J)\,.
\eeq
Next we use the basis $\Omega_a$ of $\Omega^{(1,0)}(\Sigma,N_{Z_3}\Sigma)$ to expand the section
\begin{equation} \label{eq:basisexpansion}
	\bar\partial s\lrcorner J=c^{\bar b}\bar\Omega_{\bar b}\,.
\end{equation}
The coefficients $c^{\bar{b}}$ are determined by contraction of \eqref{eq:basisexpansion} with $\Omega_{a}^{\bar\imath\bar\jmath}$, where we have raised the form indices using the hermitian metric on $Z_3$. This way we obtain a function on $\Sigma$ that we can integrate over $\Sigma$ using the volume form $\vol_{\Sigma}=\iota^*(J)$ to determine the $c^{\bar b}$ so that
\begin{equation} 
	\bar\partial s\lrcorner J=\frac{-2i}{\int\Omega\wedge \bar\Omega}\,\mathcal{G}^{a\bar b}\,\bar{\Omega}_{\bar b}\int_{\Sigma}\bar{\partial} s\lrcorner \Omega_{a}\,.
\end{equation} 
With this expansion we immediately obtain the desired form of the F-term superpotential
\begin{equation}
 	V=\frac{2e^K\mu_5}{e^{\phi}}\int_{\Sigma}\bar\partial s\lrcorner \big(\Omega_a\mathcal{G}^{a\bar b}\int_{\Sigma}\partial \bar{s}\lrcorner \bar{\Omega}_b\big)=\frac{-\mu_5e^{3\phi}}{2\mathcal{V}^2}\int_{\Sigma}\bar\partial s\lrcorner \partial \bar s\lrcorner J
	= \frac{\mu_5e^{3\phi}}{2\mathcal{V}^2}\int_{\Sigma}||\bar\partial s||^2 \iota^\ast(J)\,.
\end{equation}
This perfectly matches the potential \eqref{eq:VDBI} obtained from the reduction of the DBI-action of the D5-brane on $\Sigma$ 

\section{A Local Study of the Blow-Up Geometry $\hat{Z}_3$}
\label{App:Local}

In this appendix we study the geometry of the blow-up $\hat{Z}_3$ in more detail in a local analysis. The obtained results provide insights in the blow-up process that immediately apply for the global discussion of section \ref{5braneblowupsanddefs} and allow for a derivation of the expressions used for $\hat{Z}_3$ as a complete intersection and in particular $\hat{\Omega}$ as a residue integral.

The following discussion bases on the general lore in algebraic geometry that the process of blowing up is local in nature, i.e.~just affects the geometry near the subvariety $\Sigma$ which is blown-up leaving the rest of the ambient space invariant. Thus the geometrical properties of the blow up geometry can be studied in a completely local analysis in an open neighborhood around $\Sigma$. In particular, this applies for the case at hand, the blow-up of the curve $\Sigma$ in the compact Calabi-Yau threefold $Z_3$ into a divisor $E$ in the threefold $\hat{Z}_3$.

Starting from a given open covering of $Z_3$ by local patches $U_k\cong \C^3$ we choose a neighborhood $U$ centered around the curve $\Sigma$. Thus, this local patch can be modeled by considering just $\C^3$ on which we introduce local coordinates $x_1$,$x_2$, $x_3$. The holomorphic three-form $\Omega$ on $Z_3$ takes then simply the local form $\Omega=dx_1\wedge dx_2\wedge dx_3$. The curve $\Sigma$ is accordingly described as the complete intersection
\begin{equation}
 	D_1\cap D_2=\{h_1(x_i)=0\}\cap\{h_2(x_i)=0\}\,,
\end{equation}
for two given polynomials $h_i$ with corresponding divisor classes $D_i$ in $\C^3$. 

To construct the blow-up along $\Sigma$, denoted by $\hat{\C}^3$, we have to consider the new ambient space of the projective bundle $\mathcal{W}=\P(\mathcal{O}(D_1)\oplus\mathcal{O}(D_2))$. This is locally of the form ``$\C^3\times \P^1$'' as necessary for the blow-up procedure described in standard textbooks, see e.g.~p. 182 and 602 of \cite{Griffiths}. Next we introduce homogeneous coordinates $(l_1,l_2)$ on the $\P^1$ to obtain the blow-up $\hat{\C}^3$ as the hypersurface 
\begin{equation} \label{eq:blowuplocal}
 	Q\equiv l_1h_2-l_2h_1=0 
\end{equation}
in $\cW$ as before in \eqref{eq:blowup}. 

In the following we construct and study the pullback form $\hat{\Omega}=\pi^*(\Omega)$ that is a section of the canonical bundle $K\hat{\C}^3$, cf.~p. 187 of \cite{Griffiths}. To simplify the calculations we first perform a coordinate transformation to coordinates $y_i$ such that $y_1=h_1(x_i)$, $y_2=h_2(x_i)$ and $y_3=x_j$ for appropriate\footnote{This choice is fixed by the inverse function theorem stating that for every point $x_0\in \C^3$ with $(\partial_k h_1\partial_l h_2-\partial_l h_1\partial_k h_2)\vert_{x_0}\neq0$ for $k,l\neq j$, there exists a local parameterization of $\cC$ near $x_0$ as a graph over $x_j$. In particular, the blow-up is not independent of the coordinates used, cf.~p.~603 of \cite{Griffiths}.} $j$. For notational convenience we relabel the coordinates such that $y_3=x_3$. Thus, we obtain
\begin{equation}
 	Q=l_1y_2-l_2y_1\,,\qquad\Omega=\det{J}^{-1}dy_1\wedge dy_2\wedge dy_3
\end{equation}
in the coordinates $y_i$, where $J=\frac{\partial y_i}{\partial x_j}$ denotes the Jacobian of the coordinate transformation that is generically non-zero by assumption of a complete intersection $\Sigma$.

Now we perform the blow-up on the two local patches on the $\P^1$-fiber of $\mathcal{W}$ that are defined as usual by $U_i=\{l_i\neq0\}$ for $i=1,2$. In the patch $U_1$, for example, we introduce coordinates $z^{(1)}_1$, $z^{(1)}_2$ on $\hat{\C}^3$ as
\begin{equation} \label{eq:coordsU1}
 	z^{(1)}_1=y_1\,,\qquad z^{(1)}_2=-\frac{l_2}{l_1}=\frac{y_2}{y_1}\,,\qquad z^{(1)}_3=y_3
\end{equation}
which allows us to evaluate the pullback map as
\begin{eqnarray}\label{eq:OmegaU1}
 	\pi^*(\Omega)&=&\det J^{-1}\pi^*(dy_1\wedge dy_2\wedge dy_3)=\det J^{-1}dz^{(1)}_1\wedge d(z^{(1)}_2z^{(1)}_1)\wedge dz_3^{(1)}\nonumber\\&=& z_1^{(1)}\det J^{-1}dz^{(1)}_1\wedge dz^{(1)}_2\wedge dz_3^{(1)}\,.
\end{eqnarray}
From this expression we can read off the canonical bundle $K\hat{\C}^3$, cf.~p. 608 of \cite{Griffiths}, by determining the zero-locus of $\pi^*(\Omega)$. Indeed we obtain $K\hat{\C}^3=E$ as mentioned in section \ref{geometricblowups} and in \cite{Grimm:2008dq} since $z^{(1)}_1=0$ describes the exceptional divisor $E$ in $U_1$ and $\det J$ is non-zero by assumption. 
Analogously, we obtain a similar expression in the patch $U_2$ for local coordinates
\begin{equation} \label{eq:coordsU2}
 	z^{(2)}_1=\frac{l_1}{l_2}=\frac{y_1}{y_2}\,,\qquad z^{(2)}_2=y_2\,,\qquad z^{(2)}_3=y_3
\end{equation}
that reads
\begin{equation} \label{eq:OmegaU2}
 	\pi^*(\Omega)=z_2^{(2)}\det J^{-1}dz^{(2)}_1\wedge dz^{(2)}_2\wedge dz_3^{(2)}\,,
\end{equation}
which as well vanishes on $E$ since $E=\{z^{(2)}_2=0\}$ in $U_2$. Thus we extract the transition functions $g_{ij}$ of $\mathcal{O}(E)$ on $U_1\cap U_2$ to be given as
\begin{equation} 
 	g_{ij}=z^{(j)}_i=\frac{y_i}{y_j}=\frac{l_i}{l_j}
\end{equation}
which reflects the fact that $E=\P(O(D_1)\oplus O(D_2))$ over $\Sigma$ with fiber $\P^1$. Additionally, the transition functions $g_{ij}$, when restricted to $E$, are just those of $\mathcal{O}_{\P^1}(-1)$ on the $\P^1$ fiber in $E$ and thus we obtain $\mathcal{O}(E)\vert_E=\mathcal{O}(-1)$ as used in section \ref{su3structur} and in \cite{Grimm:2008dq,Grimm:2009ef,Grimm:2009sy}.

Let us now take a different perspective on the pullback-form $\pi^*(\Omega)$ of the blow-up $\hat{\C}^3$ that is more adapted for the global geometry of $\hat{Z}_3$ as a complete intersection $P=Q=0$ in \eqref{eq:blowup}. The key point will be the description of the blow-up $\hat{\C}^3$  as the hypersurface \eqref{eq:blowuplocal} in $\mathcal{W}$ that will allow for a residue integral representation of $\pi^*(\Omega)$. In particular, the advantage of this residue expression in contrast to the local expressions \eqref{eq:OmegaU1}, \eqref{eq:OmegaU2} is the fact that it can straight forwardly be extended to a global expression on $\hat{Z}_3$ as used in \eqref{eq:ResZhat}. 

Let us start with an ansatz $\hat{\Omega}$ for $\pi^*(\Omega)$,
\begin{equation} \label{eq:OmegaAnsatz}
	\hat{\Omega}=\int_{S^1}A(x_i,l_j)\frac{dx_1\wedge dx_2\wedge dx_3\wedge\Delta_{\P^1}}{Q}\,,
\end{equation}
where $\Delta_{\P^1}=l_1dl_2-l_2dl_1$ is the measure on $\P^1$ obtained from \eqref{eq:MeasurePn} and $S^1_Q$ denotes a loop in $\mathcal{W}$ centered around $Q=0$. The function $A(x_i,l_j)$ of the coordinates $x_i$, $l_j$ is fixed by its scaling behavior w.r.t.~the $\C^*$-action $(l_1,l_2)\mapsto \lambda(l_1,l_2)$. Since $Q\mapsto \lambda Q$ and $\Delta_{\P^1}\mapsto\lambda^2\Delta_{\P^1}$ we demand $A(x_i,\lambda l_j)=\lambda^{-1}A(x_i,l_j)$, i.e.~it is a section of $\mathcal{O}(E)$. Thus we make the ansatz 
\begin{equation}
 	A(x_i,l_j)=a_1\frac{h_1}{l_1}+a_2\frac{h_2}{l_2}=-a_1\frac{Q}{l_1l_2}+(a_2+a_1)\frac{h_2}{l_2}=a_2\frac{Q}{l_1l_2}+(a_1+a_2)\frac{h_1}{l_1}\,,
\end{equation}
which we insert in $\hat{\Omega}$ of \eqref{eq:OmegaAnsatz} to obtain
\begin{equation} \label{eq:HatOmegaLocalapp}
 	\hat{\Omega}=(a_1+a_2)\int_{S^1_Q}\frac{h_i}{l_i}\frac{dx_1\wedge dx_2\wedge dx_3\wedge\Delta_{\P^1}}{Q}\,,\quad i=1,2\,.
\end{equation}
Now we show that this precisely reproduces the local expressions \eqref{eq:OmegaU1}, \eqref{eq:OmegaU2}, this way fixing the free parameters $a_i$.

Let us perform the calculations in the local patch $U_1$. Then the measure on $\P^1$ reduces to $\Delta_{\P^1}=(l_1)^2dz_2^{(1)}$ with $z^{(1)}_2=l_2/l_1$ and we obtain, after a change of coordinates to $y_i$, 
\begin{eqnarray}
 	\hat{\Omega}&=&(a_1+a_2)\int_{S^1_Q}\det{J}^{-1} \frac{y_il_1}{l_i}\frac{dy_1\wedge dy_2\wedge dy_3\wedge dz^{(1)}_2}{-z^{(1)}_2 y_1+y_2}\\
		    &=&-(a_1+a_2) (\det{J}^{-1}\frac{y_il_1}{l_i}dy_1\wedge dy_3\wedge dz^{(1)}_2)\vert_{y_2=z^{(1)}_2y_1}\,.
\end{eqnarray}
In the last line we indicated that the residue localizes on the locus $Q=0$. This implies that $z^{(1)}_2=\frac{y_2}{y_1}=\frac{l_2}{l_1}$ as before in \eqref{eq:coordsU1} and we put $z^{(1)}_1=y_1$, $z^{(1)}_3=y_3$ as well. Next, we evaluate $\hat{\Omega}$ for $i=1$ for which the prefactor reduces to $\frac{y_1l_1}{l_1}=z^{(1)}_1$, and for $i=2$, for which we get $\frac{y_2l_1}{l_2}=z^{(1)}_1$. Thus, the two expressions in \eqref{eq:HatOmegaLocalapp} for $i=1,2$ yield one unique form $\hat{\Omega}$ after evaluating the residue integral,
\begin{equation}
 	\hat{\Omega}=(a_1+a_2)\det J^{-1}z^{(1)}_1dz_1^{(1)}\wedge dz_2^{(1)}\wedge dz_3^{(1)}\,,
\end{equation}
which agrees with $\pi^*(\Omega)$ on $U_1$ for $a_1+a_2=1$. Thus, we propose the global residue expression
\begin{equation} \label{eq:RedidueHatC3}
 	\hat{\Omega}\equiv\pi^*(\Omega)=\int_{S^1_Q}\frac{h_1}{l_1}\frac{\Omega\wedge\Delta_{\P^1}}{Q}=\int_{S^1_Q}\frac{h_2}{l_2}\frac{\Omega\wedge\Delta_{\P^1}}{Q}
\end{equation}
The transition from the local chart $U\cong\C^3$ on $Z_3$ to the global threefold $Z_3$ is then effectively performed by replacing the  three-form $\Omega=dx_1\wedge dx_2\wedge dx_3$ on $\C^3$ by the residue integral $\Omega=\text{Res}_P(\frac{\Delta_{\P_{\Delta}}}{P})$ of \eqref{eq:residueZ3} on $Z_3$ in \eqref{eq:RedidueHatC3}. This way the local analysis motivates and proves the global expression of \eqref{eq:ResZhat} for the pullback form $\hat{\Omega}\equiv\pi^*(\Omega)$ of the holomorphic three-form $\Omega$ to $\hat{Z}_3$ and its properties mentioned there.

We conclude this analysis by briefly checking that this residue also reproduces $\pi^*(\Omega)$ on $U_2$ as evaluated in \eqref{eq:OmegaU2}. We set $z^{(2)}_1=\frac{l_1}{l_2}$ for which the measure reduces as $\Delta_{\P^1}=-l_2^2dz_1^{(2)}$ which we readily insert into \eqref{eq:RedidueHatC3} to obtain
\begin{eqnarray}
 	\hat{\Omega}&=&\int_{S^1_Q}\det{J}^{-1}\frac{y_il_2}{l_i}\frac{dy_1\wedge dy_2\wedge dy_3\wedge dz_1^{(2)}}{y_1-z_1^{(2)}y_2}\\
	&=& (\det{J}^{-1}\frac{y_il_2}{l_i}dy_2\wedge dy_3\wedge dz_1^{(2)})\vert_{y_1=z^{(2)}_1y_2}=\det{J}^{-1}z^{(2)}_2  dz_1^{(2)}\wedge dz^{(2)}_2\wedge dz^{(2)}_3\,. 
\end{eqnarray}
Here we again use $z^{(2)}_1=\frac{y_1}{y_2}=\frac{l_1}{l_2}$ on $Q=0$ and introduce the local $z^{(2)}_2=y_2$, $z^{(2)}_3=y_3$ as in \eqref{eq:coordsU2}. The result is in perfect agreement with the local expression \eqref{eq:OmegaU2} on $U_2$.

\clearpage
\section{Compact Disk Instantons on $Z_3(1,1,1,6,9)$}
\label{instantonsLV11169}

\begin{table}[!ht]
\centering
$
\scriptscriptstyle
 \begin{array}{|c|rrrrrrrr|}
\hline
\rule[-0.2cm]{0cm}{0.6cm}  i&\!j=0&\!\!j=1&\!\!j=2&\!\!j=3&\!\!j=4&\!\!j=5&\!\!j=6&\!\!j=7\\
\hline
 0&\!0 &\!\! 1 &\!\! 0 &\!\! 0 &\!\! 0 &\!\! 0 &\!\! 0 &\!\! 0 \\
 1&\!2 &\!\! -1 &\!\! -1 &\!\! -1 &\!\! -1 &\!\! -1 &\!\! -1 &\!\! -1 \\
 2&\!-8 &\!\! 5 &\!\! 7 &\!\! 9 &\!\! 12 &\!\! 15 &\!\! 19 &\!\! 23 \\
 3&\!54 &\!\! -40 &\!\! -61 &\!\! -93 &\!\! -140 &\!\! -206 &\!\! -296 &\!\! -416 \\
 4&\!-512 &\!\! 399 &\!\! 648 &\!\! 1070 &\!\! 1750 &\!\! 2821 &\!\! 4448 &\!\! 6868 \\
 5&\!5650 &\!\! -4524 &\!\! -7661 &\!\! -13257 &\!\! -22955 &\!\! -39315 &\!\! -66213 &\!\! -109367 \\
 6&\!-68256 &\!\! 55771 &\!\! 97024 &\!\! 173601 &\!\! 312704 &\!\! 559787 &\!\! 989215 &\!\! 1719248 \\
 7&\!879452 &\!\! -729256 &\!\! -1293185 &\!\! -2371088 &\!\! -4396779 &\!\! -8136830 &\!\! * &\! * \\
 8&\!-11883520 &\!\! 9961800 &\!\! 17921632 &\!\! 33470172 &\!\! * &\!\! * &\!\! * &\!\! * \\
 9&\!166493394 &\!\! -140747529 &\!\! * &\!\! * &\!\! * &\!\! * &\!\! * &\!\! *\\ 
\hline
\end{array}
$
\caption{$k=0$: Disk instanton invariants $n_{i,k, i+j, k}$ on $Z_3(1,1,1,6,9)$ at large volume. $i$, $k$ label the classes $t_1$, $t_2$ of $Z_3$, where $j$ labels the brane winding. These results for $k=0$ agree with phase $I/II$ in table 5 of \cite{Aganagic:2001nx}. Entries $*$ exceed the order of our calculation.}
\end{table}
\begin{table}[!ht]
\centering
$
\scriptstyle
 \begin{array}{|c|rrrrrr|}
\hline
\rule[-0.2cm]{0cm}{0.6cm}  i&j=1&j=2&j=3&j=4&j=5&j=6\\
\hline
 0& ** & 0 & 0 & 0 & 0 & 0 \\
 1& 300 & 300 & 300 & 300 & 300 & 300 \\
 2& -2280 & -3180 & -4380 & -5880 & -7680 & -9780 \\
 3& 24900 & 39120 & 61620 & 95400 & 144060 & 211800 \\
 4& -315480 & -526740 & -892560 & -1500900 & -2477580 & -3996780 \\
 5& 4340400 & 7516560 & 13329060 & 23641980 & 41421000 & 71240400 \\
 6& -62932680 & -111651720 & -204177600 & -375803820 & -686849280 & * \\
 7& 946242960 & 1707713040 & 3192621180 & * & * & *      \\
\hline
\end{array}
$
\caption{$k=1$: Compact disk instanton invariants $n_{i,k, i+j, k}$ for brane phase $I/II$ on $Z_3(1,1,1,6,9)$ at large volume. $i$, $k$ label the classes $t_1$, $t_2$ of $Z_3$, where $j$ labels the brane winding. The entry $**$ as well as the $j=0$ column could not be fixed by our calculation. In overlapping sectors these results agree with table 3.b of \cite{Alim:2009rf}.}
\end{table}
\begin{table}[!ht]
\centering
$
 \begin{array}{|c|rrrrrr|}
\hline
\rule[-0.2cm]{0cm}{0.6cm}  i&\!j=2&\!\!j=3&\!\!j=4&\!\!j=5&\!\! j=6&\!\! j=7\\
\hline
 0&\! ** &\!\! 0 &\!\! 0 &\!\! 0 &\!\! 0 &\!\! 0 \\
 1&\! -62910 &\!\! -62910 &\!\! -62910 &\!\! -62910 &\!\! -62910 &\!\! -62910 \\
 2&\!778560 &\!\! 1146690 &\!\! 1622580 &\!\! 2206530 &\!\! 2898240 &\!\! 3698010 \\
 3&\! -12388860 &\!\! -20596140 &\!\! -33454530 &\!\! -52626780 &\!\! -80081460 &\!\! -118092960 \\
 4&\! 208471080 &\!\! 368615070 &\!\! 645132360 &\!\! 1103916150 &\!\! 1838367780 &\!\! 2976756210 \\
 5&\! -3588226470 &\!\! -6587809920 &\!\! -12083913000 &\!\! -21840712470 &\!\! * &\!\! * \\
 6&\! 62538887280 &\!\! 117754228980 &\!\! * &\!\! * &\!\! * &\!\! *\\
\hline
\end{array}
$
\caption{$k=2$: Compact disk instanton invariants $n_{i,k, i+j, k}$ for brane phase $I/II$ on $Z_3(1,1,1,6,9)$ at large volume. $i$, $k$ label the classes $t_1$, $t_2$ of $Z_3$, where $j$ labels the brane winding. The entry $**$ as well as the $j=0,\,1$ columns could not be fixed by our calculation. In overlapping sectors these results agree with table 3.c of \cite{Alim:2009rf}.}
\end{table}
\begin{table}[!ht]
\centering
$
 \begin{array}{|c|rrrrrrrr|}
\hline
\rule[-0.2cm]{0cm}{0.6cm}  i&j=0&j=1&j=2&j=3&j=4&j=5& j=6&j=7\\
\hline
  0&0 & -1 & 0 & 0 & 0 & 0 & 0 & 0 \\
  1&** & 2 & 1 & 1 & 1 & 1 & 1 & 1 \\
  2&**& -5 & -4 & -3 & -4 & -5 & -7 & -9 \\
  3&**& 32 & 21 & 18 & 20 & 26 & 36 & 52 \\
  4&**& -286 & -180 & -153 & -160 & -196 & -260 & -365 \\
  5&** & 3038 & 1885 & 1560 & 1595 & 1875 & 2403 & 3254 \\
  6&**& -35870 & -21952 & -17910 & -17976 & -20644 & -25812 & -34089 \\
  7&** & 454880 & 275481 & 222588 & 220371 & 249120 & * & * \\
  8&**& -6073311 & -3650196 & -2926959 & * & * & * & * \\
  9&**& 84302270 & * & * & * & * & * & *\\
\hline
\end{array}
$
\caption{$k=0$: Disk instanton invariants $n_{i,k, i, k+j}$ on $Z_3(1,1,1,6,9)$ at large volume. $i$, $k$ label the classes $t_1$, $t_2$ of $Z_3$, where $j$ labels the brane winding. The results for $k=0$ agree with phase $III$ in table 6 of \cite{Aganagic:2001nx}.}
\end{table}
\begin{table}[!ht]
\centering
$
 \begin{array}{|c|rrrrrrr|}
\hline
\rule[-0.2cm]{0cm}{0.6cm}  i&j=1&j=2&j=3&j=4&j=5& j=6&j=7\\
\hline
 0&\! 0 &\! 0 &\! 0 &\! 0 &\! 0 &\! 0 &\! 0 \\
 1&\! -540 &\! -300 &\! -300 &\! -300 &\! -300 &\! -300 &\! -300 \\
 2&\! 2160 &\! 1620 &\! 1680 &\! 2280 &\! 3180 &\! 4380 &\! 5880 \\
 3&\! -18900 &\! -12960 &\! -12300 &\! -15000 &\! -21060 &\! -31200 &\! -47220 \\
 4&\! 216000 &\! 140940 &\! 126240 &\! 142380 &\! 185280 &\! 261300 &\! 386160 \\
 5&\! -2800980 &\! -1775520 &\! -1535160 &\! -1653900 &\! -2046060 &\! -2750280 &\! -3896760 \\
 6&\! 39087360 &\! 24316200 &\! 20544720 &\! 21489780 &\! 25725600 &\! * &\! * \\
 7&\! -572210460 &\! -351319680 &\! -292072920 &\! * &\! * &\! * &\! * \\
 8&\! 8663561280 &\! * &\! * &\! * &\! * &\! * &\! *\\
\hline
\end{array}
$
\caption{$k=1$: Compact disk instanton invariants $n_{i,k, i, k+j}$ for brane phase $III$ on $Z_3(1,1,1,6,9)$ at large volume. $i$, $k$ label the classes $t_1$, $t_2$ of $Z_3$, where $j$ labels the brane winding. The $j=0$ column could not be fixed by our calculation.}
\end{table}
\begin{table}[!ht]
\centering
$
 \begin{array}{|c|rrrrrr|}
\hline
\rule[-0.2cm]{0cm}{0.6cm}  i&\!j=1&\!j=2&\!j=3&\!j=4&\!j=5&\! j=6\\
\hline
 0&\! 0 &\! 0 &\! 0 &\! 0 &\! 0 &\! 0 \\
 1&\! ** &\! 62910 &\! 62910 &\! 62910 &\! 62910 &\! 62910 \\
 2&\! -430110 &\! -413640 &\! -557010 &\! -836340 &\! -1223730 &\! -1718880 \\
 3&\! 5190480 &\! 3923100 &\! 4415580 &\! 6237810 &\! 9720180 &\! 15561180 \\
 4&\! -76785570 &\! -52941600 &\! -52475850 &\! -65786040 &\! -93752550 &\! -143003760 \\
 5&\! 1227227760 &\! 806981670 &\! 747944550 &\! 869842800 &\! 1154721060 &\! *\\
 6&\! -20387141100 &\! -13027278600 &\! -11592978930 &\! * &\! * &\! * \\
 7&\! 346430247840 &\! * &\! * &\! * &\! * &\! *\\
\hline
\end{array}
$
\caption{$k=2$: Compact disk instanton invariants $n_{i,k, i, k+j}$ for brane phase $III$ on $Z_3(1,1,1,6,9)$ at large volume. $i$, $k$ label the classes $t_1$, $t_2$ of $Z_3$, where $j$ labels the brane winding. The entry $**$ and the $j=0$ column could not be fixed by our calculation.}
\end{table}
\clearpage
\end{appendix}

\providecommand{\href}[2]{#2}\begingroup\raggedright

\end{document}